\documentclass[fleqn,usenatbib]{mnras}

\usepackage[T1]{fontenc}
\usepackage{ae,aecompl}

\usepackage{graphicx}	% Including figure files
\usepackage{amsmath}	% Advanced maths commands
\usepackage{amssymb}	% Extra maths symbols

\title[Chemical Tagging of Clusters in the GALAH Survey]{The GALAH survey: Chemical Tagging of Star Clusters and New Members in the Pleiades}

\author[J. Kos et al.]{
Janez Kos,$^{1}$\thanks{E-mail: janez.kos@sydney.edu.au}
Joss Bland-Hawthorn,$^{1}$
Ken Freeman,$^{2}$
Sven Buder,$^{3}$\newauthor
Gregor Traven,$^{4}$
Gayandhi M. De Silva,$^{5,1}$
Sanjib Sharma,$^{1}$\newauthor
Martin Asplund,$^{2}$
Ly Duong,$^{2}$
Jane Lin,$^{2}$
Karin Lind,$^{3,6}$
Sarah Martell,$^{7}$\newauthor
Jeffrey D. Simpson,$^{8}$
Dennis Stello,$^{7,9,1}$
Daniel B. Zucker,$^{8}$
Toma\v{z} Zwitter,$^{4}$\newauthor
Borja Anguiano,$^{10}$
Gary Da Costa,$^{2}$
Jonathan Horner,$^{11}$
Prajwal R. Kafle,$^{12}$\newauthor
Geraint Lewis,$^{1}$
Ulisse Munari,$^{13}$
David M. Nataf,$^{14}$
Melissa Ness,$^{3}$
Warren Reid,$^{8,15}$\newauthor
Katie Schlesinger,$^{2}$
Yuan-Sen Ting,$^{2}$
Rosemary Wyse$^{14}$
\\
$^{1}$Sydney Institute for Astronomy, School of Physics, A28, The University of
Sydney, NSW, 2006, Australia\\
$^{2}$Research School of Astronomy \& Astrophysics, Australian National University, ACT 2611, Australia\\
$^{3}$Max-Planck-Institut for Astronomy, K\"onigstuhl 17, D-69117 Heidelberg, Germany\\
$^{4}$Faculty of Mathematics and Physics, University of Ljubljana, Jadranska 19, 1000 Ljubljana, Slovenia\\
$^{5}$Australian Astronomical Observatory, North Ryde, NSW 2133, Australia\\
$^{6}$Department of Physics and Astronomy, Uppsala University, Box 516, SE-751 20 Uppsala, Sweden,\\
$^{7}$School of Physics, UNSW, Sydney, NSW 2052, Australia\\
$^{8}$Department of Physics and Astronomy, Macquarie Univeristy, Sydney, NSW 2109, Australia\\
$^{9}$Stellar Astrophysics Centre, Department of Physics and Astronomy, Aarhus University, DK-8000, Aarhus C, Denmark\\
$^{10}$Department of Astronomy, University of Virginia, Charlottesville, VA 22904-4325, USA\\
$^{11}$Computational Engineering and Science Research Centre, University of Southern Queensland, Toowoomba, Queensland 4350, \\Australia\\
$^{12}$International Centre for Radio Astronomy Research (ICRAR), The University of Western Australia (M468), \\35 Stirling Highway, CRAWLEY WA 6009, Australia\\
$^{13}$INAF National Institute of Astrophysics, Astronomical Observatory of Padova, 36012 Asiago, Itay\\
$^{14}$Center for Astrophysical Sciences and Department of Physics and Astronomy, The Johns Hopkins University, \\Baltimore, MD 21218,\\
$^{15}$Western Sydney University, Locked Bag 1797, Penrith South DC, NSW 2751, Australia}

\date{Accepted XXX. Received YYY; in original form ZZZ}

\pubyear{2017}

\begin{document}
\label{firstpage}
\pagerange{\pageref{firstpage}--\pageref{lastpage}}
\maketitle

% Abstract of the paper
\begin{abstract}
The technique of chemical tagging uses the elemental abundances of stellar atmospheres to `reconstruct' chemically homogeneous star clusters that have long since dispersed. The GALAH spectroscopic survey --which aims to observe one million stars using the Anglo-Australian Telescope -- allows us to measure up to 30 elements or dimensions in the stellar chemical abundance space, many of which are not independent. How to find clustering reliably
in a noisy high-dimensional space is a difficult problem that remains largely unsolved. Here we explore {\it t-distributed stochastic neighbour embedding} (t-SNE) -- which identifies an optimal mapping of a high-dimensional space into fewer dimensions -- whilst conserving the original clustering information. Typically,
the projection is made to a 2D space to aid recognition
of clusters by eye.
We show that this method is a reliable tool for chemical tagging because it can:
(i) resolve clustering in chemical space alone,
(ii) recover known open and globular clusters with high efficiency and low contamination, and
(iii) relate field stars to known clusters.
t-SNE also provides a useful visualization of a high-dimensional space.
We demonstrate the method on a dataset of 13 abundances measured in the spectra of 187,000 stars by the GALAH survey. We recover 7 of the 9 observed clusters (6 globular and 3 open clusters) in chemical space with minimal contamination from field stars and low numbers of outliers.
With chemical tagging,
we also identify two Pleiades supercluster members (which we confirm kinematically), one as far as 6$^\circ$ -- one tidal radius away from the cluster centre.
\end{abstract}

% Select between one and six entries from the list of approved keywords.
% Don't make up new ones.

%%%%%%%%%%%%%%%%%%%%%%%%%%%%%%%%%%%%%%%%%%%%%%%%%%

%%%%%%%%%%%%%%%%% BODY OF PAPER %%%%%%%%%%%%%%%%%%

\section{Introduction} \label{sec:intro}

The use of chemical tagging in Galactic archaeology was first proposed by \citet{freeman02}, who suggested that the abundances of elements in stars could be used as unique signatures over their lifetime to
`reconstruct' stellar groups that have long
since dissolved. Theoretical arguments 
indicate that chemical homogeneity (with the exception of light elements)
is guaranteed in open clusters up to $10^5\ \mathrm{M_{\sun}}$ and in globular clusters up to a limit of $10^7\ \mathrm{M_{\sun}}$ \citep{jbh10a}. 
In practice,
the degree of homogeneity may depend on
the initial abundance spread in the collapsing cloud from which the cluster forms \citep{feng14}. But, to
date, essentially all open 
clusters appear to be chemically homogeneous to better than 0.1~dex
\citep{desilva06, sestito07, bovy16}.
%and the vast majority of stars are indeed born in clusters of such mass \citep{fall01, lada03, wong08}. 
Both young and ancient (up to $\sim$9 Gyr) open clusters
appear to be chemically homogeneous
\citep{desilva06,desilva07} indicating that
pollution from the interstellar medium
does not wipe out this information.
For chemical tagging to be feasible for field stars, a large amount of high quality data has to be collected, i.e.
on the order of $10^6$ observed stars and $\sim$30 measured elements \citep{jbh04,ting15}. Indeed, these are the design goals
of the GALAH\footnote{GALAH survey webpage is \texttt{https://galah-survey.org}} survey on the HERMES instrument
at the Anglo-Australian Telescope \citep{barden10,desilva15,martell17}.
This requirement can be much lower for `soft' chemical tagging if there
is additional information (e.g. kinematics,
location) to associate the stars.

To chemically tag stars, one has to 
search for clustering in chemical space ($\mathcal{C}$-space), i.e.
an $N$-dimensional space determined by the
measured number of elemental abundances.
Strictly speaking, these dimensions are 
unlikely to be independent, e.g. iron-peak
elements are strongly coupled. Different elements also experience a different cosmic spread \citep[e.g.][]{bensby14}, so they cannot all be treated equally. \citet{bensby14} only give cosmic spreads for the solar neighbourhood. Sample used in our study covers much larger region, as well as more stellar types. Cosmic spreads that would be useful for our work are therefore largely unknown.
In the GALAH survey, there are up to 30
elements for which abundances can be determined in each
star, but in our study we will concentrate on a smaller
number ($N=13$) of elements with well determined abundances.

How are we to find substructures in a 
high dimensional space?
The human brain is excellent at detecting
clustering in three or fewer dimensions,
but falls short for problems in more dimensions.
Most work to date has focussed on finding clusters in the original $N$-dimensional space. For example, \citet{hogg16} searched for chemical groups in the APOGEE data by the $k$-means algorithm and showed that some clusters correspond to groups in phase space. \citet{blanco15} utilized PCA to distinguish between known clusters.
\citet{jbh10b} used a 
density-based hierarchical clustering
algorithm and introduced the $S$-statistic to show that clustering exists in a simulated dwarf galaxy.
 \citet{mitschang14, quillen15} used a probabilistic approach to resolve chemical groups in a blind chemical tagging study, but were unable to determine whether the groups found were, in fact, co-natal (born together), or simply had nearly identical abundances.  When key chemical signatures can be confined to two or three dimensions, this classification becomes straightforward. \citet{martell16} were able to identify halo stars that originated in globular clusters, and \citet{desilva11} placed Hyades supercluster members in one chemical group. \citet{desilva13} used chemical tagging to relate the Argus association to IC 2391. A more advanced algorithm, that also provides visualisation, was used by \citet{jofre16}, who applied a method of evolutionary trees to stellar abundances and produced a phylogenetic tree for 21 solar twins and the Sun.

One of the most successful methods in recent
years exploits the huge computational power
now available in desktop computers. The
so-called t-distributed stochastic neighbour
embedding (t-SNE) algorithm is a remarkable
technique for reducing the dimensions of
a problem \citep{maaten08}.
That technique embeds each high-dimensional data point into a two dimensional `visualisation' space where `similar' points are kept together and `dissimilar' points are moved apart. Once the problem is reduced into two dimensions, the clustering can be identified by eye. We
find that this method is highly effective
in identifying known and unknown cluster
members. The method has some limitations:
(i) it is a black box that
is difficult to tune or control; and
(ii) the abundance
measurement errors are not used in the 
present application. But the method is
extraordinarily powerful as demonstrated
in recent papers, for example, 
to efficiently identify peculiar stars and stellar populations
in large surveys with a high
level of completeness \citep{matijevic15, lochner16, valentini16, traven16}.

%A problem where clustering in more dimensions has to be found must therefore be solved by a computer following one of the two options (or combination of both): 
 
%clusters are found in a high-dimensional space, or mapping into a low (usually 2) dimensional space is calculated, where clusters can be identified by eye. In this paper we explore the latter option, where mapping and visualization is done by the t-distributed stochastic neighbour embedding (t-SNE) \citep{maaten08}. 

We describe our data in Section \ref{sec:data} and the method in Section \ref{sec:method}. In Section \ref{sec:cluster} we explore the efficiency of our method on 9 known clusters and in Section \ref{sec:new} we present a more detailed analysis of the Pleiades cluster. In Section \ref{sec:dis} we discuss the implications of this work and future development of the field.

\section{The Data}%done
\label{sec:data}
The data set analysed here has
been drawn from three programs:
the GALAH pilot program, the K2-HERMES survey \citep{sharma17}, and the main GALAH survey \citep{desilva15,martell17}. These three programs have different selection functions, but share the same observing procedures, reduction pipeline, and analysis pipeline
\citep{kos17}. Stars from all programs are analysed together so the stellar parameters and the abundances are comparable. 

All stars used in this paper have abundances measured by The Cannon \citep{ness15}, applying a data-driven approach to estimate stellar parameters and abundances using linear algebra to combine the spectral flux for each pixel. The quadratic spectral model of The Cannon was trained on a representative set of spectra, consisting of benchmark stars \citep{heiter15, jofre15}, K2 stars with known seismic gravities \citep{stello16b, stello16} as well as stars with high-quality spectra covering the parameter space. For the training set, stellar parameters and abundances were estimated using the spectrum synthesis code SME \citep{valenti96, piskunov16}.

The complete data set consists of 187,640 stars, mostly dwarfs, observed between November 2013 and January 2016. 15,601 stars have unreliable stellar parameters (e.g. because of a peculiar spectrum, strong cosmic rays, etc.) and were excluded from the study. As this is the first time The Cannon has been used with GALAH data and represents the first internal release of abundances, the uncertainties of the measured abundances have not been validated yet. A map of the observed fields on the celestial sphere is given in \citet{martell17}.

Our data-set consists of abundances of 13 elements (Na, Mg, Al, Si, K, Ca, Sc, Ti, Cr, Fe, Ni, Cu, Ba) representing groups of light, light odd-Z, alpha, iron peak and s-process elements. All abundances were measured by The Cannon. The number of lines used to measure the abundances of each element is given in Table \ref{tab:n}. Stellar parameters ($T_{\mathrm{eff}}$, $\log g$, $[\mathrm{Fe}/\mathrm{H}]$, and radial velocity) were measured by fitting synthetic models of stellar atmospheres to one dimensional GALAH spectra \citep{kos17}. Abundances for all 13 elements were measured for each of 187,640 stars.

\begin{table}
\caption{Number of lines used to measure the abundances of each element.}
\label{tab:n}
\begin{minipage}{0.49\columnwidth}
\begin{tabular}{lc}
\hline
Element&N of lines\\
\hline\hline
Na&3\\
Mg&2\\
Al&4\\
Si&5\\
K&2\\
Ca&5\\
Sc&10\\
\hline
\end{tabular}
\end{minipage}
\begin{minipage}{0.49\columnwidth}
\begin{tabular}{lc}
\hline
Element&N of lines\\
\hline\hline
Ti&20\\
Cr&9\\
Fe&52\\
Ni&7\\
Cu&2\\
Ba&2\\
\hline
 & \\%aligns both tables
\end{tabular}
\end{minipage}
\end{table}

Proper motions from UCAC4 are also available for all stars, and parallaxes from Gaia TGAS for some. For most stars we calculated the photometric distances following \citep{zwitter10} using APASS and 2MASS photometry \citep{martell17} and our stellar parameters.

\section{\lowercase{t}-distributed stochastic neighbour embedding (\lowercase{t}-SNE)}
\label{sec:method}
t-SNE is an algorithm from a family of manifold learning algorithms. It has been extensively used in data science and made a break into astronomy as a classification algorithm \citep{matijevic15, lochner16,  valentini16, traven16}, along with other manifold learning algorithms \citep[e.g.][]{vanderplas09, daniel11, bu14}. We extend its use as a pure manifold learning algorithm to find structure in a 13-dimensional $\mathcal{C}$-space.

t-SNE's input is a set of N high-dimensional objects $\mathbf{x}_1, \ldots, \mathbf{x}_N$. In our case each $\mathbf{x}_i$ will be a collection of 13 abundances for a single star:
\begin{equation}
\mathbf{x}_i=\left( \left[ \frac{\mathrm{Na}}{\mathrm{Fe}} \right]_i, \left[ \frac{\mathrm{Mg}}{\mathrm{Fe}} \right]_i, \ldots, \left[ \frac{\mathrm{Ba}}{\mathrm{Fe}} \right]_i, \left[ \frac{\mathrm{Fe}}{\mathrm{H}} \right]_i \right)
\end{equation}
 Following \citet{maaten08}, we first calculate similarities $\mathbf{p}_{ij}$ of the input set:
\begin{equation}
\mathbf{p}_{ij}=\frac{\mathbf{p}_{i|j}+\mathbf{p}_{j|i}}{2N}, \  \mathbf{p}_{j|i}=\frac{\exp{(-||\mathbf{x}_i-\mathbf{x}_j||^2/2\sigma_i^2)}}{\sum_{k\neq i} \exp{(-||\mathbf{x}_j-\mathbf{x}_k||^2/2\sigma_i^2)}}
\label{eq:p1}
\end{equation}
where $\sigma_i$ is a parameter calculated by t-SNE depending on the perplexity (see \citet{tsne16} for a demonstration of how perplexity works) and local density of the data-set. Distances $||\mathbf{x}_i-\mathbf{x}_j||$ and $||\mathbf{x}_j-\mathbf{x}_k||$ in the original implementation are Euclidean. We modified the code to use Manhattan distances, as they are less sensitive to sporadic outliers. Manhattan distance between two points $\mathbf{p}$ and $\mathbf{q}$ in $n$ dimensions is defined as:
\begin{equation}
d(\mathbf{p}, \mathbf{q})=\parallel \mathbf{p} - \mathbf{q} \parallel_1 = \sum_{i=1}^n |p_i-q_i|
.\end{equation}
We aim to produce a lower dimensional map with objects $\mathbf{y}_1, \ldots, \mathbf{y}_N$ with similarities:
\begin{equation}
\mathbf{q}_{ij}=\frac{(1+||\mathbf{y}_i-\mathbf{y}_j||^2)^{-1}}{\sum_{k\neq i}(1+||\mathbf{y}_k-\mathbf{y}_i||^2)^{-1}}
\label{eq:p2}
\end{equation}

To find the optimal mapping where $\mathbf{q}_{ij}$ reflects $\mathbf{p}_{ij}$ as well as possible, we minimize the Kullback-Leibler divergence:
\begin{equation}
KL(P||Q)=\sum_{i\neq j} \mathbf{p}_{ij} \log{\frac{\mathbf{p}_{ij}}{\mathbf{q}_{ij}}}.
\label{eq:p3}
\end{equation}
Kullback-Leibler divergence is a non-convex function that is minimised by gradient descent initialised randomly. Different runs of t-SNE, even when the same parameters are used, can therefore result in a different mapping. The algorithm is usually run several times and the mapping with the lowest Kullback-Leibler divergence is used.

The scale of the map produced by t-SNE is irrelevant. Only the relative relations between objects and groups on the map hold any information. We therefore refrain from plotting the coordinate system on the maps.

The above algorithm has a time dependence of $\mathcal{O}(N^2)$, because we have to calculate similarities for every pair of objects. This is impractical for most applications, so we employ the Barnes-Hut algorithm to calculate sparse similarities in $\mathcal{O}(N \log(N))$ time \citep{maaten13}. With such optimization we can analyse our whole dataset on an average desktop computer in less than one hour.

t-SNE is presented in a more intuitive way in Appendix {sec:appB}, together with a comparisson with some other algorithms.

\section{Recovering known clusters in chemical space}
\label{sec:cluster}
The GALAH survey targeted only a small number of
individual clusters (47 Tuc, NGC 288, NGC 1851, M 30, $\omega$ Cen, NGC 362, and M 67) as part of the pilot survey. An additional two
clusters (NGC 2516 and the Pleiades) were observed because their stars
happen to fall into the magnitude range of the main survey or K2-HERMES survey. A list of the observed clusters is given in Table \ref{tab:clusters}. In the pilot survey only probable members were observed. To confirm the membership of such stars, as well as members of serendipitously observed clusters, we made additional cuts in radial velocity, position, and in some instances, their proper motion. Details are given in Appendix \ref{sec:membership}. The exceptions are three clusters for which we found members in the literature. There are also three more clusters (Hyades, NGC 2243, and NGC 6362) in the observed fields, but we could not find any of their members among survey stars with valid parameters and abundances.

\begin{table}
\setlength\tabcolsep{2.2pt}
\caption{Clusters with observed members in the GALAH and K2 surveys. Six clusters with no given literature for membership have members identified by us (See Appendix \ref{sec:membership}).}
\begin{tabular}{l p{0.6cm}lp{5.2cm}}
\hline
Cluster & N of stars & Type & Notes\\
\hline\hline
47~Tuc & 90 & GC & Membership from \citet{tucholke92}.\\
%Hyades & 0 & OC & Membership from \citet{roeser11}. We observed and analysed 7 members, but none have reliable parameters.\\
M30 & 4 & GC & \\
M67 & 113 & OC & Membership from \citet{geller15}. 404 spectra of 113 unique stars.\\
NGC288 & 14 & GC\\
NGC362 & 27 & GC\\
NGC1851 & 7 & GC\\
NGC2516 & 3 & OC & Membership from \citet{jeffries01}.\\
Pleiades & 27 & OC &\\
$\omega$ Cen & 230 & GC & 246 spectra of 230 unique stars.\\
\hline
\end{tabular}
\newline
GC=globular cluster\\
OC=open cluster
\label{tab:clusters}
\end{table}

\begin{figure}
\centering
\includegraphics[width=\columnwidth]{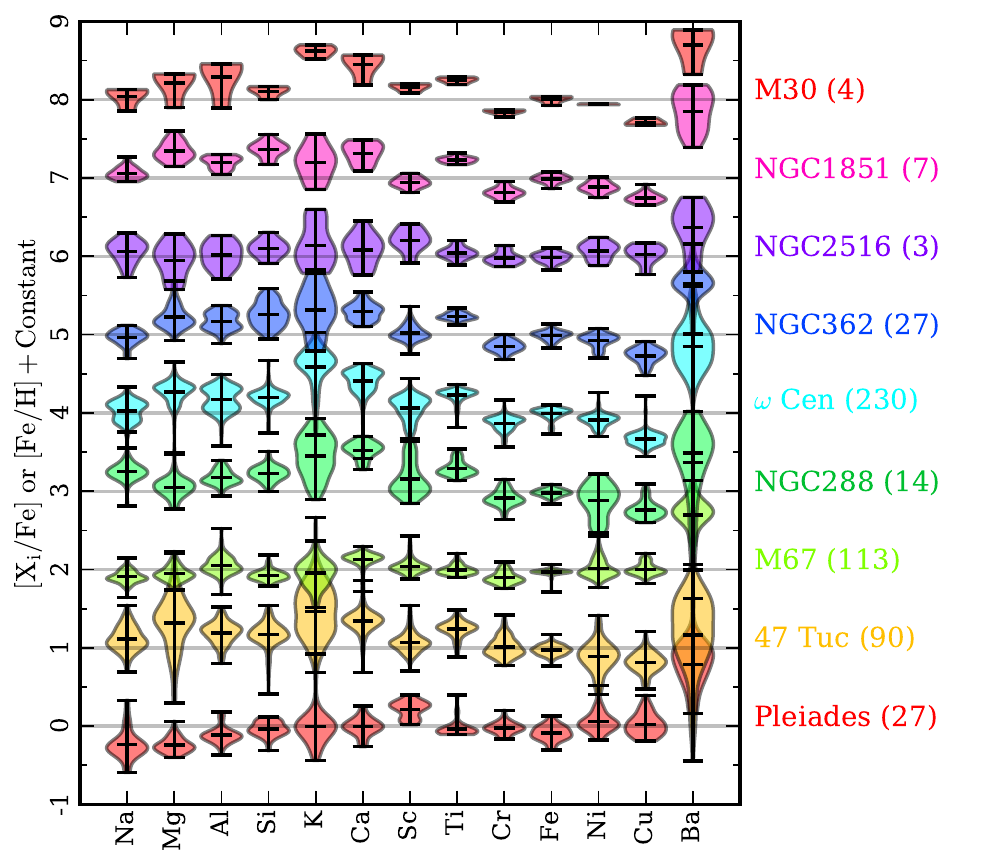}
\caption{Abundances of 13 elements in 9 studied clusters. A violin plot represents the distribution of measured abundances for all stars that we identified as cluster members. The number of all stars in each cluster is given next to the cluster name. Note how some elements have consistently more scattered distributions.}
\label{fig:cluster_abund}
\end{figure}

Figure \ref{fig:cluster_abund} shows the abundances for all 13 elements in the observed clusters. Note that the scatter for some elements is consistently high, regardless of the cluster. Around half of the scatter is statistical noise (see Table \ref{tab:weights}). We demonstrate this by measuring the uncertainties of the abundances from repeated observations of field stars and M67. Only measurements made from spectra with SNR$\geq45$ per pixel were used in order to distinguish between repeats done with the purpose of quality estimation and repeats done to boost the SNR of some lower quality data. The rest of the scatter is systematic, arising from the abundance determination pipeline being sensitive to temperature variations or dwarf--giant distinction. 

\begin{table}
\caption{Uncertainties measured from 1579 repeated observations in the whole GALAH sample and 377 repeated observations of M67 stars compared to scatter observed in Figure \ref{fig:cluster_abund}, and weights assigned to each element. Uncertainties and scatters are expressed as standard deviations.}
\begin{tabular}{l p{1.5cm} p{1.5cm} p{1.3cm} p{0.9cm}}
\hline
Element & Uncertainty from all repeats & Uncertainty from M67 repeats & Scatter in all clusters & Weight\\\hline
& dex & dex & dex & \\\hline\hline
Na & 0.063 & 0.063 & 0.122 & 1.0\\
Mg & 0.078 & 0.079 & 0.162 & 0.5\\
Al & 0.066 & 0.063 & 0.129 & 1.0\\
Si & 0.053 & 0.053 & 0.106 & 1.0\\
K & 0.099 & 0.129 & 0.228 & 0.25\\
Ca & 0.065 & 0.056 & 0.131 & 0.5\\
Sc & 0.050 & 0.054 & 0.115 & 1.0\\
Ti & 0.044 & 0.048 & 0.071 & 2.0\\
Cr & 0.047 & 0.049 & 0.081 & 2.0\\
Fe & 0.024 & 0.021 & 0.060 & 2.0\\
Ni & 0.056 & 0.057 & 0.112 & 1.0\\
Cu & 0.049 & 0.036 & 0.095 & 2.0\\
Ba & 0.114 & 0.135 & 0.230 & 0.25\\\hline
\end{tabular}
\label{tab:weights}
\end{table}

\begin{figure*}
\includegraphics[width=0.24\textwidth]{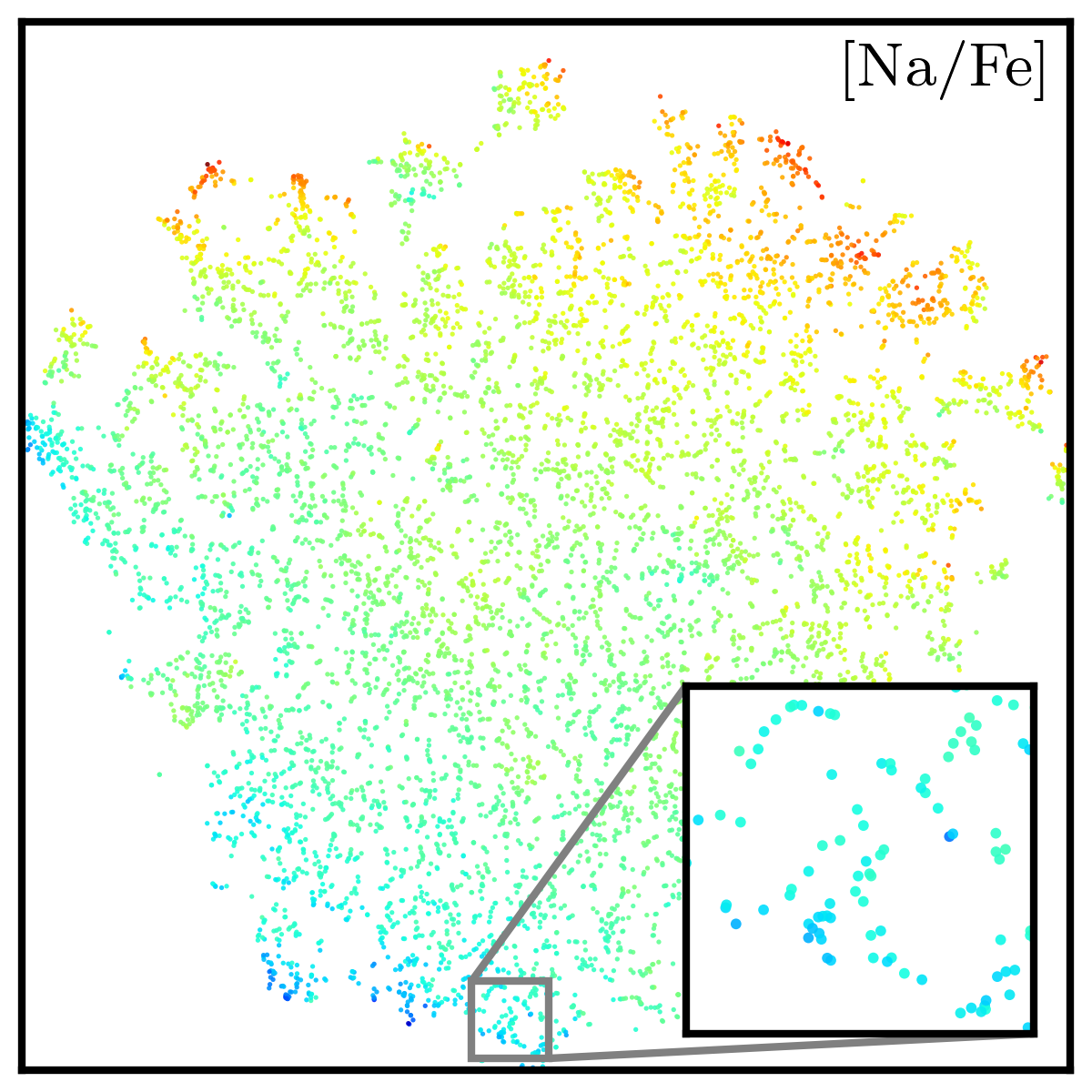}
\includegraphics[width=0.24\textwidth]{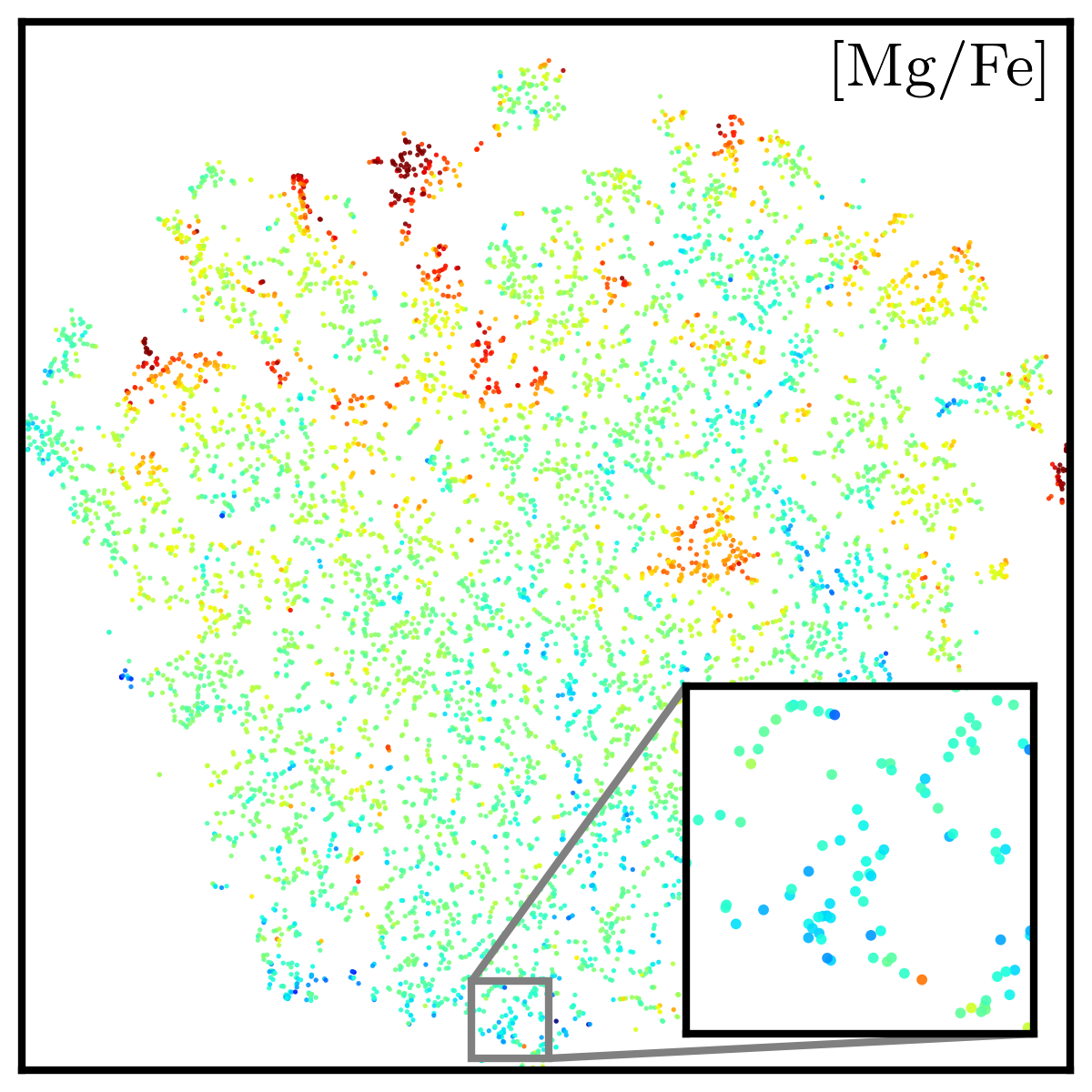}
\includegraphics[width=0.24\textwidth]{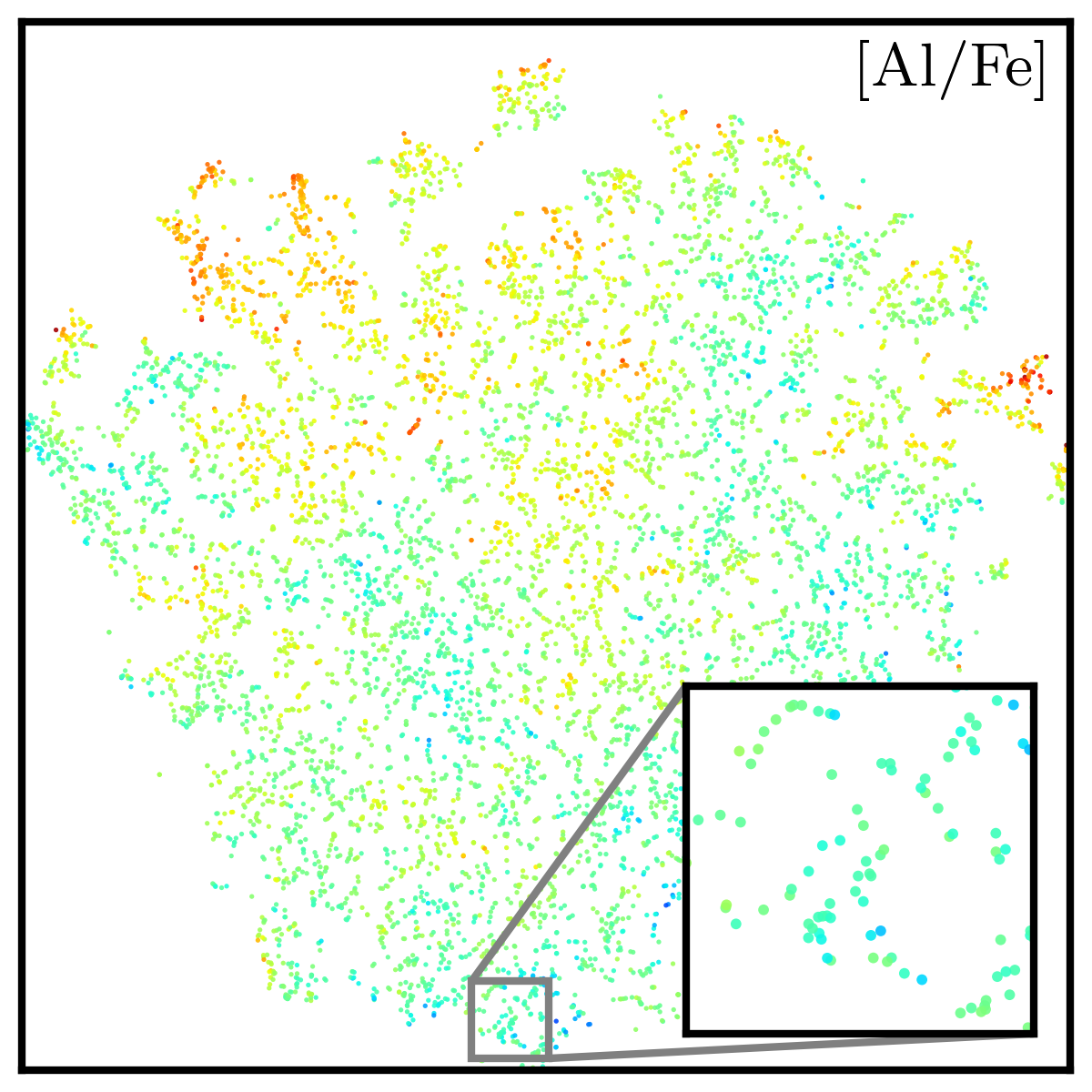}
\includegraphics[width=0.24\textwidth]{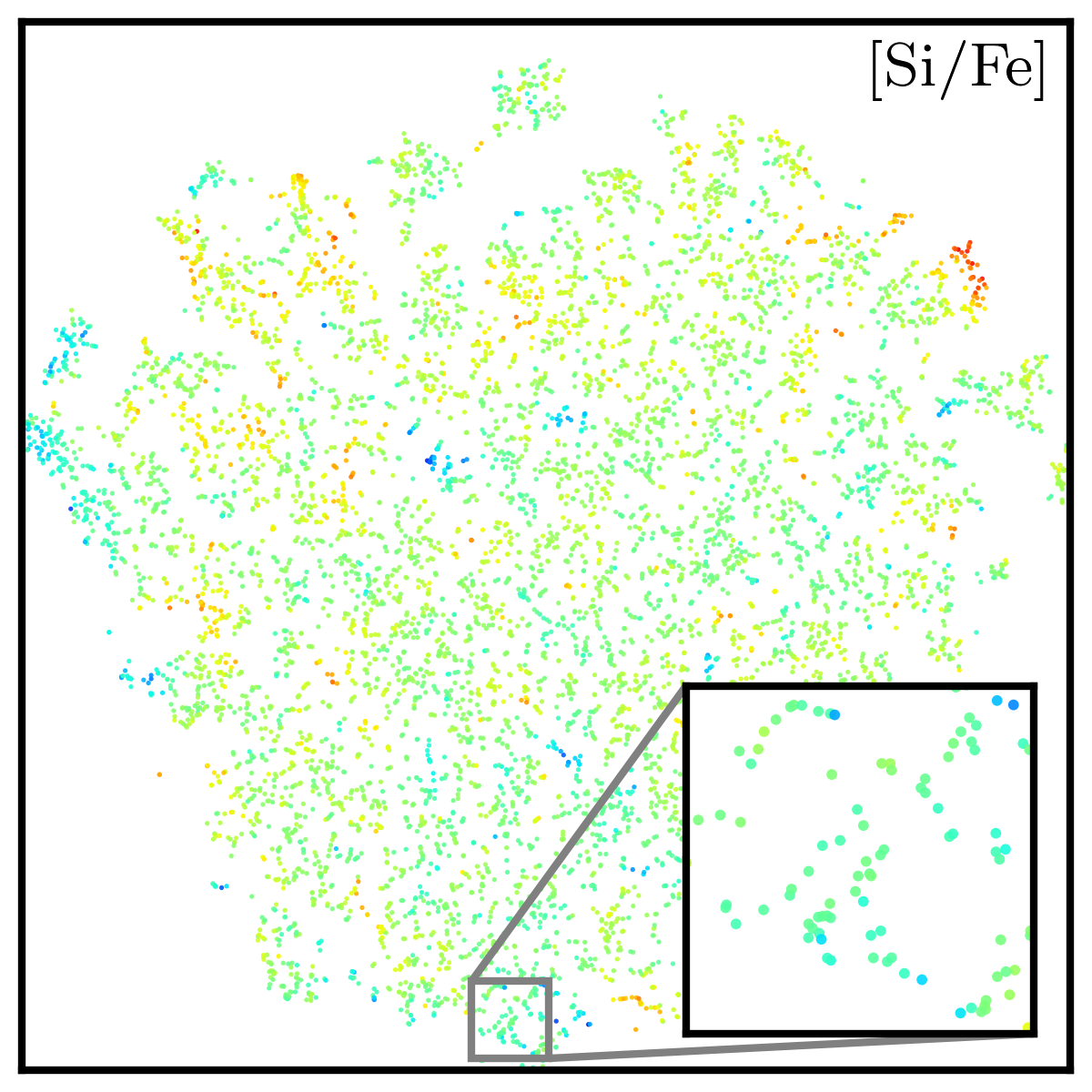}\\
\includegraphics[width=0.24\textwidth]{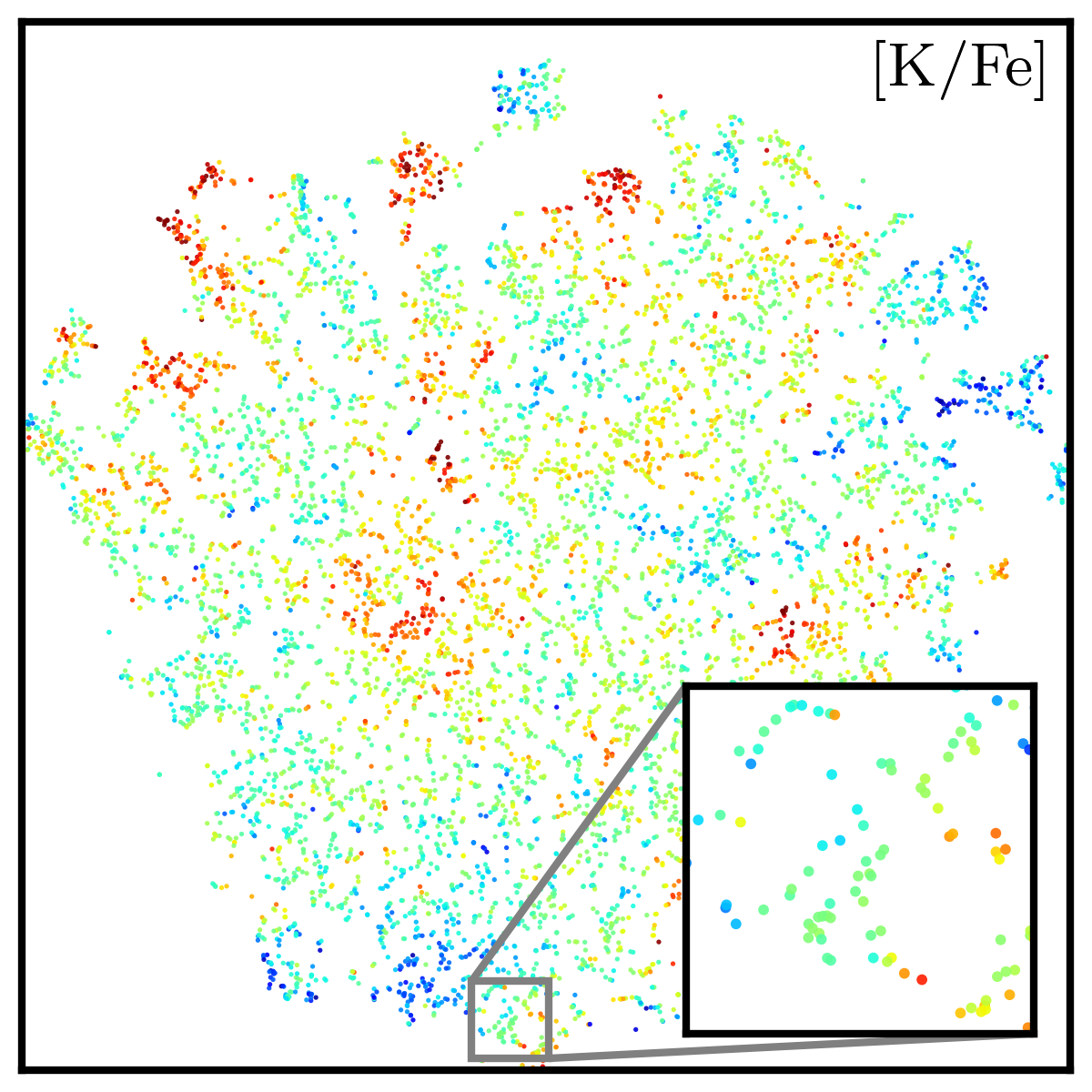}
\includegraphics[width=0.24\textwidth]{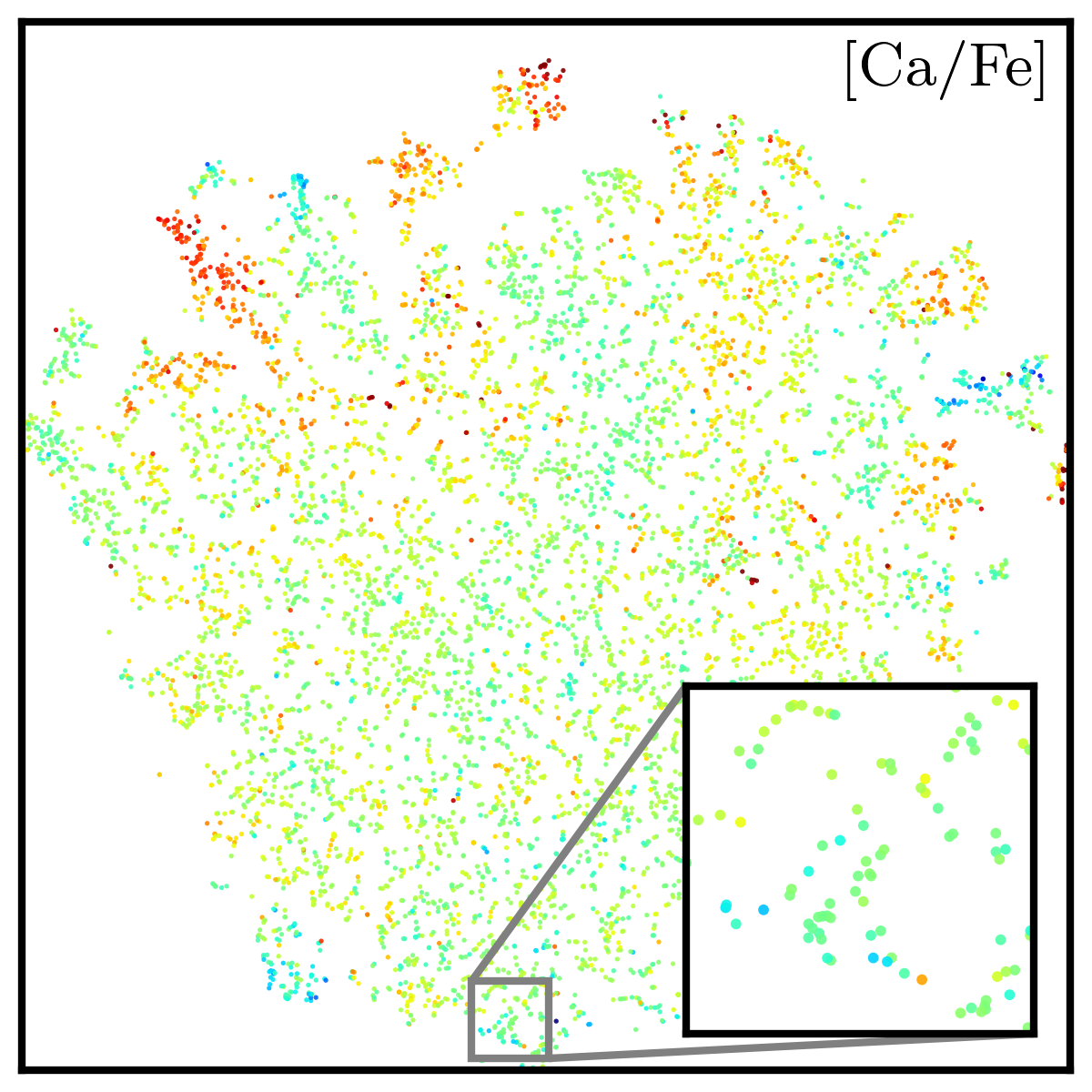}
\includegraphics[width=0.24\textwidth]{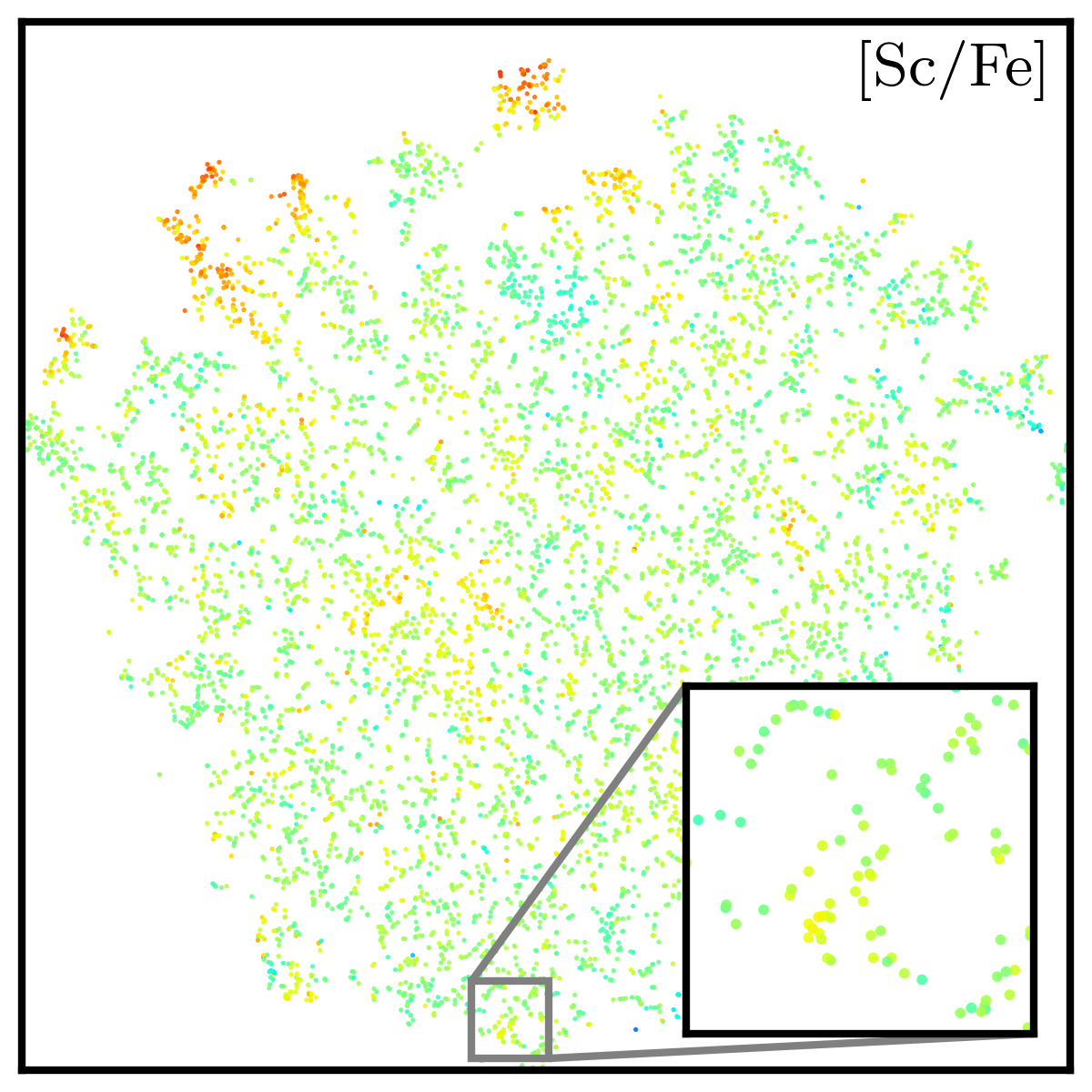}
\includegraphics[width=0.24\textwidth]{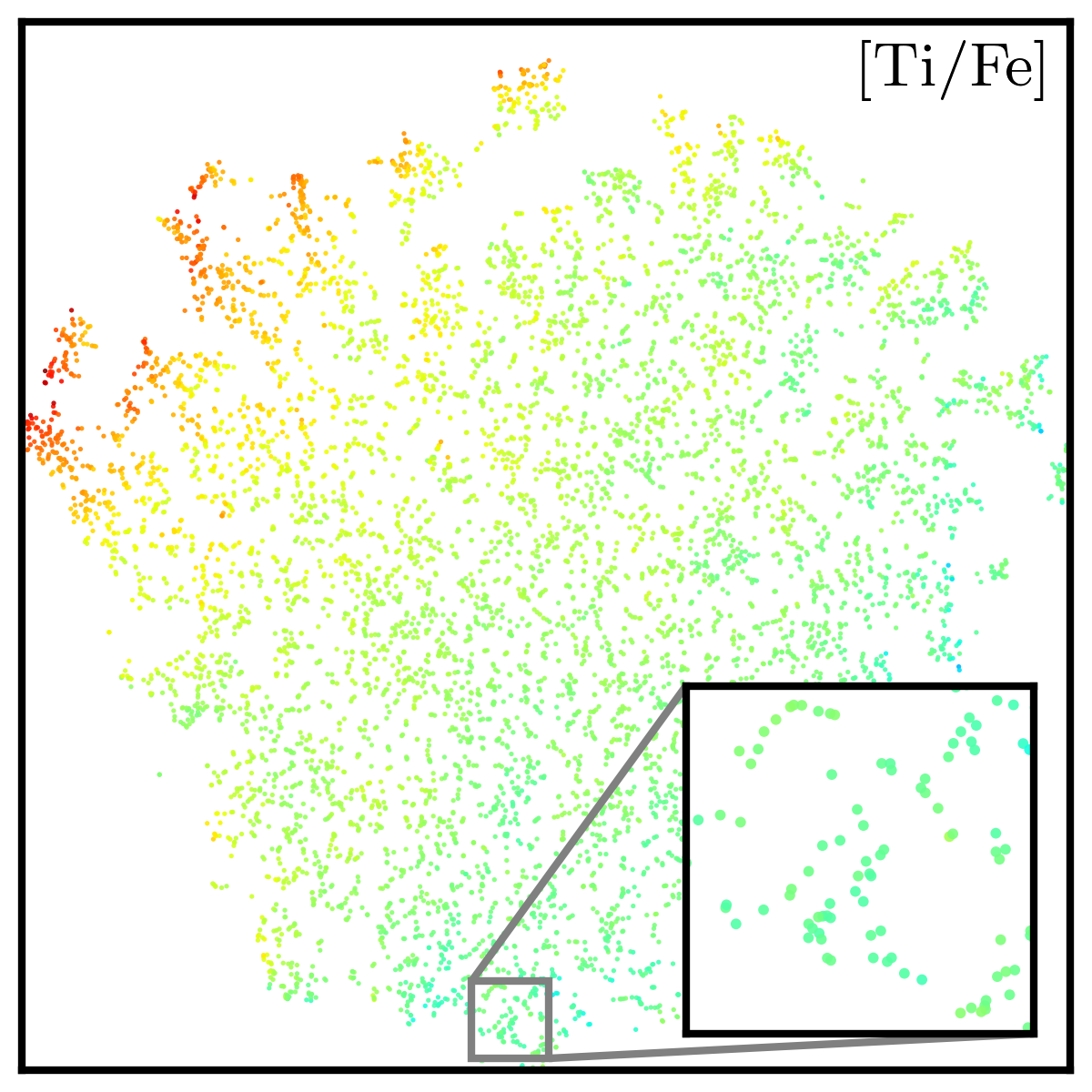}\\
\includegraphics[width=0.24\textwidth]{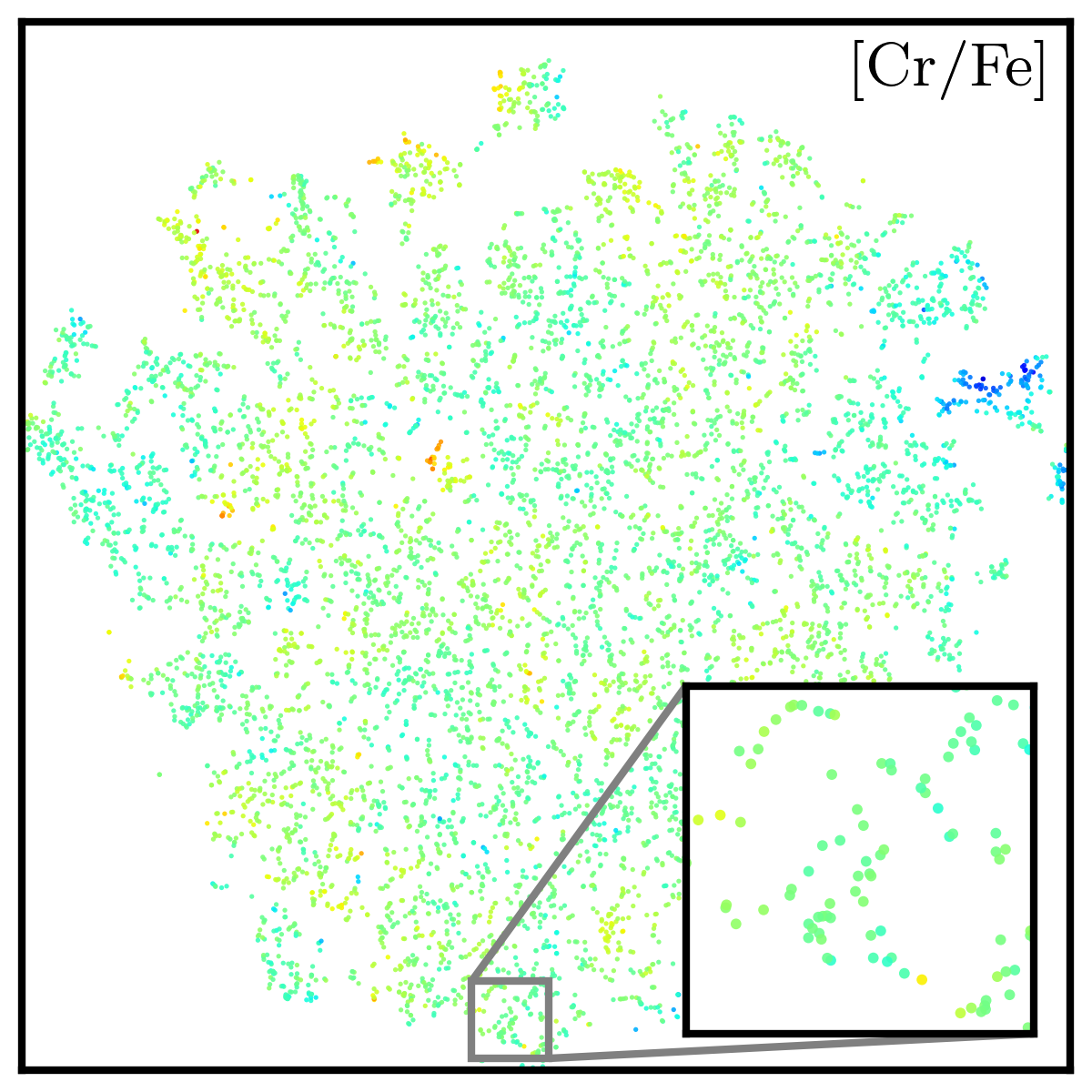}
\includegraphics[width=0.24\textwidth]{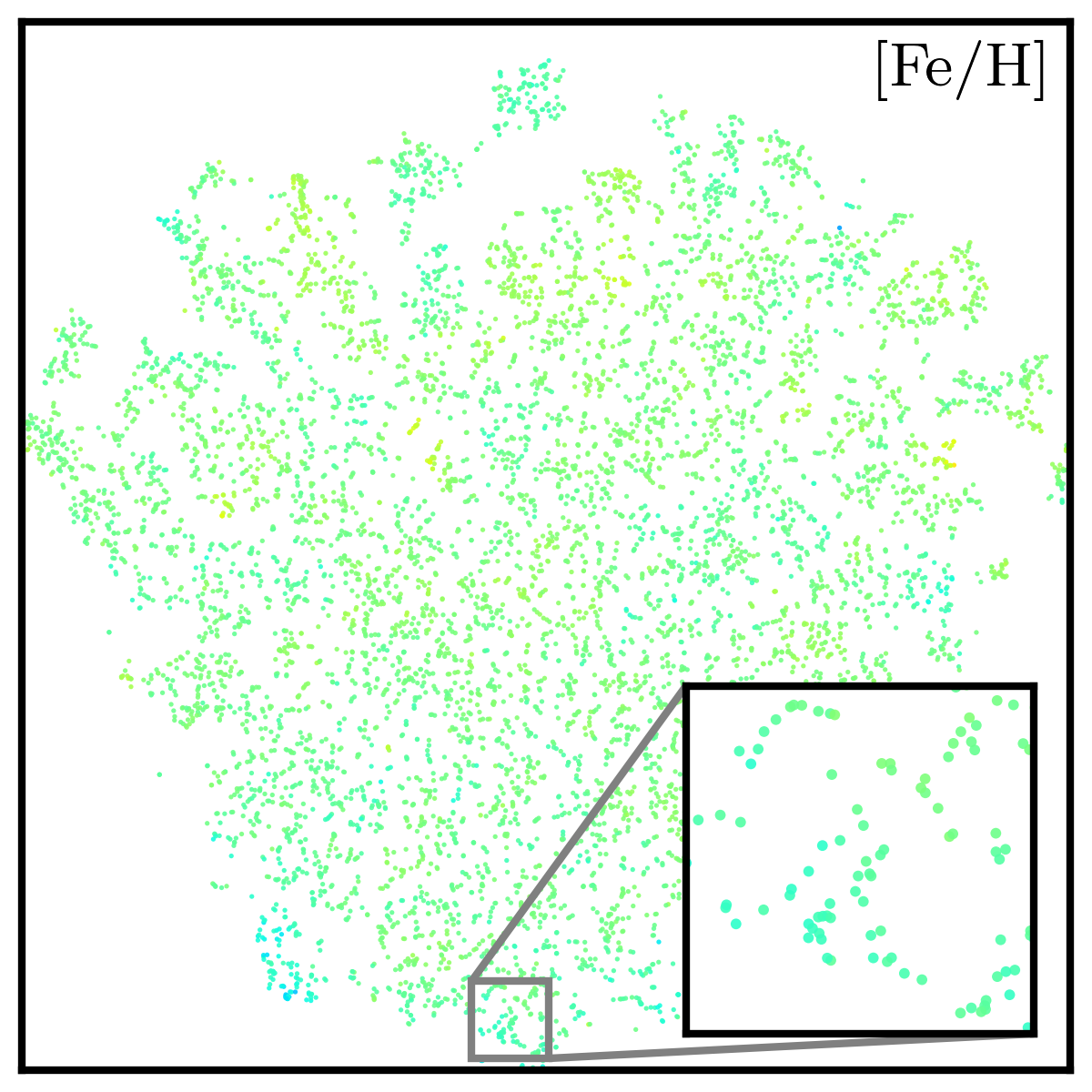}
\includegraphics[width=0.24\textwidth]{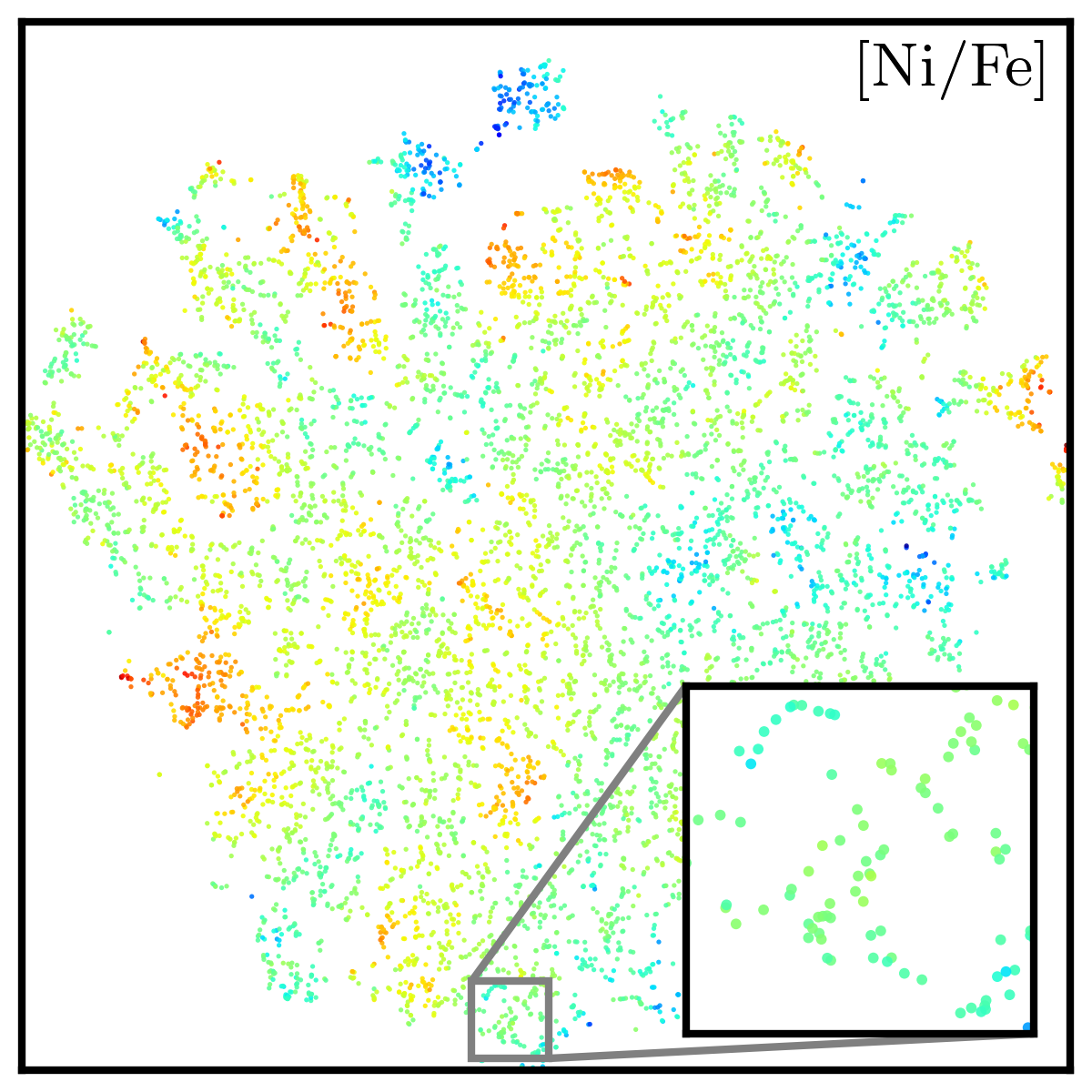}
\includegraphics[width=0.24\textwidth]{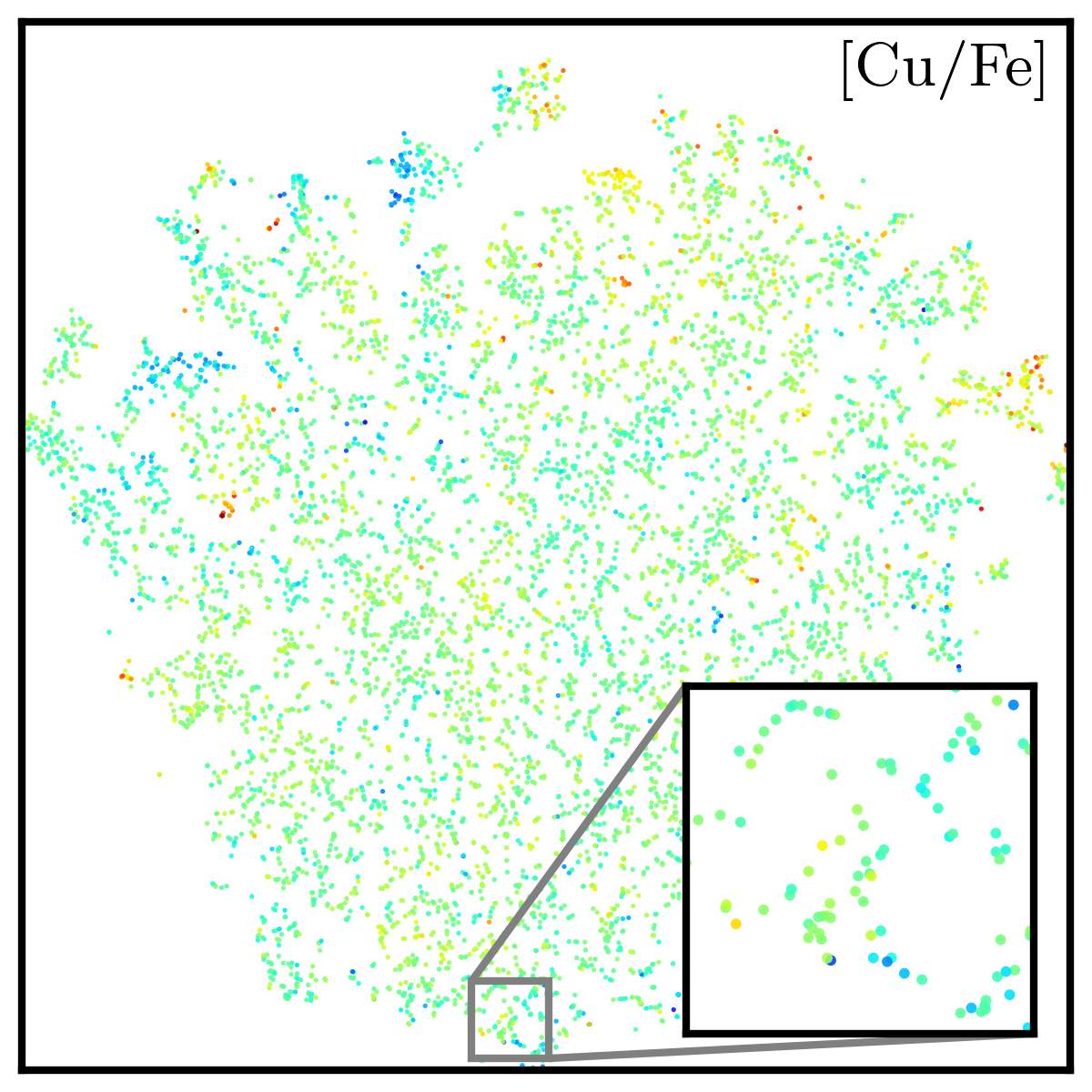}\\
\includegraphics[width=0.24\textwidth]{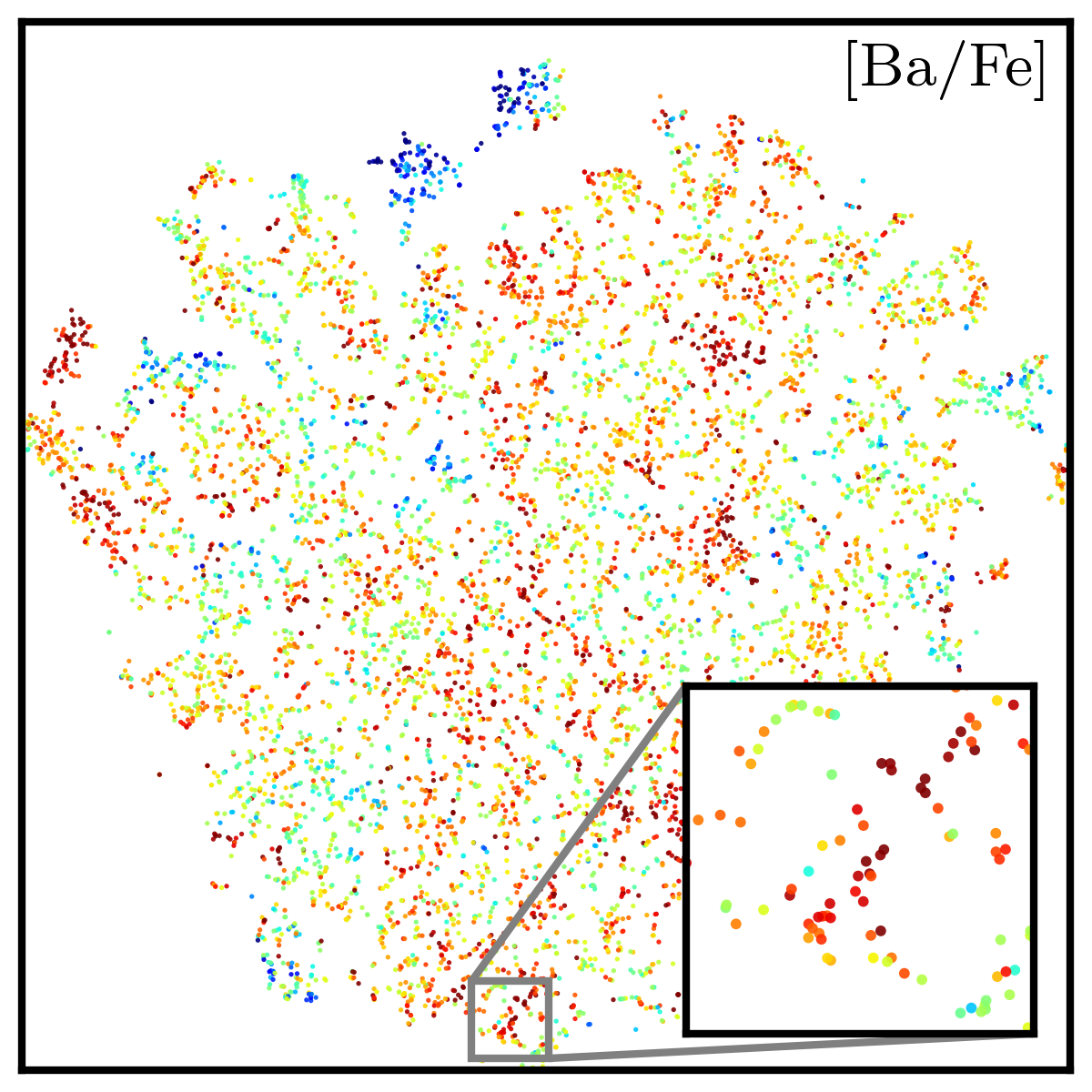}
\includegraphics[width=0.24\textwidth]{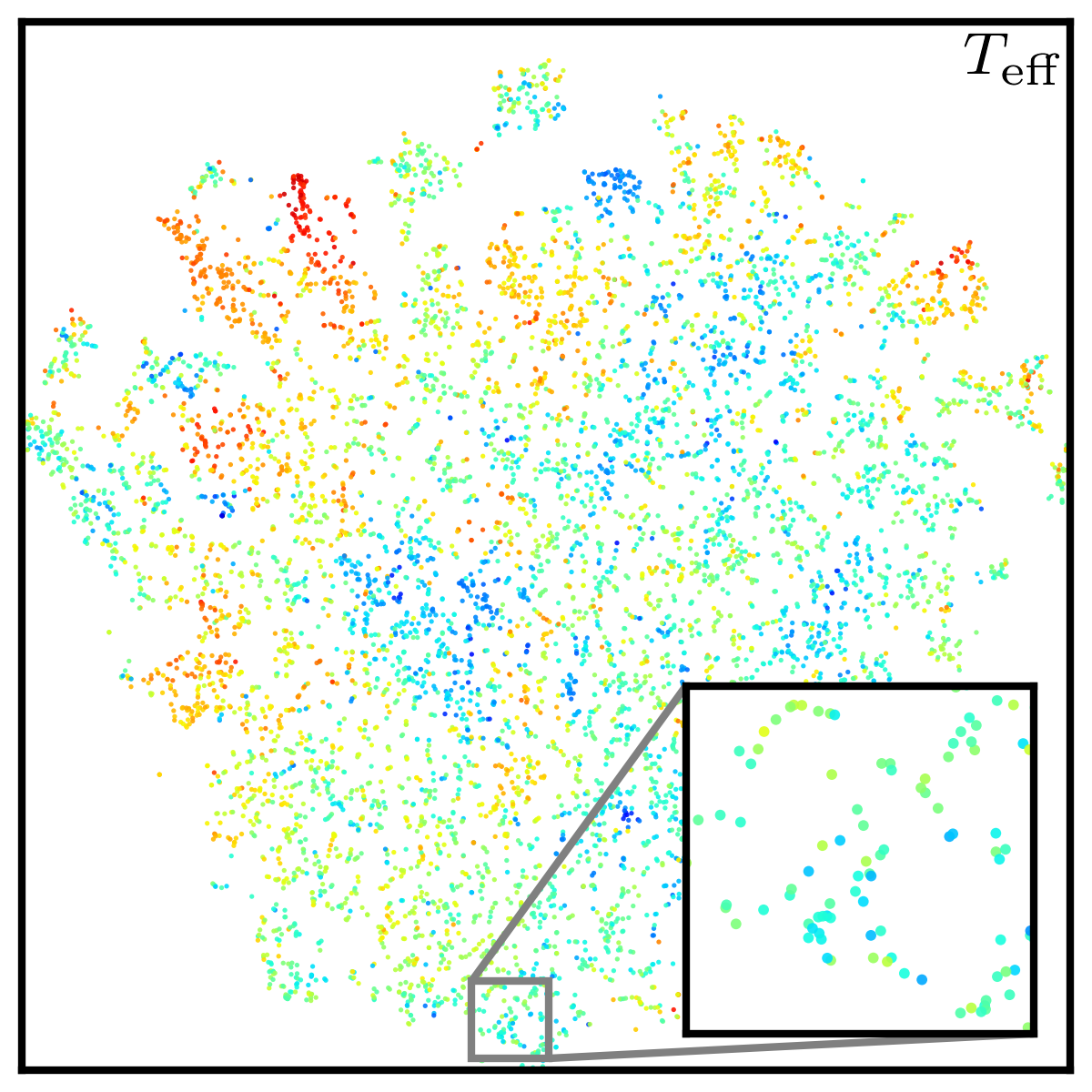}
\includegraphics[width=0.24\textwidth]{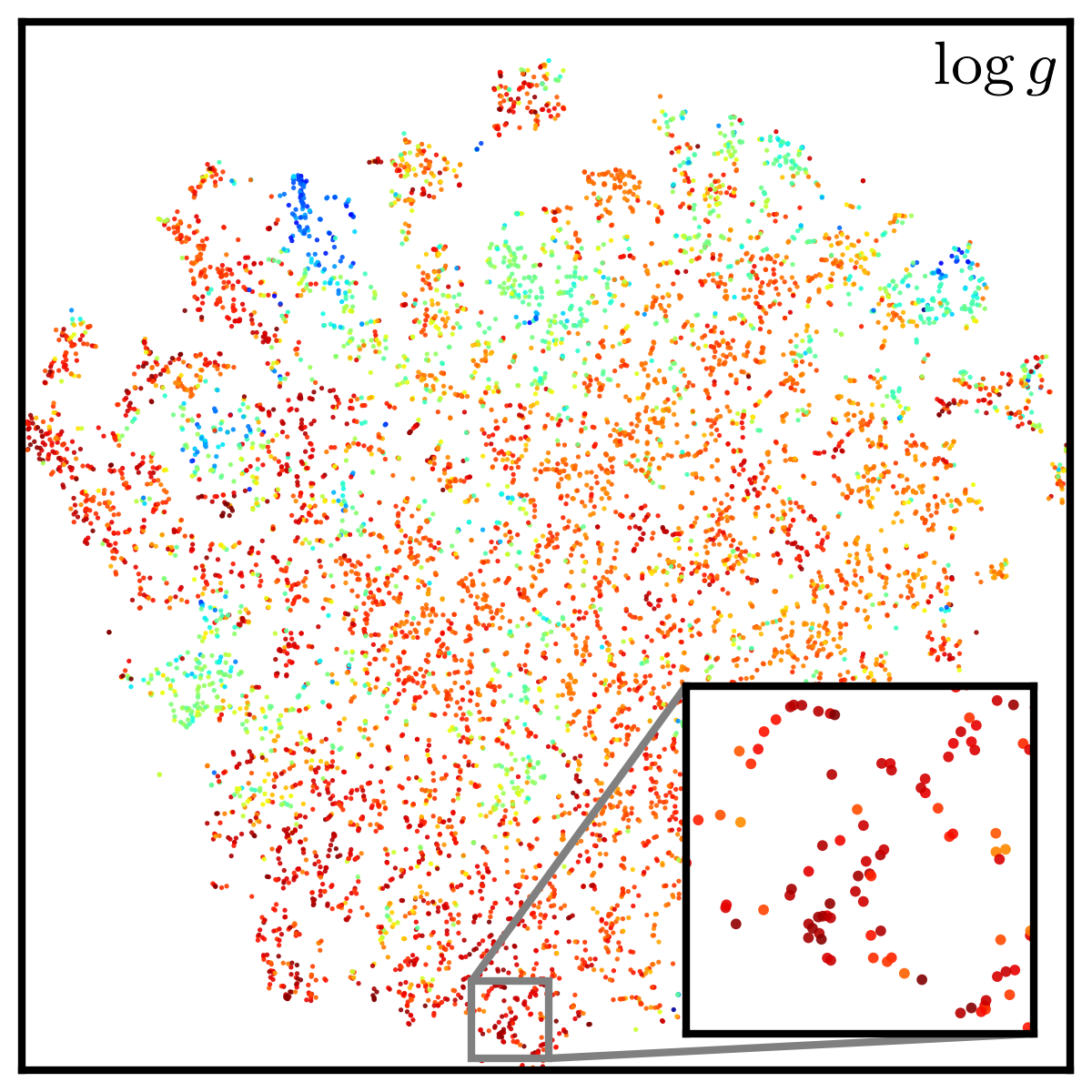}
\includegraphics[width=0.24\textwidth]{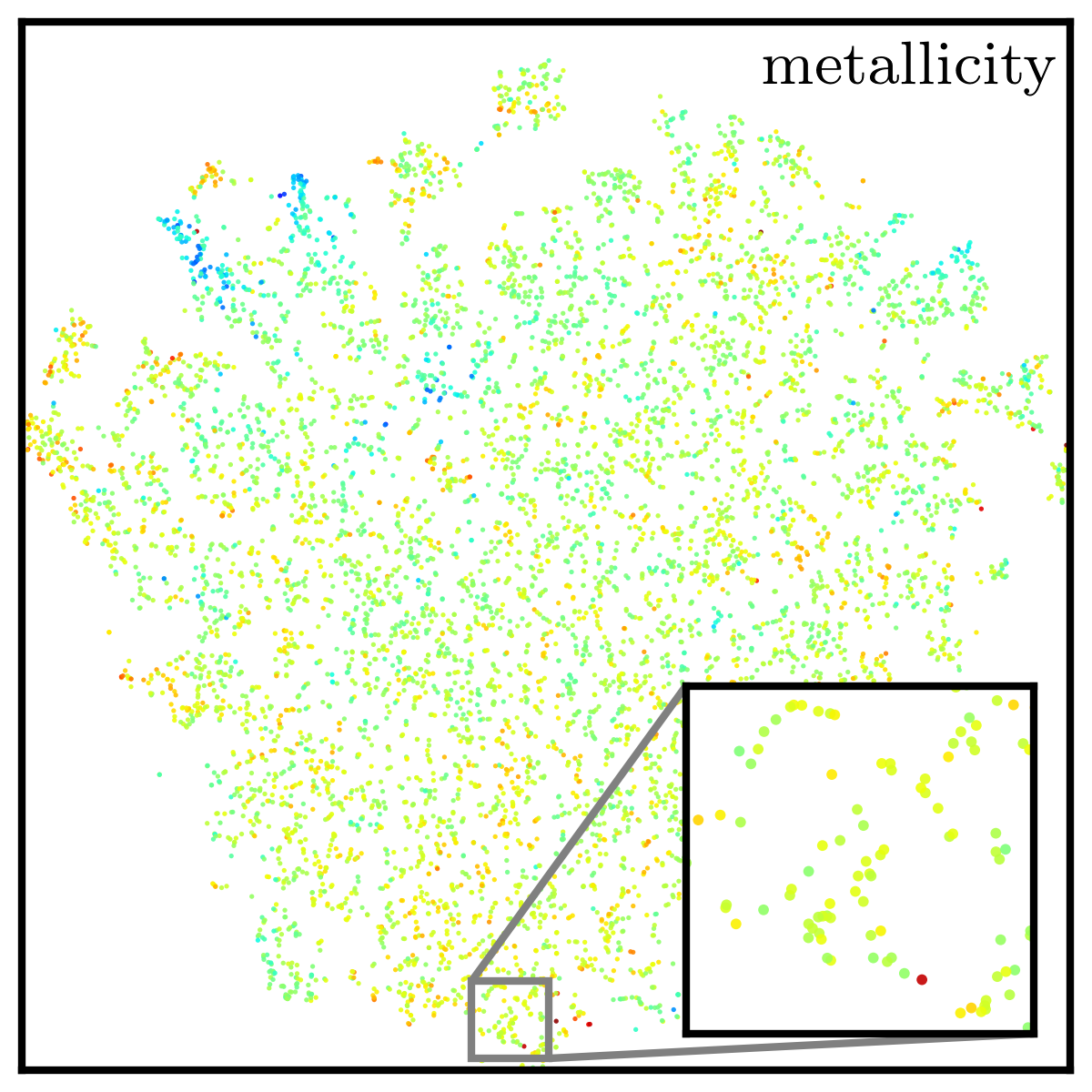}\\
\includegraphics[width=0.24\textwidth]{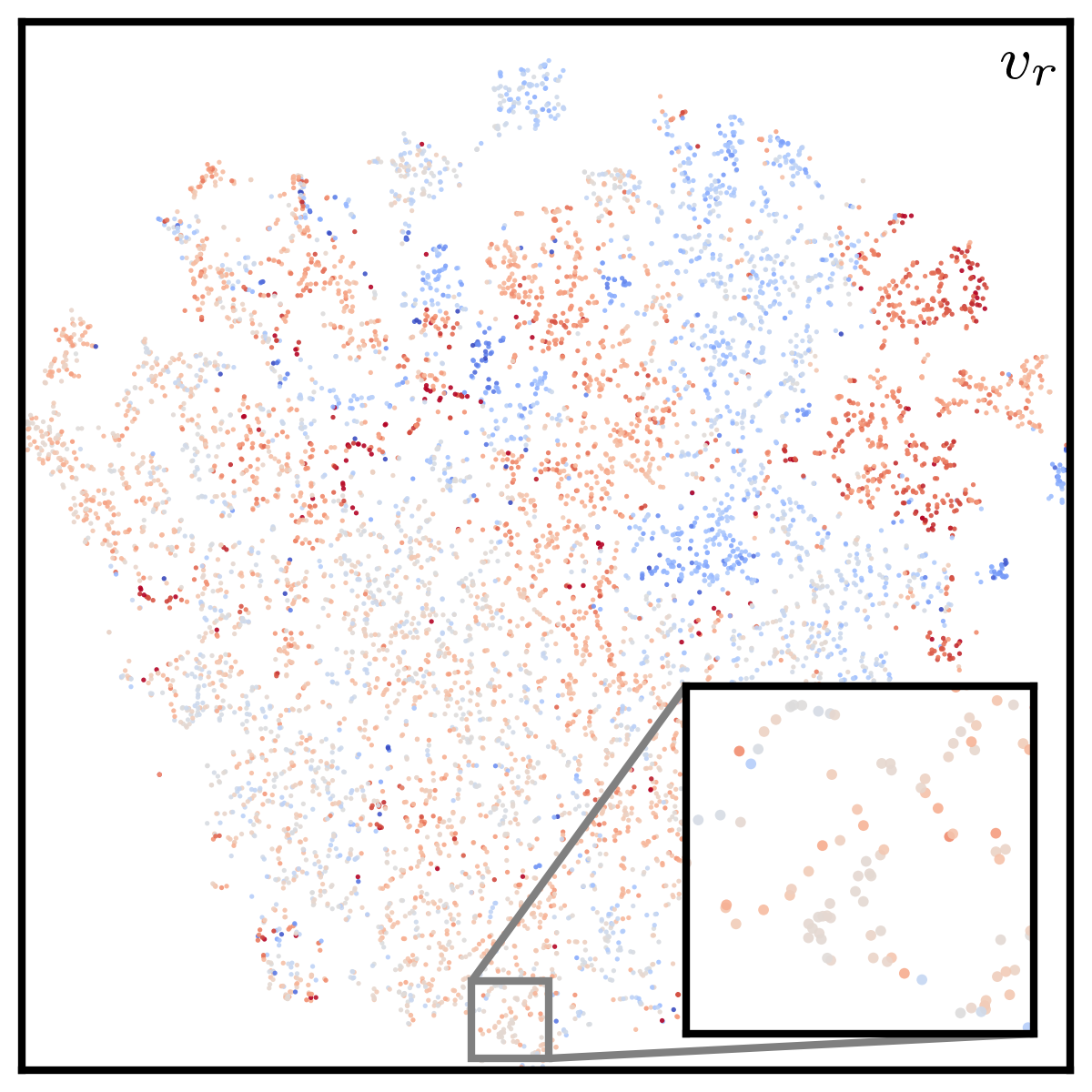}
\includegraphics[width=0.24\textwidth]{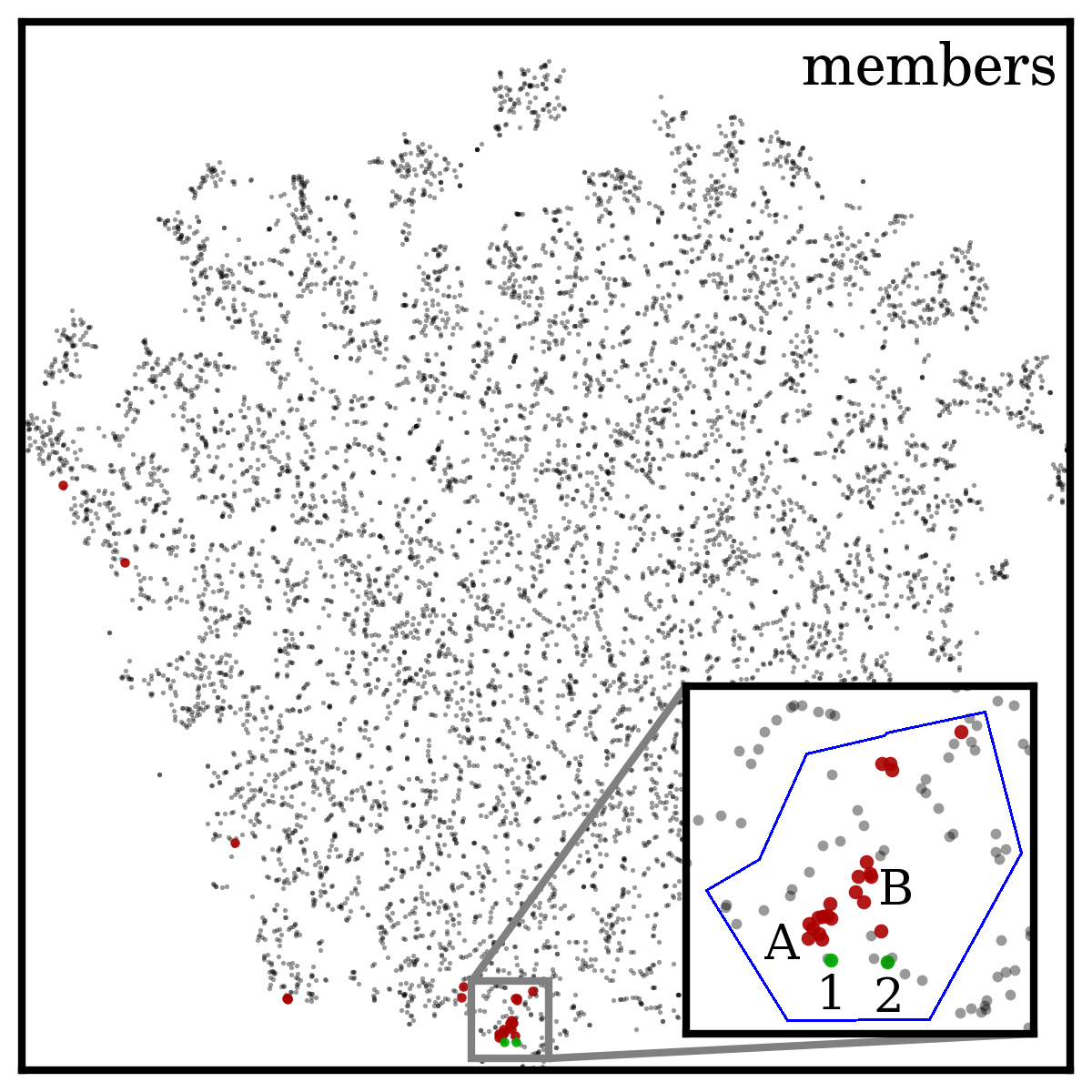}\hspace{1.2em}
\includegraphics[width=0.06\textwidth]{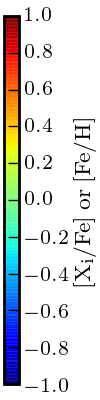}\hspace{2em}
\includegraphics[width=0.06\textwidth]{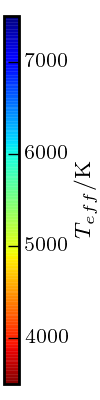}\hspace{2em}
\includegraphics[width=0.06\textwidth]{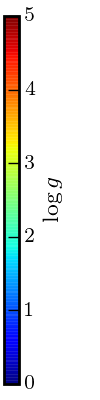}\hspace{2em}
\includegraphics[width=0.06\textwidth]{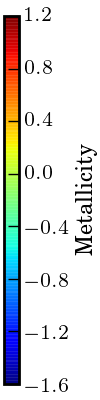}\hspace{2em}
\includegraphics[width=0.06\textwidth]{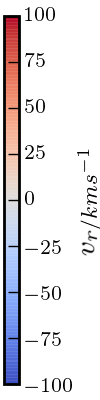}
\caption{t-SNE projection of 9408 stars in a 40$^\circ$ radius around the Pleiades. Each panel shows the same projection with different colour-codes for different quantities (given in the top-right corner of each panel). Abundances of 13 elements, as well as $T_\mathrm{eff}$, $\log g$, metallicity, and radial velocity are colour coded. The panel labeled ``members'' shows the stars that belong to the cluster in red and field stars in grey. Two stars marked in green and numbered 1  and 2 are newly discovered Pleiades members discussed in Section \ref{sec:new}. They lie slightly away from the rest of the Pleiades because of their slightly different abundances, more clearly illustrated in Figure \ref{fig:pl_det}. 17 out of 27 Pleiades stars lie in the two tight groups in the bottom of the map marked A and B. The blue polygon marks the Pleiades' chemical group (see Section \ref{sec:new} and Figure \ref{fig:pl} for details). Colour version of this figure is available online.}
\label{fig:proj_Pleiades}
\end{figure*}

The scatter in the Ba and K abundances is highest. Elements like Fe, Ti, and Cr have lower uncertainties. It is therefore not fair to treat elements with different uncertainties as equally important dimensions in the $\mathcal{C}$-space. Before we use the abundances in t-SNE, we standardize them so that the distribution of abundances of every element has a zero median and a standard deviation of unity. Standardization is done once for the complete data-set (187,640 stars). Then we change the standard deviation of the standardized set based on the weights that are proportional to the scatter we observe in clusters. We are confident that there are no misidentified members contributing to the scatter (see Appendix \ref{sec:membership}). Elements with more scatter will have a narrower distribution, so the distances in those dimensions will always be damped and will not carry as much importance as those for less scattered elements. Because it is hard to quantitatively determine the weight for each element, we will distribute the elements into 4 groups. Ba and K have by far the highest scatter, so they will be given a weight of 0.25. Mg and Ca also have high scatter, so they will have weights equal to 0.5. Fe, Ti, Cr, and Cu have the smallest scatter and will have a weight of 2.0, and the rest of the elements will have a weight of 1.0. Weights, uncertainties measured from the repeated observations, scatter in clusters, and related weights are collected in Table \ref{tab:weights}. Weights are a way to implement uncertainties into the t-SNE, as in our case the uncertainties of individual measurements have not been estimated. Without these weights, there would be fewer groups in the t-SNE map and the stars from known clusters would end up scattered over a larger area. 

We use the weighted abundances to produce a t-SNE projection for a region around Pleiades (Figure \ref{fig:proj_Pleiades}) and other clusters (Appendix \ref{sec:appA}). In cases where we have more than one measurement for a star, as for most M67 stars and some $\omega$~Cen stars, we first calculated the average abundances for each star and used those in the t-SNE. Stars with repeated observations are therefore only plotted once in the t-SNE maps. No other information is used in the projection, eventhough other stellar parameters (i.e. $T_{\mathrm{eff}}, \log g, v_r, \ldots$) are displayed in the color-coded t-SNE maps. One can pick out many groups in the t-SNE map, some more pronounced than the others. Groups associated with each cluster are marked and we leave a detailed analysis of other pronounced groups for a later study.

\begin{figure*}
\includegraphics[width=0.24\textwidth]{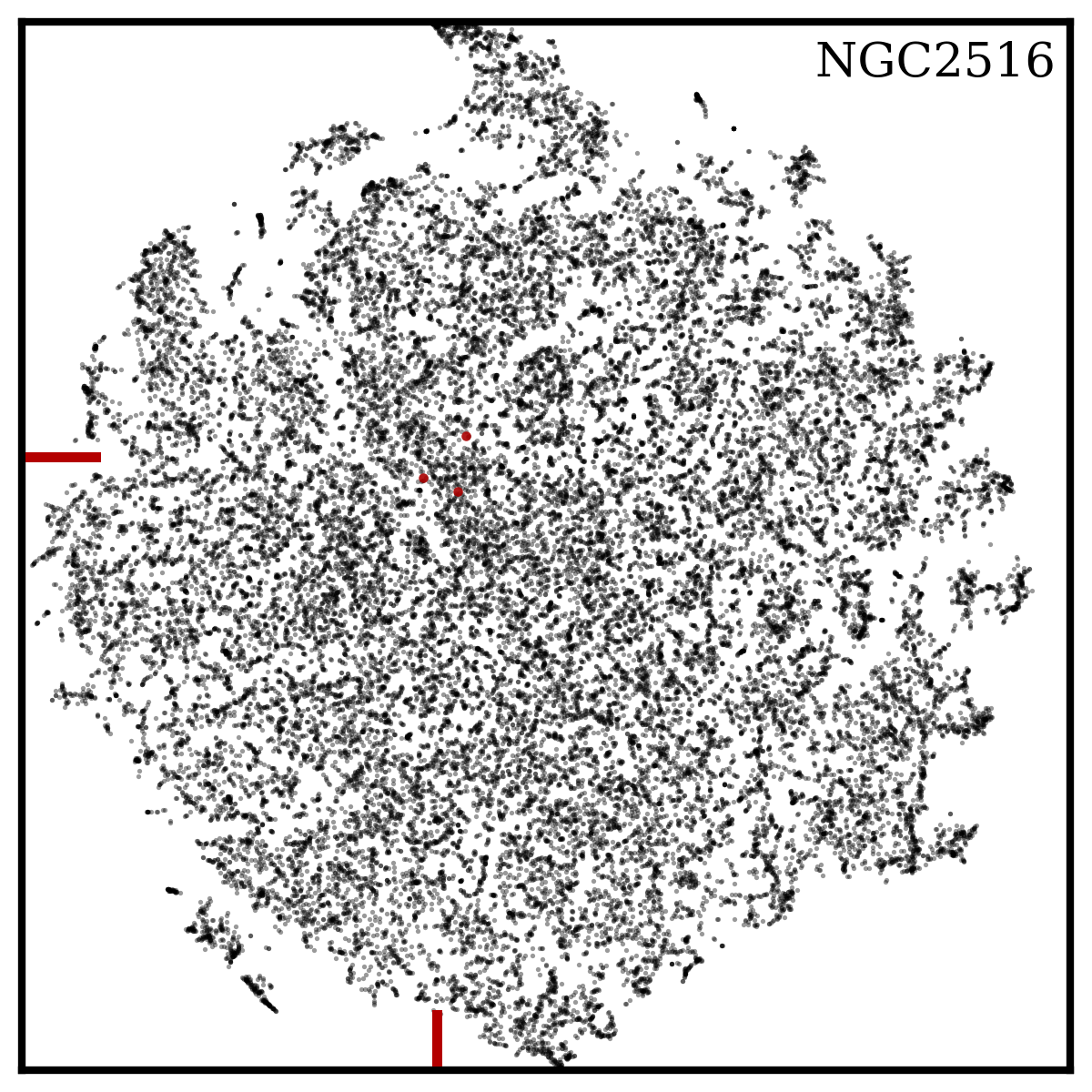}
\includegraphics[width=0.24\textwidth]{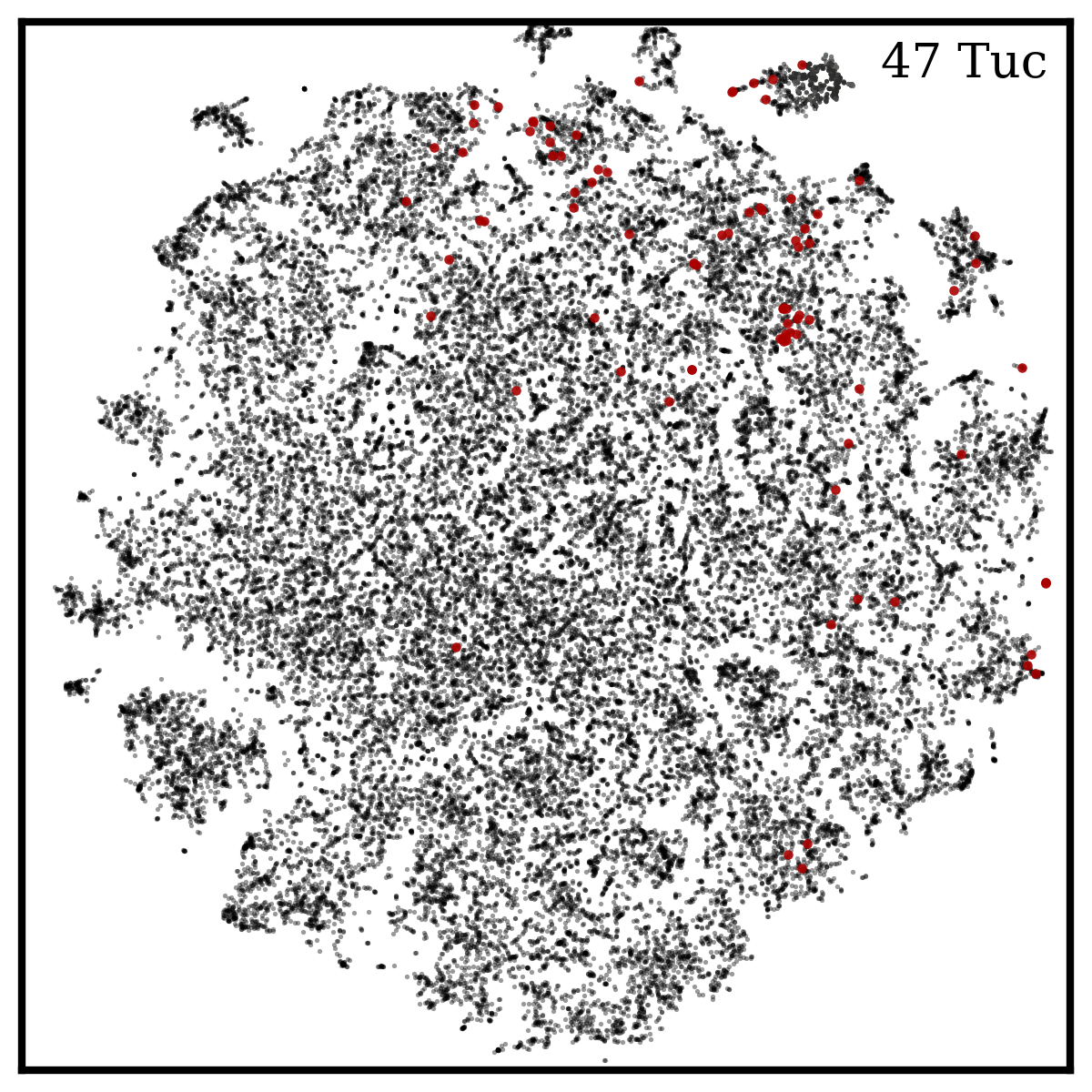}
\includegraphics[width=0.24\textwidth]{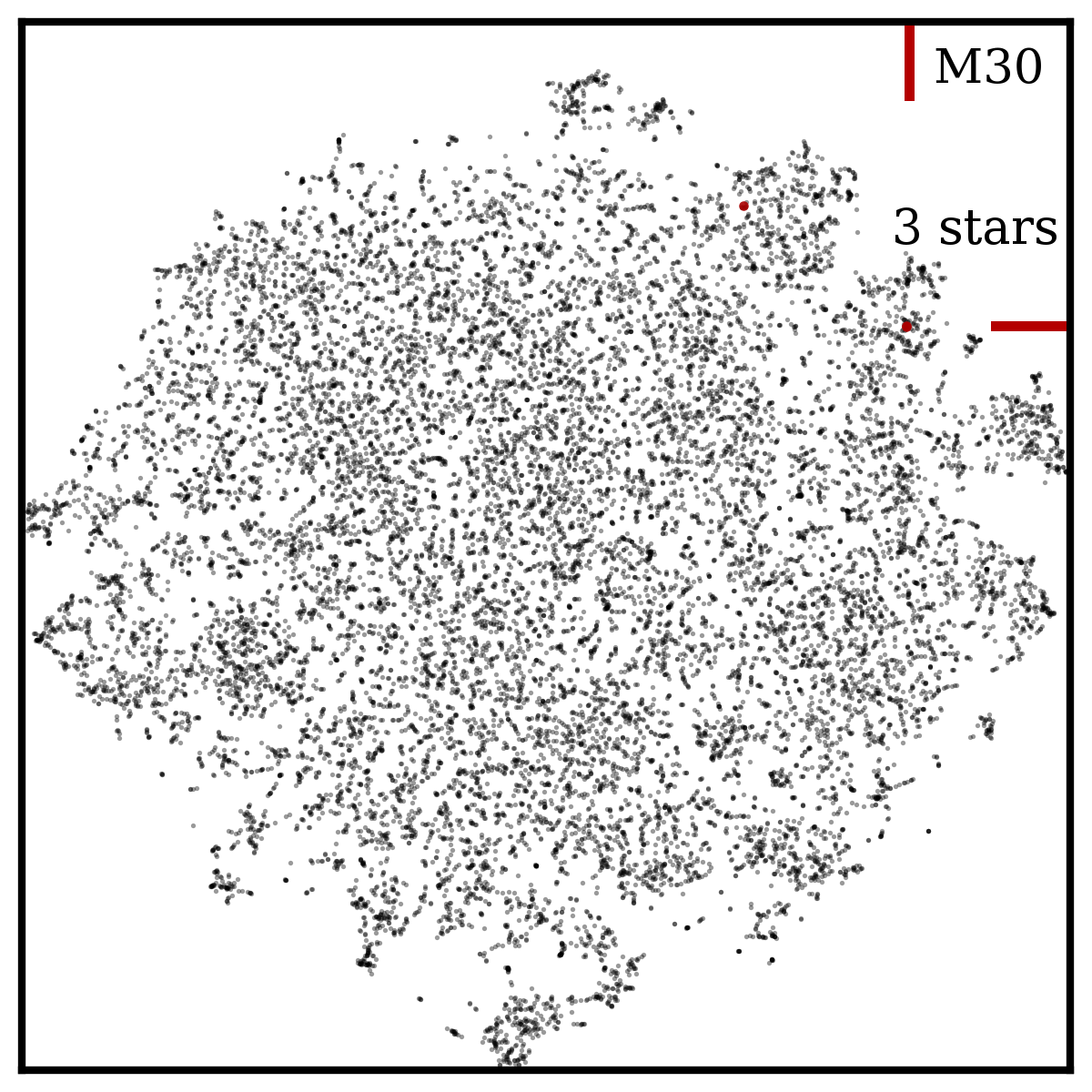}
\includegraphics[width=0.24\textwidth]{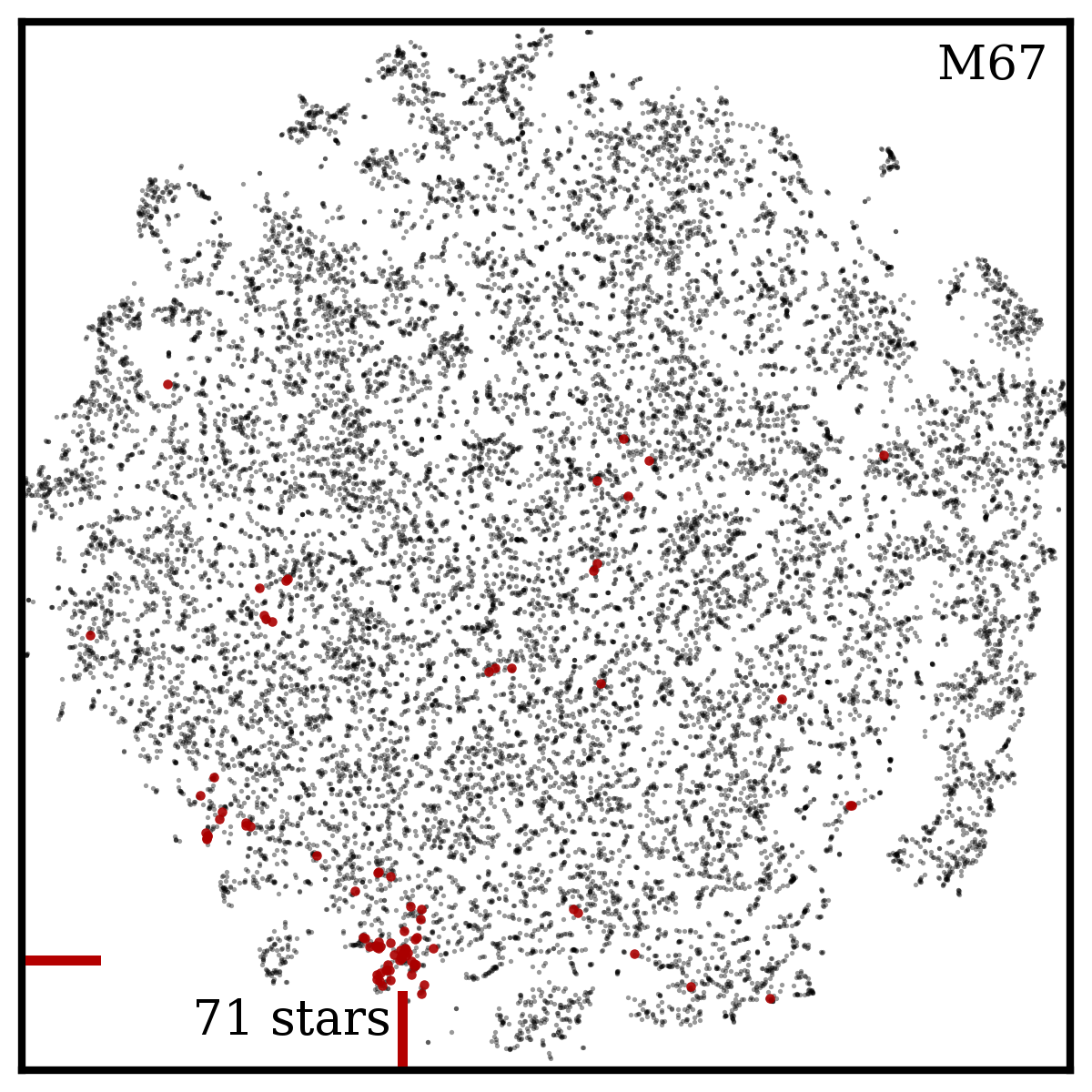}\\\hspace{0.1em}
\includegraphics[width=0.24\textwidth]{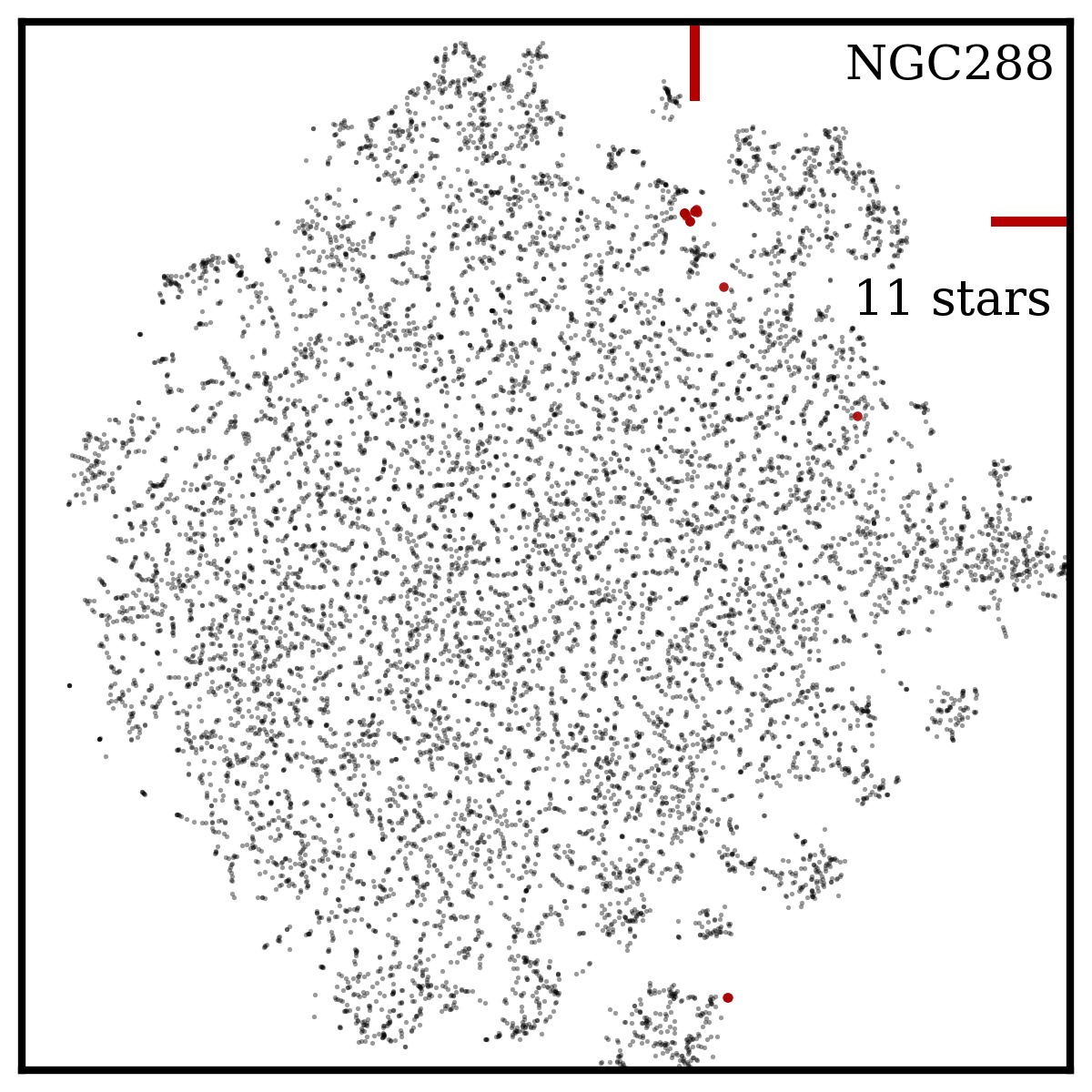}
\includegraphics[width=0.24\textwidth]{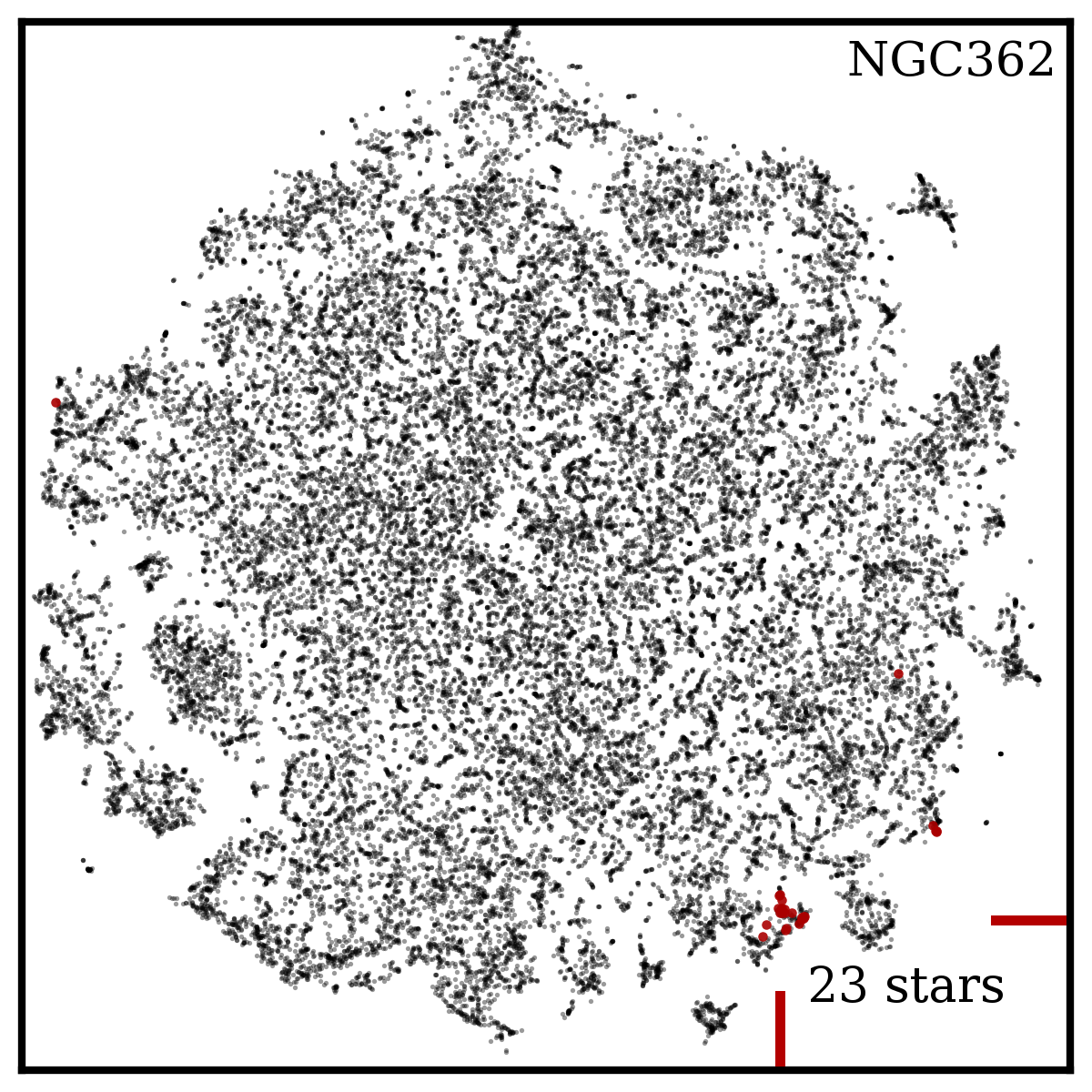}
\includegraphics[width=0.24\textwidth]{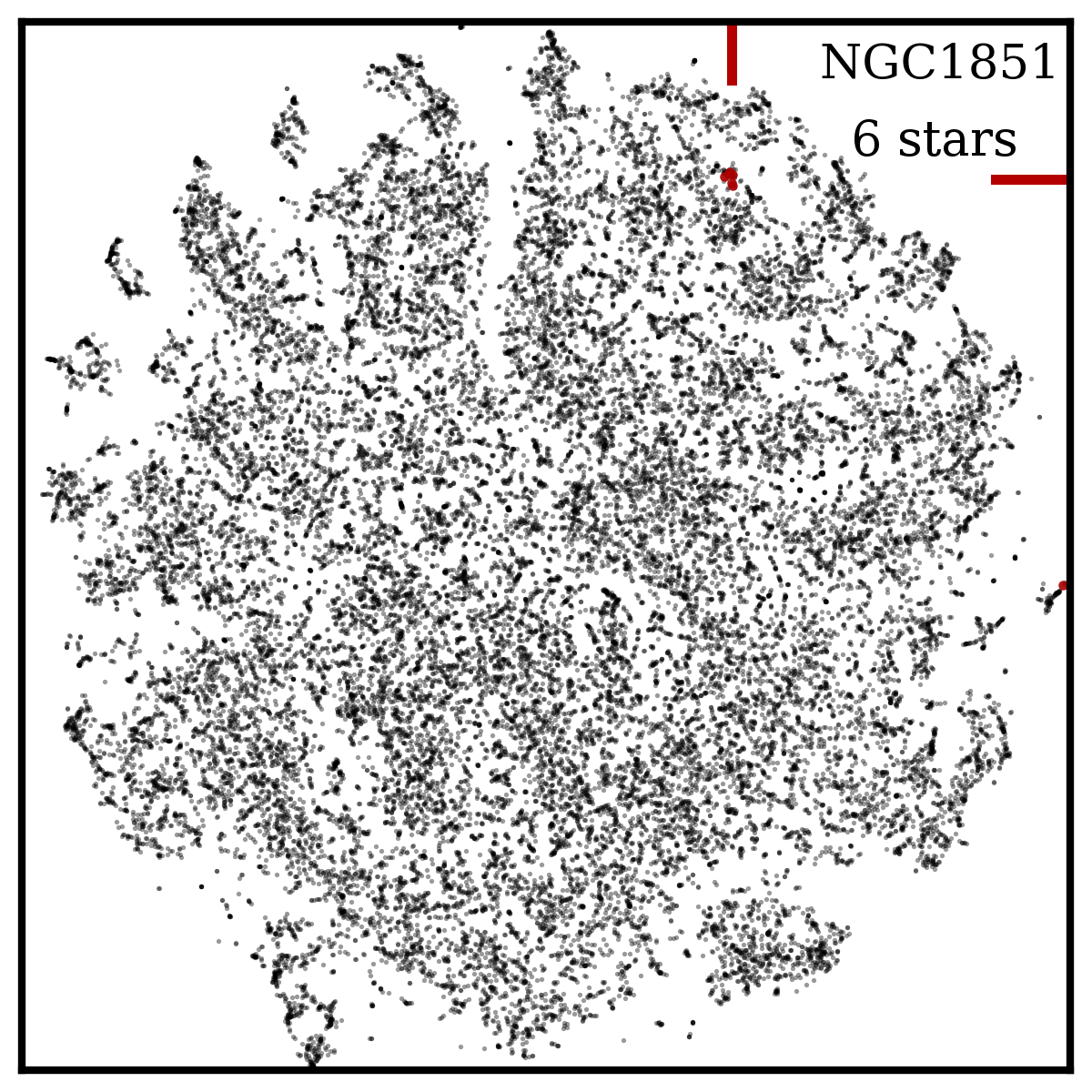}
\includegraphics[width=0.24\textwidth]{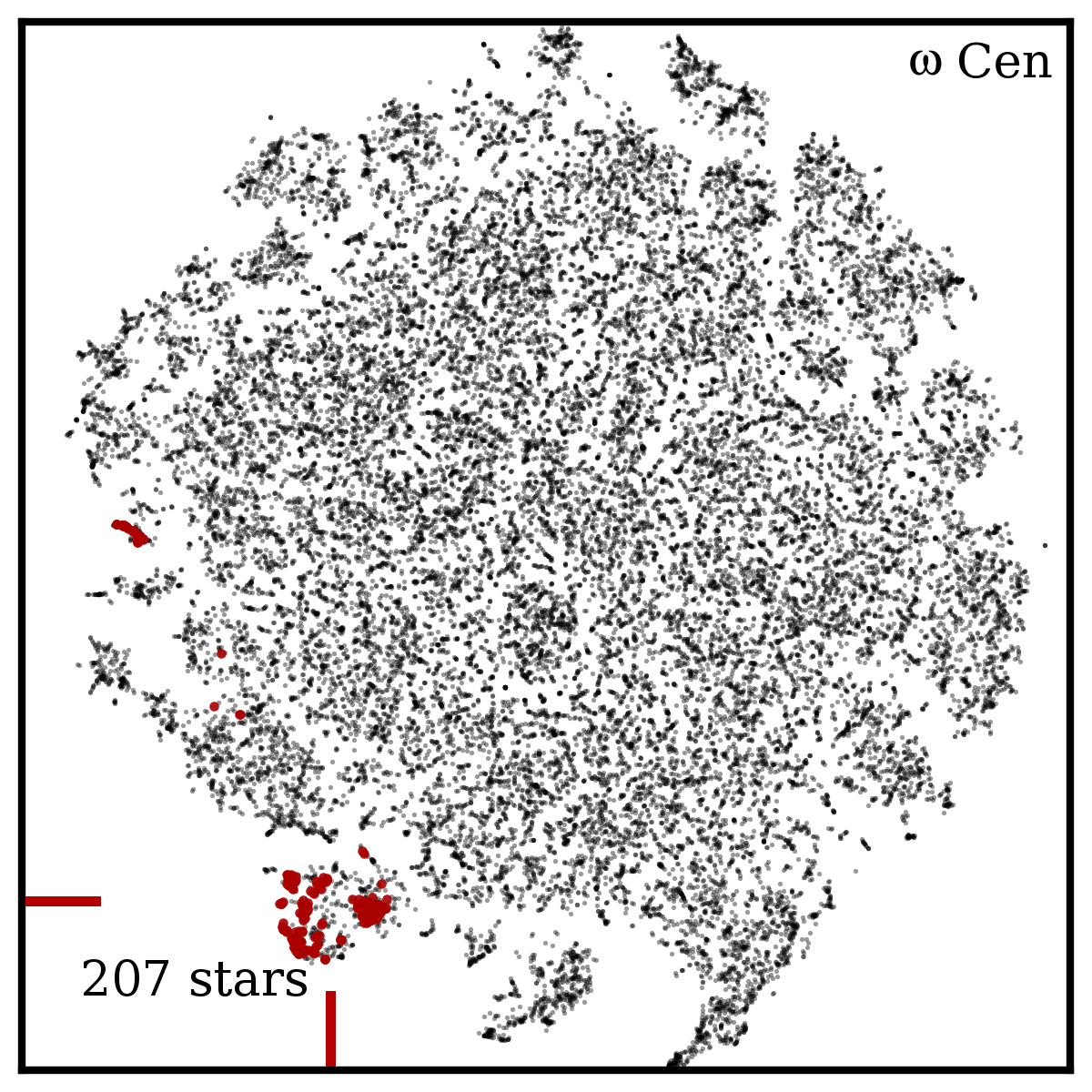}
\caption{t-SNE projections for regions around the 8 remaining clusters. Known members are marked in red and the number in some panels tells the number of the stars in the main group, as the points often overlap. Top-left, to bottom-right the following clusters are shown: NGC2516 (41,106 stars in a 30$^\circ$ radius), 47~Tuc (44,037 stars in a 35$^\circ$ radius), M30 (20,254 stars in a 35$^\circ$ radius), M67 (25,648 stars in a 45$^\circ$ radius), NGC288 (11,535 stars in a 45$^\circ$ radius), NGC362 (41,578 stars in a 35$^\circ$ radius), NGC1851 (33,882 stars in a 35$^\circ$ radius), and $\omega$~Cen (33,281 stars in a 30$^\circ$ radius). Red marks on the edge of each panel point toward the position of the main group of members. Colour version of this figure is available online.}
\label{fig:members}
\end{figure*}

Figure \ref{fig:members} shows only the members in the t-SNE maps for the eight remaining clusters. Out of all nine clusters, we claim that t-SNE gives good results for all but two of them. In NGC2516 the membership of three stars is not completely certain, so we can not base our conclusions on this cluster. Note that we only cover an edge of the cluster in one of the observed fields, so a low number of members is expected. We did not find any large groups in $\mathcal{C}$-space for 47~Tuc, so the chemical tagging of this cluster was unsuccessful. It is not clear why only 47 Tuc was so resistant to chemical tagging, since all globular clusters have inhomogeneities in their light-element abundances  \citep[e.g.][]{thygesen14}, and we were able to successfully tag stars from the other globular clusters. Perhaps the abundances in other globular clusters are distinct enough from field stars that they still form an isolated group. 

In most of the tagged clusters we find a small number of outliers: stars that are cluster members, but do not lie in the same chemical group as the rest of the stars. This is due to our measured abundances being significantly different, so t-SNE did not associate them with the main part of the cluster. Some outliers are expected, as our reduction and analysis pipelines do not produce perfect results. We also expected to see traces of fibre numbers in the t-SNE map (there are two sets of 392 fibres in the fibre positioner, so different stars could be observed with a given fibre with a problematic PSF, which might manifest itself as a systematic error in the measured abundance). With the exception of a few ill-performing fibres, we see no relation between measured abundances and fibre used.

One can also notice that every cluster's chemical group is populated with some stars that are not members. This contamination is expected, as 13 abundances are not enough to completely isolate the cluster \citep{ting12, mitschang12, ting15}. We explore these stars further in the case of Pleiades in the next section. We chose the Pleiades for this experiment because it is the only young cluster with distinct kinematics that we can use to verify potential new member candidates.

\section{Chemical populations and new members of the Pleiades}
\label{sec:new}

The Pleiades is a young \citep{brandt15} cluster for which we expect to find some members well away from the centre of the cluster, yet close enough that we can focus only on a small region around the cluster \citep{kroupa02}. The tidal radius of the Pleiades is $\sim6^\circ$ \citep{adams01}, and we do not expect to find any members at distances much larger than this, considering a low number of observed stars in the broader Pleiades region. Despite this, we focus our effort into an area of radius 40$^\circ$ around the Pleiades to demonstrate the method on a larger number of stars. The Pleiades are one of the most northerly objects that GALAH has explored, in one of the K2 fields, so the 40$^\circ$ radius region includes mostly K2-HERMES survey fields, a few pilot survey fields and some regular survey fields at $\delta<+10^\circ$ that have been observed, but most of the 40$^\circ$ region has no observations at all. The Pleiades members were identified by us using cuts in the position, radial velocity, and proper motions (see Appendix \ref{sec:membership}). In this way we identified 27 members. 

\begin{figure}
\includegraphics[width=0.98\columnwidth]{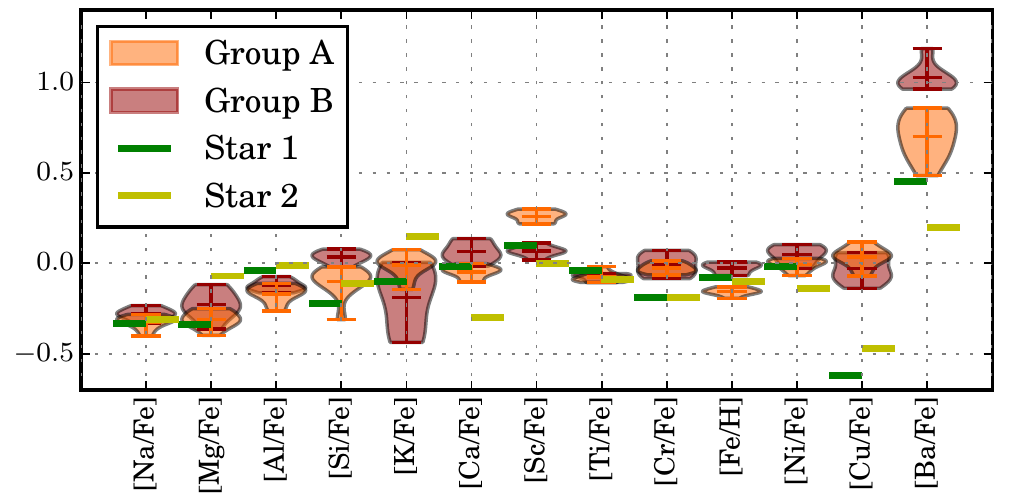}
\includegraphics[width=\columnwidth]{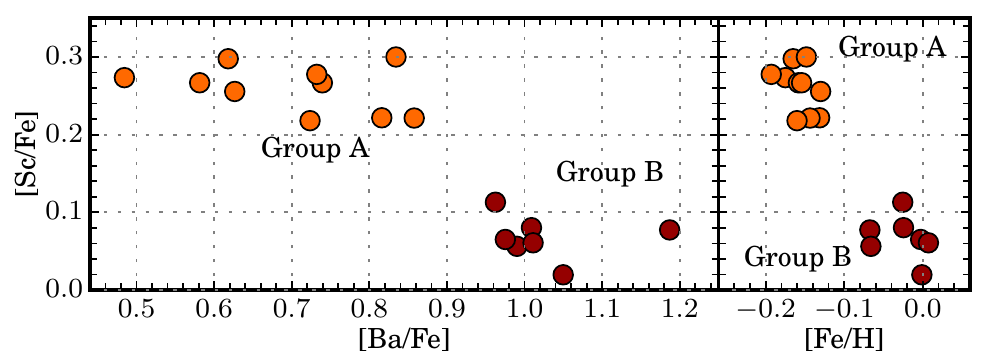}
\includegraphics[width=\columnwidth]{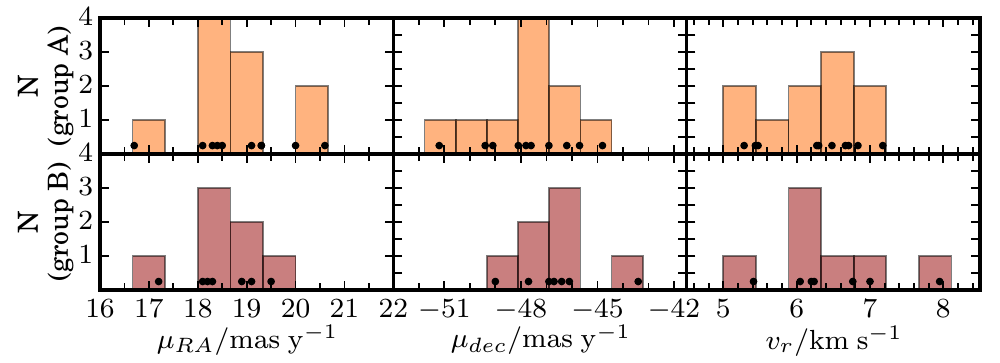}
\caption{Top: The abundances for all 13 elements in groups marked A and B in Figure \ref{fig:proj_Pleiades} and abundances for two new members. Middle: Groups A and B are separated in $\mathrm{[Sc/Fe]}$, $\mathrm{[Ba/Fe]}$, and $\mathrm{[Fe/H]}$ abundances. Abundances for each star are plotted. Bottom: Kinematics for each group.}
\label{fig:pl_det}
\end{figure}

\begin{figure}
\includegraphics[width=0.98\columnwidth]{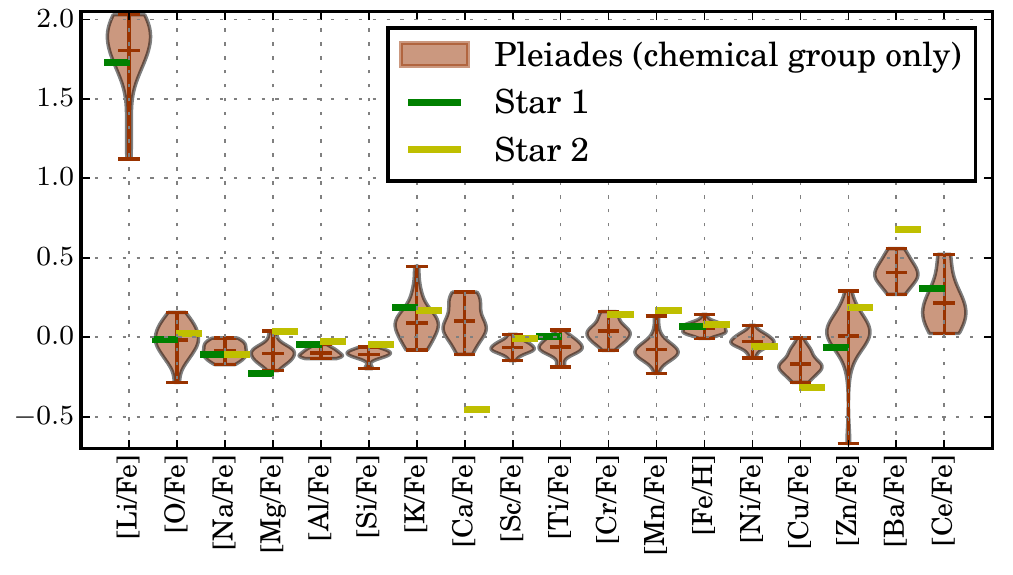}
\caption{SME abundances for the Pleiades stars from the Pleiades chemical group. Note that there is no bimodality in $\mathrm{[Sc/Fe]}$, $\mathrm{[Ba/Fe]}$, and $\mathrm{[Fe/H]}$ like we see with The Cannon abundances.}
\label{fig:pl_det2}
\end{figure}

After making the t-SNE map we can see in Figure \ref{fig:proj_Pleiades} that most Pleiades stars fall into one clump with few contaminating field stars. Closer inspection shows, that the clump consists of two parts (groups marked A and B on Figure \ref{fig:members}) with slightly different abundances of $\mathrm{[Sc/Fe]}$, $\mathrm{[Ba/Fe]}$, and $\mathrm{[Fe/H]}$ (Figure \ref{fig:pl_det}). Stars from both groups are well mixed within 0.5~dex in $\log g$ and within 1000~K in $T_{\mathrm{eff}}$, where stars from group A are on average hotter than stars from group B with a large overlap. We tried to verify the two chemical groups by analysing the Pleiades spectra with SME \citep{valenti96, piskunov16}. We analysed the members in the Pleiades chemical groups, as well as the two new member candidates. With SME we managed to measure more elements (18), but not for every star. Figure \ref{fig:pl_det2} shows the SME results. We cannot confirm the existence of two chemical groups we see in The Cannon abundances, so they are most probably an artefact induced by The Cannon or selection of the training set. The scatter in $\mathrm{[Sc/Fe]}$, $\mathrm{[Ba/Fe]}$, and $\mathrm{[Fe/H]}$ when calculated by the SME is similar to what we get with The Cannon, though. Any decisive conclusions will require more data and more careful analysis of abundances. There are similar observations in the literature \citep{gebran08} matching our large scatter in $\mathrm{[Sc/Fe]}$ and $\mathrm{[Ba/Fe]}$ abundances, but not confirming two separate groups, which could be due to the low number of observed stars in those studies. Binary clusters do exist \citep{slesnick02} and we can speculate that the Pleiades might be a binary or merged cluster, if two separate chemical groups existed. The idea of two populations in Pleiades has even been proposed before \citep{stello01}. In any case we show that features like split chemical groups can be picked out by t-SNE whilst still conserving the hierarchy and putting both groups close together.

We define the Pleiades chemical group by combining all small groups with at least one known Pleiades member that are close to the main group. This decision is arbitrary but conservative. The chemical group is marked with a blue polygon in Figure \ref{fig:proj_Pleiades}.

\begin{figure*}
\includegraphics[width=\textwidth]{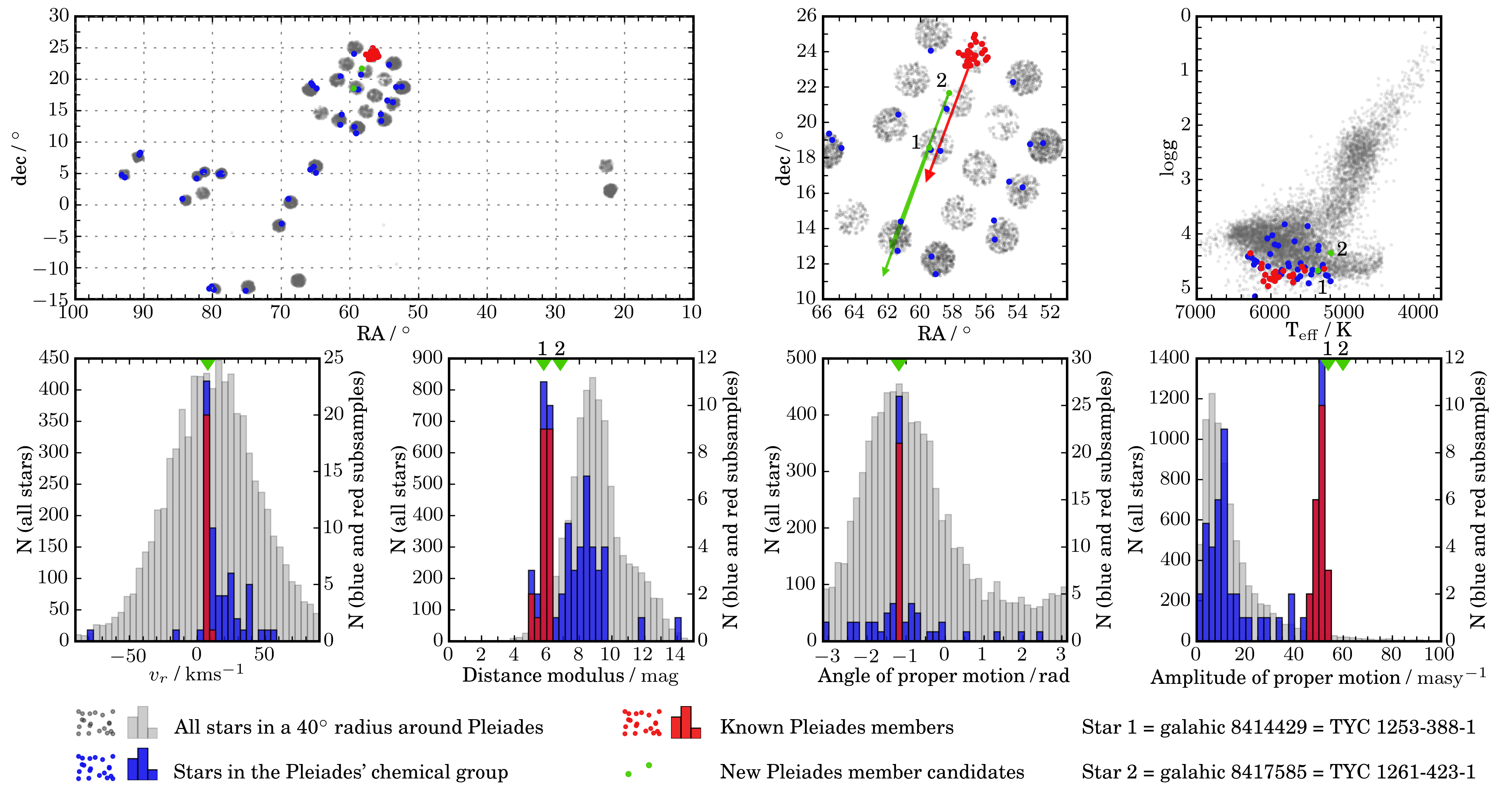}
\caption{Position of the analysed stars on the sky (top-left and top-middle panels), on the HR diagram (top-right panel) and on the radial velocity, distance, and proper motion histograms (bottom row). In grey are plotted all the analysed stars in the 40$^\circ$ radius around Pleiades. In blue are the stars that belong to the Pleiades chemical group (inside the blue polygon in Figure \ref{fig:proj_Pleiades}). In red are known Pleiades members also marked with red symbols in Figure \ref{fig:proj_Pleiades}. Green are two new Pleiades stars that we discovered by chemical tagging. Colour version of this figure is available online.}
\label{fig:pl}
\end{figure*}

We find a small number of contaminating stars in the Pleiades chemical vicinity. Some contamination is expected, therefore we can not claim that all the stars in the chemical group are Pleiades members. For clusters with adequate kinematic information, however, 13 abundances are enough, as we can use independent dimensions: radial velocity, amplitude of the proper motion and direction of the proper motion. We also have photometric information that we combine into a single parameter: the distance \citep{zwitter10}. These four additional dimensions are enough to select only those stars with kinematics and distances that match the Pleiades'. This leaves us with two stars that we claim are candidate Pleiades members. The process of reducing $\sim30$ contaminating stars into two candidate members is illustrated in Figure \ref{fig:pl}. This can also be confirmed with the SME abundances (see Figure \ref{fig:pl_det2}).

There are actually two more stars in the whole 40$^\circ$ radius region that match the Pleiades' kinematics. They are both $>20^\circ$ away from the cluster and do not fall near the Pleiades' chemical group in the t-SNE map. This means that after reducing the number of stars from $\sim9400$ and $\sim2$ coincidental stars to $\sim30$ stars by chemically tagging the cluster, we expect to find $2\frac{30}{9400}\simeq0.0064$ stars that by chance have the same kinematics as the Pleiades and that fall into the cluster's chemical group. We found two, which are therefore Pleiades members with a high degree of certainty. It must be noted, that one of the newly discovered members (star number 2) is a known supercluster candidate \citep{mermilliod97} that escaped our cluster membership determination for being too far from the cluster centre. Star number 1, however, has no relation to the Pleiades in the literature.

\section{Discussion}
\label{sec:dis}
We show that t-SNE is an appropriate algorithm to search for clustering in $\mathcal{C}$-space by demonstrating its performance on all clusters observed in GALAH and K2-HERMES surveys. With reasonable exceptions the method performs well, which we further demonstrate by discovering previously unknown members of Pleiades. 

Perhaps an even more important conclusion is that extremely precise abundances are not always needed to successfully chemically tag cluster members or maybe even field stars. It turns out that the ability to find structures in $\mathcal{C}$-space is more valuable than extreme accuracy of the data. Precise abundances help by reducing the number of outliers and increasing the prominence of the chemical groups, but we show that the groups exist and can be isolated even when chemical homogeneity is only of the order of 0.1~dex. This is best shown by being able to match the two new Pleiades members to the cluster, even when the abundances do not always agree with the abundances of Pleiades. Even with large discrepancy in $\mathrm{[Cu/Fe]}$, the two stars still lie close to the Pleiades manifold and are therefore correctly identified as members. 

We do see some pollution from field stars in the mapped groups associated with the clusters. This is largely due to only 13 elements being used for chemical tagging, two of which (K and Ba) were also given low weights. There are lines of up to 30 elements in the observed wavelength ranges, so in the future we plan to use more elements and reduce the level of pollution. It must be noted, that for some clusters and dissolved groups chemical tagging might remain unfeasible, if their abundances are not distinct enough from the observed population of stars. With the current data a large fraction of the stars are untaggable. In t-SNE maps they are collected in the middle, with no convincing structure visible that would separate them from each other. Again, this is something that more observed elements can solve. Despite the mentioned limitations we conclude that clusters tagged by t-SNE experience low pollution and fairly high efficiency, especially when compared to competing methods.

A further demonstration of the power and robustness of t-SNE is the hierarchy seen in some clusters where more than one population is found. Different populations form different groups, but they all compose one larger group that includes the majority of the cluster members. In a two-dimensional map one can easily see and correctly interpret these structures. This is very hard to do in a higher-dimensional space without a good visualization of all relevant dimensions. Built-in hierarchy also saves us from tagging dwarfs and giants separately, as some studies in the literature do. It can be seen from the $\log g$ panel in Figure \ref{fig:proj_Pleiades}, and even better from the figures in Appendix \ref{sec:appA}, that giants are mostly separated from predominantly dwarf-populated t-SNE maps. t-SNE knows nothing about $\log g$ and the result is purely a consequence of abundances being dependent on gravity. This can be either an abundance pipeline issue or a result of different stellar populations observed.

Adopting the Pleiades parallax of $\varpi=7.48\pm0.03\ \mathrm{mas}$ \citep{gaial17}, median proper motion of our known Pleiades members ($\mu=48.96\ \mathrm{mas\ y^{-1}}$), and proper motion of the newly discovered members, we can calculate, that these stars were scattered out of the cluster $7.9\pm^\infty_{4.6}\ \mathrm{Myr}$ (star 1) and $0.68\pm0.05\ \mathrm{Myr}$ (star 2) ago, at a velocity significantly larger than the escape velocity of the cluster. Considering where in the HR diagram the two stars lie, they are indeed good candidates to be ejected from the cluster due to their low mass.

It was unexpected that our method would work well for globular clusters. Globular clusters have large scatter in light elements are and often chemically inhomogeneous. Results like ones for 47~Tuc were therefore expected. Other globular clusters (with the exception of NGC~2516, where the results are inconclusive) performed well, especially $\omega$~Cen. These globular clusters are interesting targets for further studies with chemical tagging as they are obviously easiest and most reliable to tag. Stars from these globular clusters have abundances distinct enough from fields stars that they were successfully tagged. It is possible that t-SNE is robust enough that with more measured element in the future the tagging will work for 47~Tuc as well.

This is an exploratory study in the early years of
the GALAH survey to demonstrate that it is feasible
to extract homogeneous clusters from a huge
stellar survey. 
There are still numerous improvements
to be made to the stellar abundance determinations,
including 3D non-LTE
atmospheric corrections \citep{lind17},
better absorption line measurements using a photonic
comb \citep{jbh17} and
new data driven techniques to ensure abundance
uniformity across the survey \citep{ness15}.
Thus the efficacy of chemical tagging will only 
improve in the years to come.

One can see that the maps in Figure \ref{fig:members} show many more structures than we have analysed in this paper. We explored other chemical groups and observed some regularities and patterns when kinematics and positions on the sky were inspected. There are, however, some contaminating stars in these groups as well and decisive conclusions are hard to make. We leave the topic of pure blind chemical tagging of field stars for future work. Blind chemical tagging will also be much easier on the set of 30 abundances and $>$300,000 stars soon to be produced by the GALAH collaboration. More observed elements mean much less contamination of chemical groups, so we might soon be able to find long-lost relationships between field stars for the first time with good reliability. We also expect to find many more distinct clusters than we can see in the presented t-SNE maps \citep{jbh16}.

\section*{Acknowledgements}
JK is supported by a Discovery Project grant from the Australian Research Council (DP150104667) awarded to J. Bland-Hawthorn and T. Bedding. TZ acknowledge the financial support from the Slovenian Research Agency (research core funding No. P1-0188). SLM acknowledges support from the Australian Research Council through grant DE140100598. DMN is supported by the Allan C. and Dorothy H. Davis Fellowship. D.B.Z. acknowledges the financial support of the Australian Research Council through grant FT110100793

%%%%%%%%%%%%%%%%%%%%%%%%%%%%%%%%%%%%%%%%%%%%%%%%%%

%%%%%%%%%%%%%%%%%%%% REFERENCES %%%%%%%%%%%%%%%%%%

% The best way to enter references is to use BibTeX:

\bibliographystyle{mnras}
\bibliography{bib}

%%%%%%%%%%%%%%%%%%%%%%%%%%%%%%%%%%%%%%%%%%%%%%%%%%

%%%%%%%%%%%%%%%%% APPENDICES %%%%%%%%%%%%%%%%%%%%%
\clearpage
\appendix

\section{Introduction to \lowercase{t}-SNE}
\label{sec:appB}

Dimensionality reduction methods aim to reduce the number of dimensions while preserving the structure of the data that we are interested in. Here we want to identify groups of data points in a 13-dimensional space. This can be done in the original 13 dimensions, but the visualization would remain a problem. Also, we know that chemical groups are not very distinct or isolated from each other (there are field stars with very similar abundances), so fine-tuning an algorithm in 13 dimensions is nearly impossible. 

\subsubsection*{Linear and non-linear dimensionality reduction}

\begin{figure}
\centering
\includegraphics[width=0.95\columnwidth]{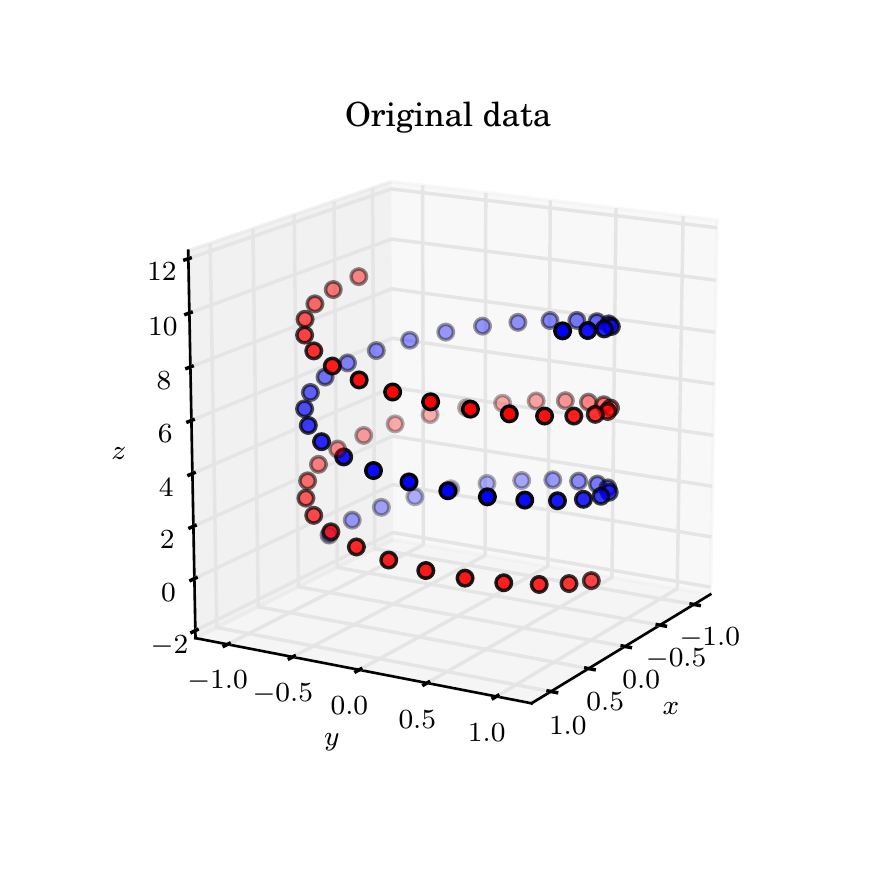}
\includegraphics[width=0.9\columnwidth]{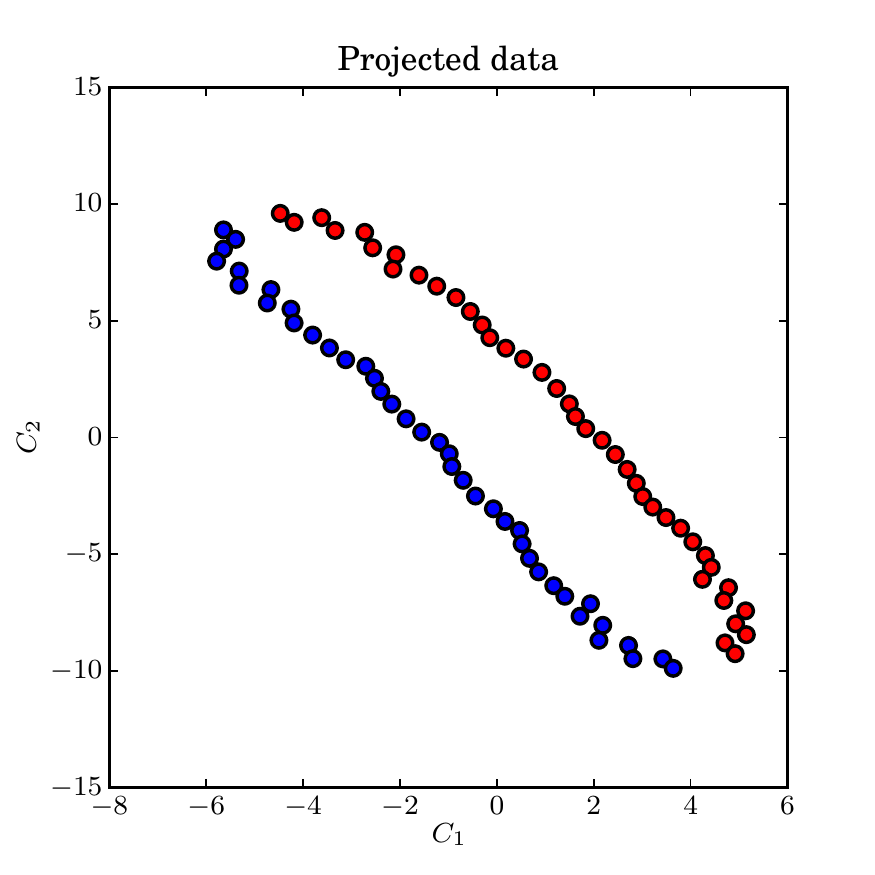}
\caption{t-SNE projection of a double helix into two dimensions. All information about the shape of the structure is lost, but two strands become separated. Colours are used for the illustration purposes only and are not part of the dataset, meaning that t-SNE only knows the coordinates of the points, regardless the colour.}
\label{fig:demo1}
\end{figure}

Linear algorithms, like PCA, are not very suitable for chemical tagging, especially if we first intend to project the data into two dimensions. Even though the abundances of different elements are correlated, the relations are not linear, so the projection into only two dimensions is unable to conserve the structure from the high dimensional space. We illustrate the problem in Figure \ref{fig:demo1}. A double helix constructed in three dimensions is projected into two dimensions with t-SNE. One can see that both strands of the double helix become separated in the t-SNE projection. A linear method would not be able to produce that. Any linear projection will result in either a ring or two interlocking ``waves'' of points. But most non-linear dimensionality reduction algorithms will be able to deal with this example. 

\subsubsection*{Local and global algorithms}

Most differences between non-linear algorithms are in the mapping of local and global details. Imagine the previous example, but with added outliers somewhere far away from the double helix. Global algorithms will try to preserve the structure at all scales. Points that are close together will remain close together in the two dimensional projection and distant outliers will be placed far away. The double helix structure, however, might not be resolved as the distances between the points in the double helix are negligible compared to the distances to the outliers. With local algorithms, the position of each point in the two-dimensional space is determined only by its nearest neighbours. Local algorithms can ``see'' the two strands of the double helix, but will fail to map the outliers as their nearest neighbours are points on the double helix. Positions of outliers on the two-dimensional map will therefore be meaningless. 

t-SNE is able to adapt itself to local density. It can map datasets with a high variation in density, so it is able to resolve small, local details, as well as the global picture.

\begin{figure*}
\includegraphics[width=\textwidth]{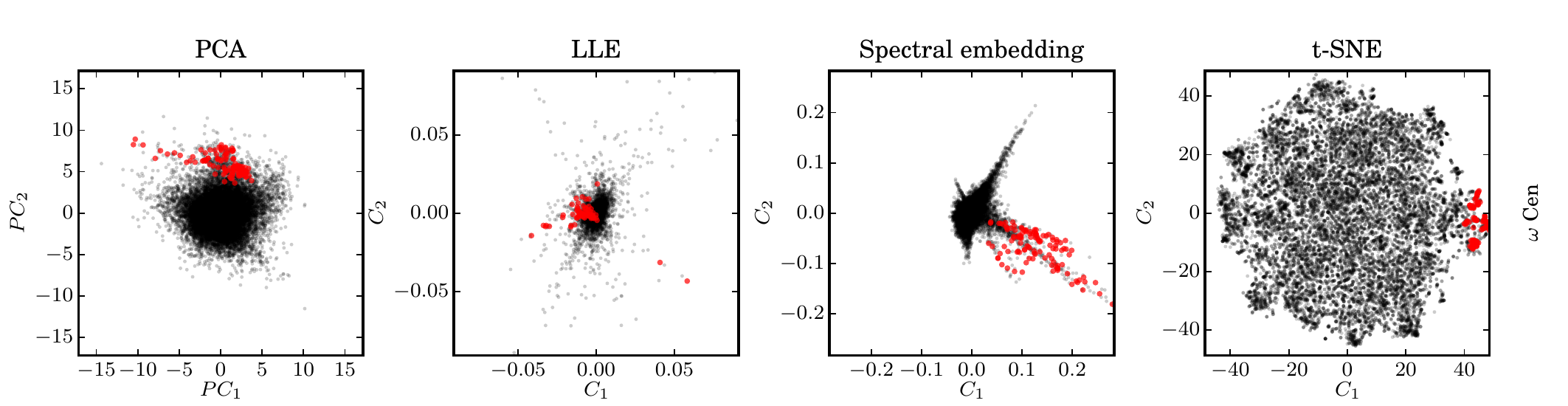}
\includegraphics[width=\textwidth]{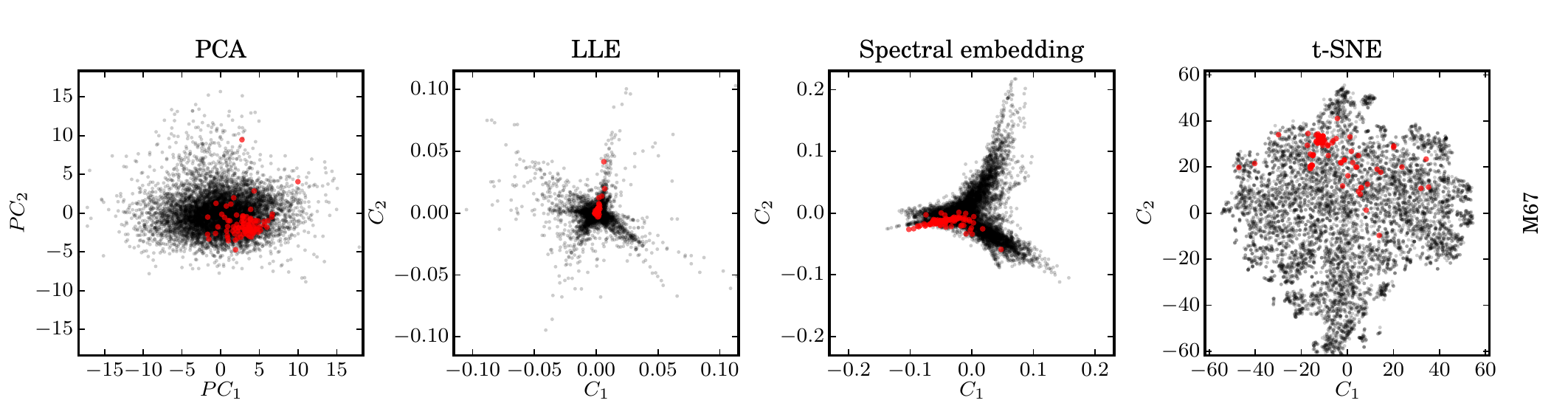}
\caption{Comparison of the performance of PCA, Locally linear embedding (LLE), Spectral embedding and t-SNE. Methods follow from the most to the least linear (left to right). In the top row, we compare the four methods on the case of $\omega$~Cen (an easy case) and in the bottom row for M67 (a harder case). Known cluster members are marked in red. Notice how efficiently t-SNE covers the plane and how many more distinct groups one can see.}
\label{fig:demo2}
\end{figure*}

Two examples in Figure \ref{fig:demo2} show a projection of elemental abundances for 13 elements for stars around $\omega$~Cen and M67 made with four different algorithms. $\omega$~Cen stars have peculiar enough abundances that they stand out from rest of the stars. M67 stars have abundances that are much closer to field stars, so it is one of the hardest clusters to chemically tag in our sample. One can see in Figure \ref{fig:demo2} that only t-SNE is able to create a map where cluster stars lie in the region not densely populated by field stars. LLE and spectral embedding are able to identify some groups of stars that are mapped into rays extending from the central region. They can not identify structures inside the central region, while t-SNE distributes stars pretty evenly, identifying many groups where LLE and spectral embedding fail. PCA and spectral embedding are able to tag $\omega$~Cen stars, but the pollution from the field stars is much higher than in t-SNE maps.

\subsubsection*{t-SNE algorithm}

The main difference between t-SNE and other methods discussed here is that t-SNE does not use a fixed number of nearest neighbours to determine the position of a point in the two dimensional map. Instead, a neighbour is only used with a probability that depends on the distance between the data-points (Equation \ref{eq:p1}) under a Student's t-distribution. This way, even points far away can be considered to calculate the position of a point in two dimensions. Student's t-distribution is heavy-tailed compared to a Gaussian, so data points that are far away from each other can actually be used as ``nearest neighbours''. Therefore t-SNE is sensitive to fine and global structures. If we imagine a two-dimensional projection, not necessarily an optimal one, a similar probability can be calculated in two dimensions (Equation \ref{eq:p2}). An optimal projection is the one, where both probabilities are as close to each other as possible. Such projection is found by minimizing the sum of Kullback-Leibler divergences for each data-point. This is computationally slow. 

\subsubsection*{Role of perplexity}

$\sigma_i$ in Equation \ref{eq:p1} is a variance of the Student's t distribution. It should be smaler for data-points in denser regions of the high dimensional space and larger in sparse regions, if regions of different densities should be mapped simultaneously. t-SNE finds $\sigma_i$ for every data-point, such that the perplexity
\begin{equation}
Perp_i=2^{-\sum_j p_{j|i} \log_2 p_{j|i}}
\end{equation}
equals to the perplexity specified by the user. The specified perplexity is typically in the range of 5 to 50.

Perplexity therefore controls whether t-SNE is more sensitive to large or small structures. Its role is similar to the number of nearest neighbours used by most other dimensionality reduction methods. See \citet{tsne16} for an excellent interactive demonstration of the role of perplexity, as well as other caveats of the t-SNE method.

\subsubsection*{Visualization}

t-SNE does not retain distances but probabilities. It is also highly non-linear, so the values for the coordinates of data points in the projected map are meaningless. So are the units. They can be tought of as coordinates of an image. In Figures \ref{fig:demo1} and \ref{fig:demo2}, we show the two axes with corresponding numerical values to spare the reader any confusion. In the main text we omit them altogether, so the projected map is actually treated as an image.

Kullback-Leibler divergence is minimized by a gradient descent method initialiezd by random sampling, so t-SNE will produce a slightly different map on every run, even with the same data and same perplexity.  In the case of chemical tagging, the only visible difference was a random rotation of the map. In general, one can run t-SNE many times and use the projection with the lowest Kullback-Leibler divergence.

\section{\lowercase{t}-SNE projections of the remaining clusters}
\label{sec:appA}

t-SNE maps for the 8 clusters, not presented in the main text, are collected here.

\begin{figure*}
\begin{center}
M30\\
\end{center}
\includegraphics[width=0.24\textwidth]{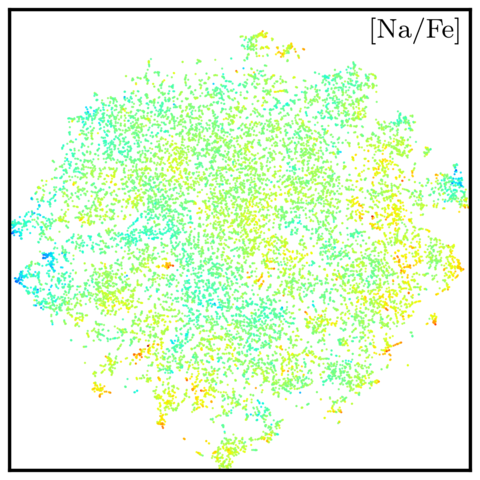}
\includegraphics[width=0.24\textwidth]{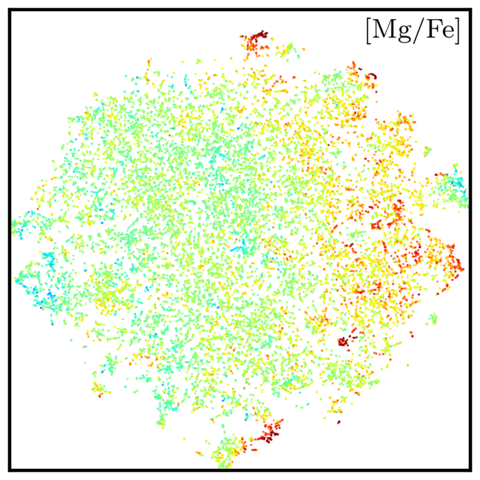}
\includegraphics[width=0.24\textwidth]{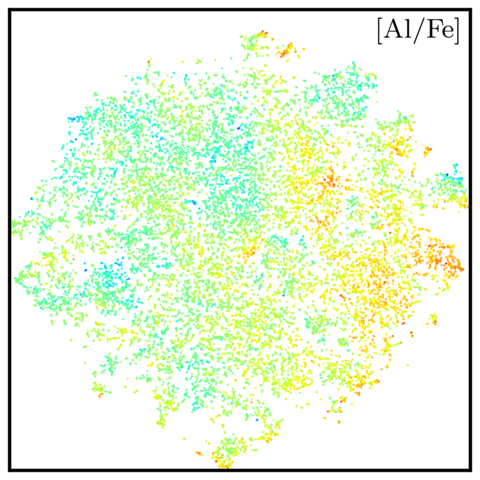}
\includegraphics[width=0.24\textwidth]{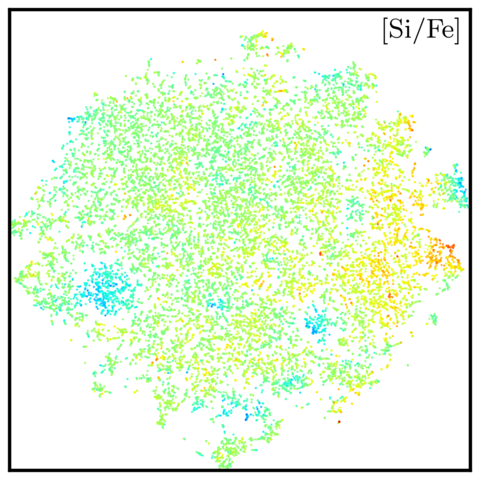}\\
\includegraphics[width=0.24\textwidth]{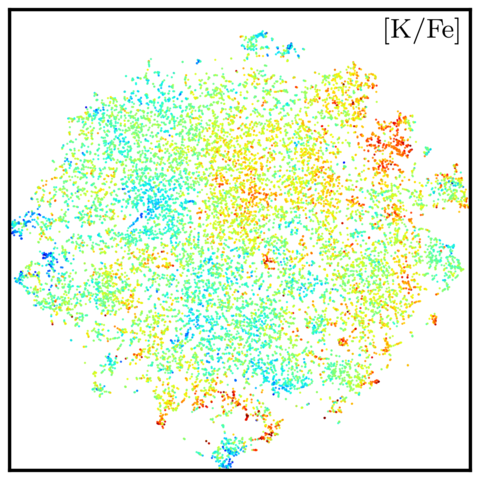}
\includegraphics[width=0.24\textwidth]{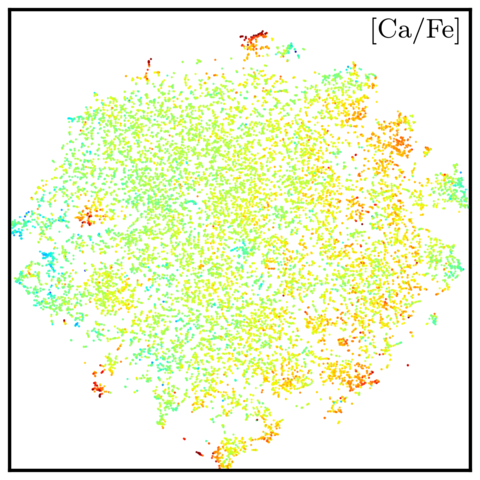}
\includegraphics[width=0.24\textwidth]{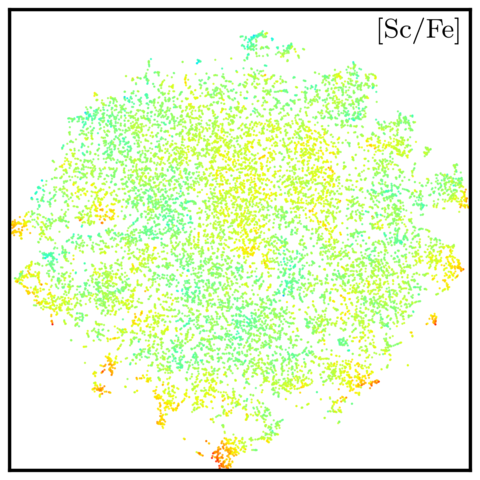}
\includegraphics[width=0.24\textwidth]{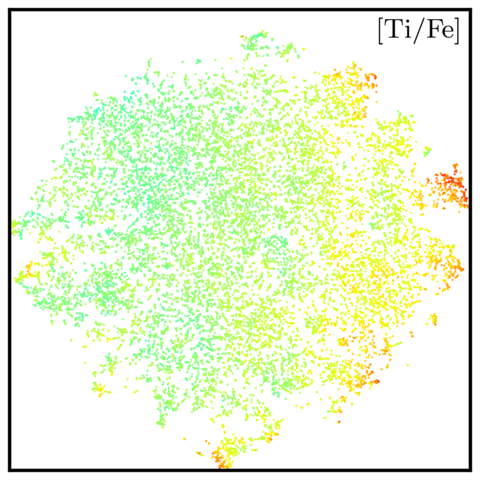}\\
\includegraphics[width=0.24\textwidth]{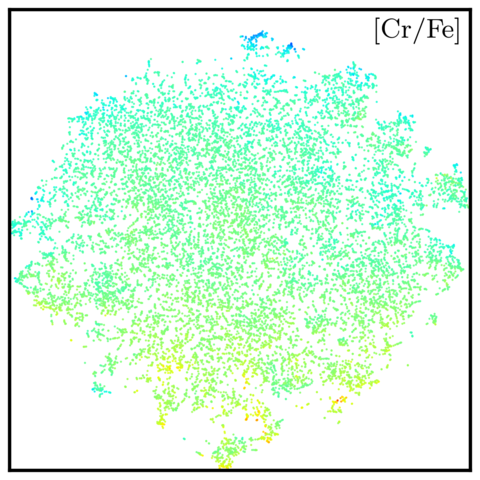}
\includegraphics[width=0.24\textwidth]{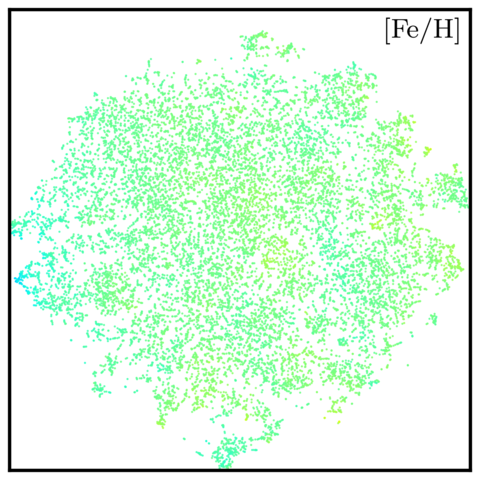}
\includegraphics[width=0.24\textwidth]{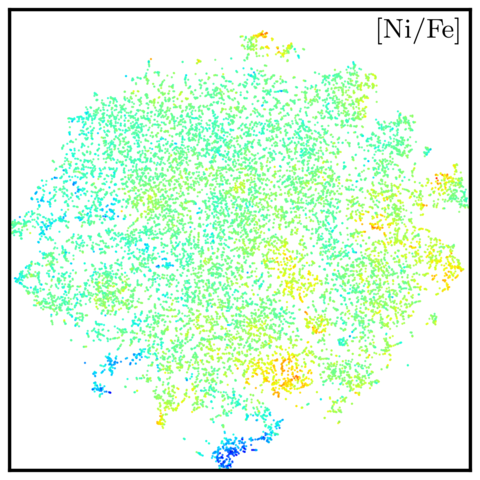}
\includegraphics[width=0.24\textwidth]{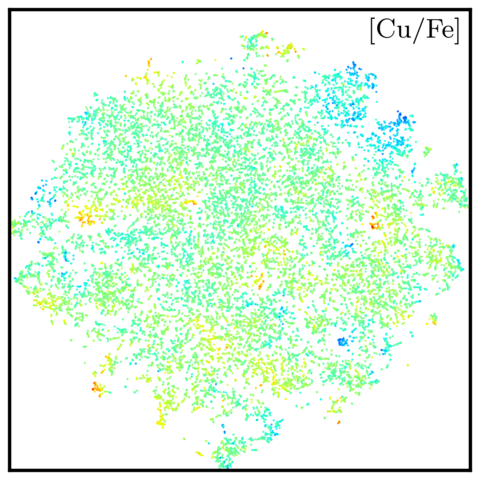}\\
\includegraphics[width=0.24\textwidth]{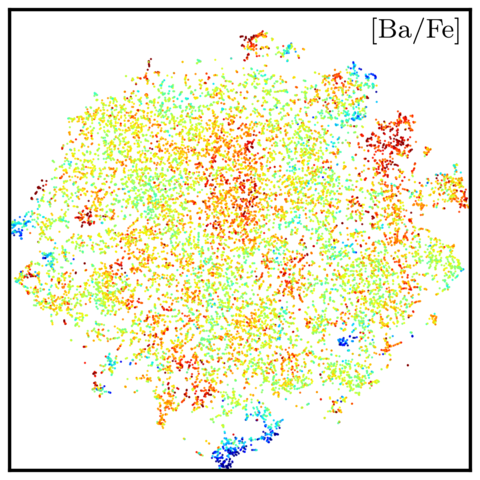}
\includegraphics[width=0.24\textwidth]{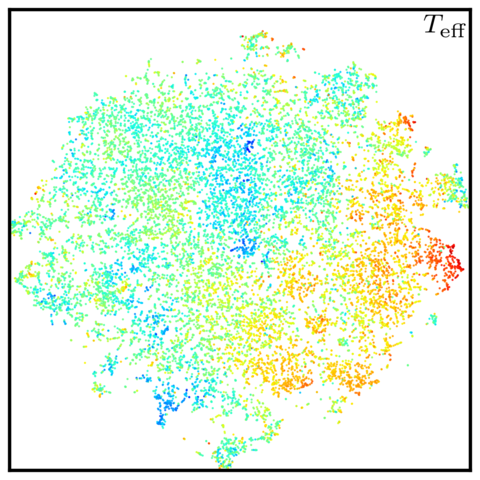}
\includegraphics[width=0.24\textwidth]{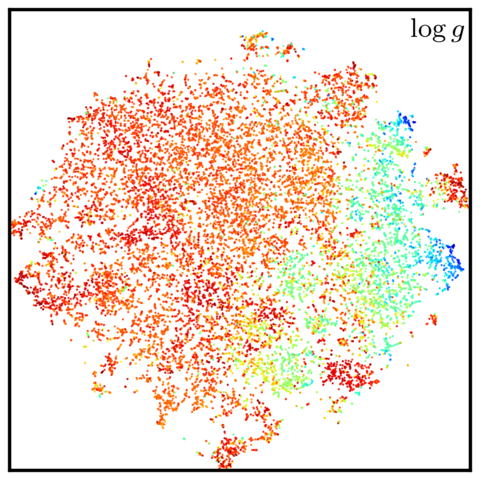}
\includegraphics[width=0.24\textwidth]{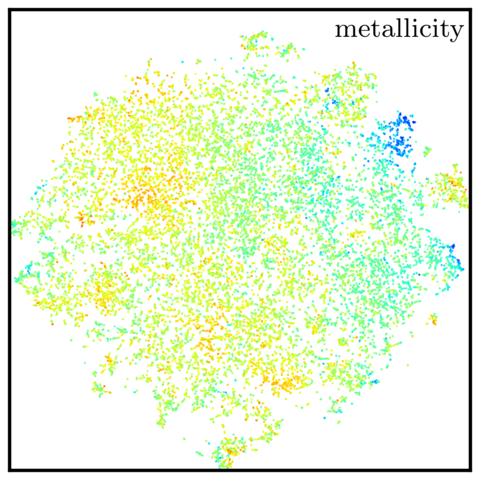}\\
\includegraphics[width=0.24\textwidth]{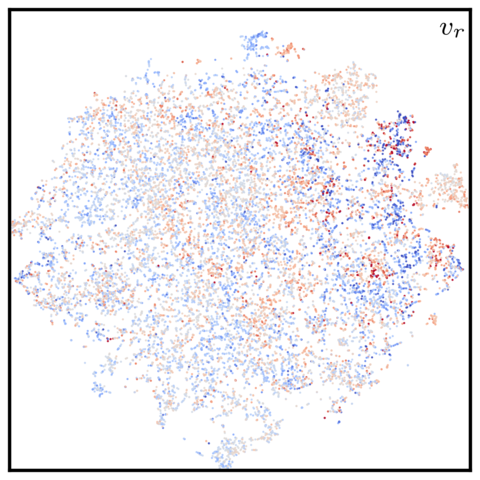}
\includegraphics[width=0.24\textwidth]{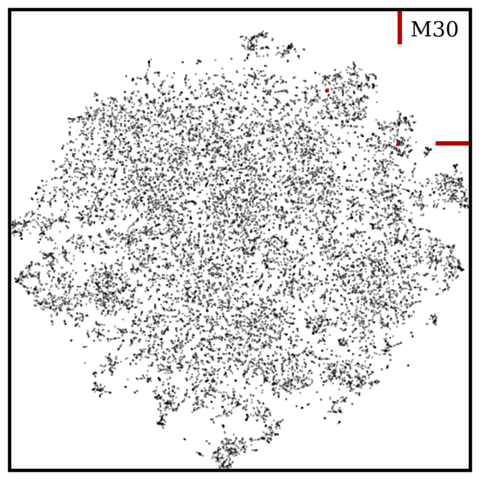}\hspace{0.5em}
\includegraphics[width=0.06\textwidth]{colorbar_abund.png}\hspace{2em}
\includegraphics[width=0.06\textwidth]{colorbar_teff.png}\hspace{2em}
\includegraphics[width=0.06\textwidth]{colorbar_logg.png}\hspace{2em}
\includegraphics[width=0.06\textwidth]{colorbar_feh.png}\hspace{2em}
\includegraphics[width=0.06\textwidth]{colorbar_rv.png}
\caption{t-SNE projection of 20,254 stars in a 35$^\circ$ radius around M30. Abundances of 13 elements used to create the projection are colour-coded. $T_\mathrm{eff}$, $\log g$, metallicity and radial velocity colour-codes are also plotted. The last panel shows the stars that belong to the cluster in red and field stars in grey. 3 out of 4 M30 stars lie in a tight group in the top-right part of the map (marked with dashes at the edge of the plot).}
\label{fig:proj_M30}
\end{figure*}

\begin{figure*}
\begin{center}
$\omega$ Cen\\
\end{center}
\includegraphics[width=0.24\textwidth]{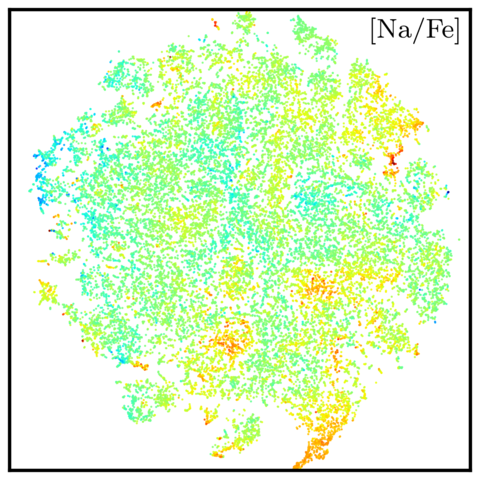}
\includegraphics[width=0.24\textwidth]{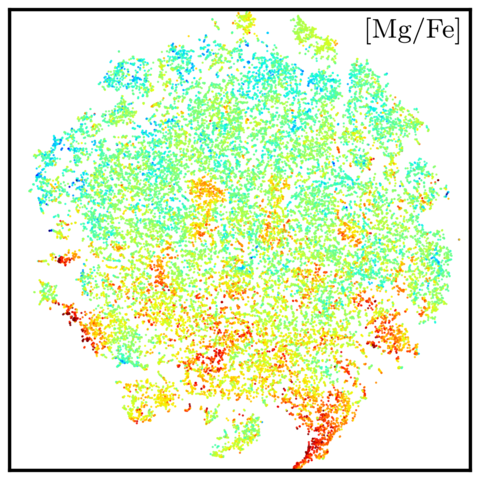}
\includegraphics[width=0.24\textwidth]{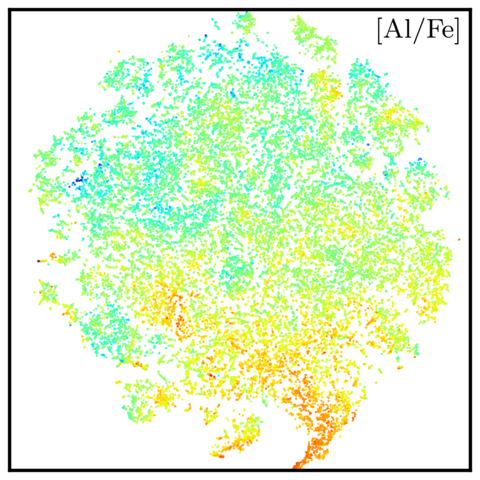}
\includegraphics[width=0.24\textwidth]{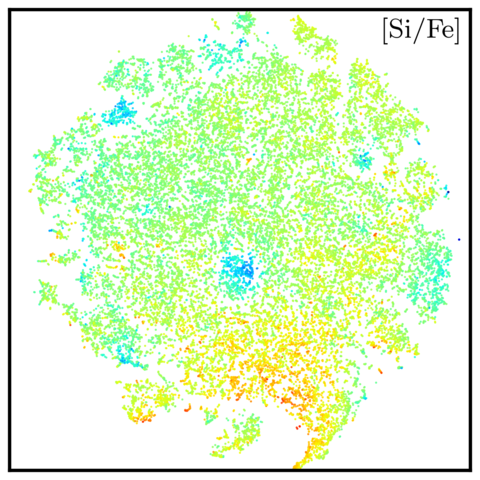}\\
\includegraphics[width=0.24\textwidth]{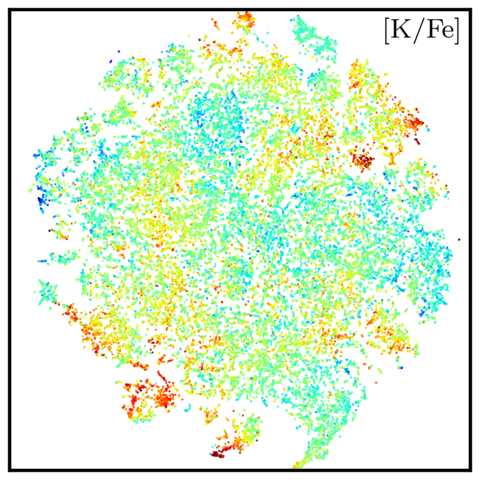}
\includegraphics[width=0.24\textwidth]{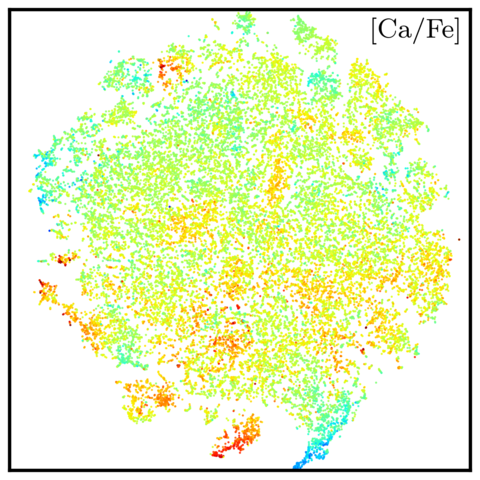}
\includegraphics[width=0.24\textwidth]{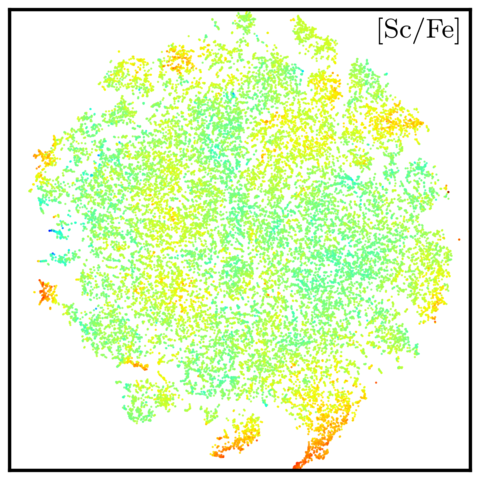}
\includegraphics[width=0.24\textwidth]{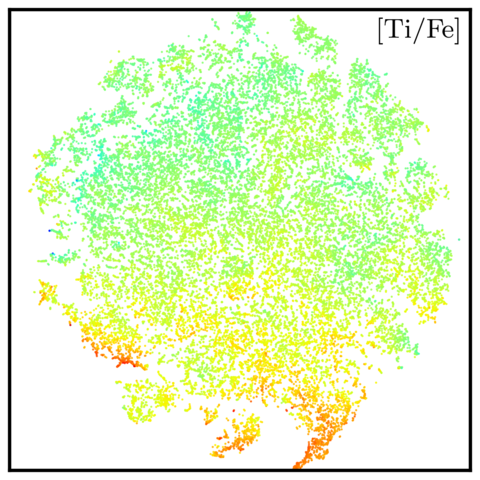}\\
\includegraphics[width=0.24\textwidth]{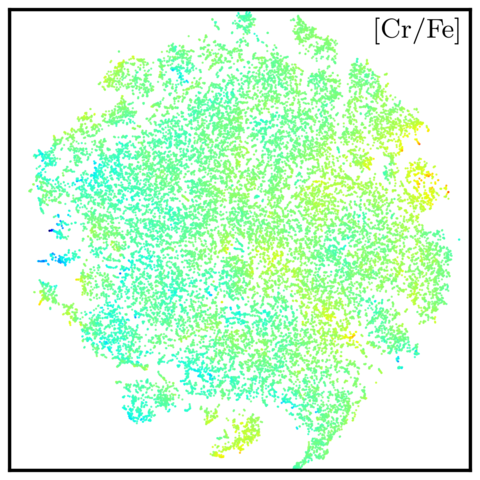}
\includegraphics[width=0.24\textwidth]{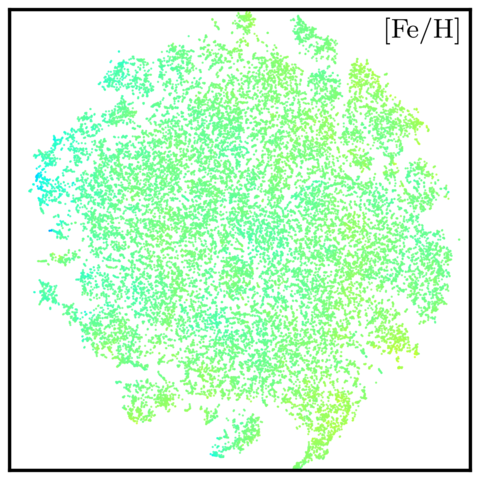}
\includegraphics[width=0.24\textwidth]{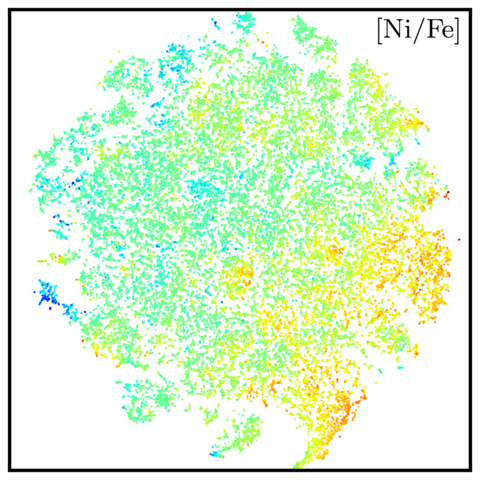}
\includegraphics[width=0.24\textwidth]{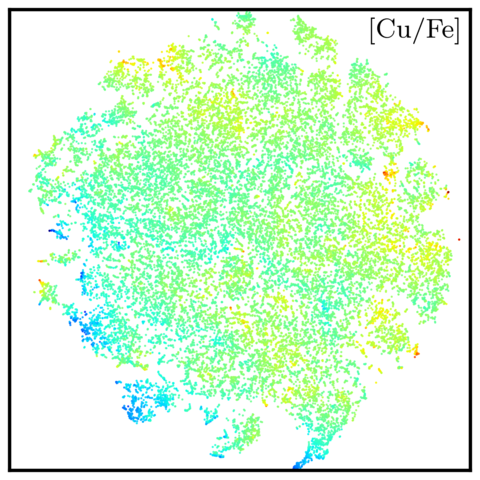}\\
\includegraphics[width=0.24\textwidth]{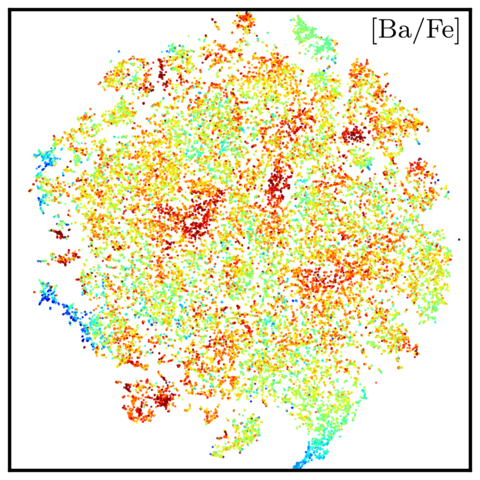}
\includegraphics[width=0.24\textwidth]{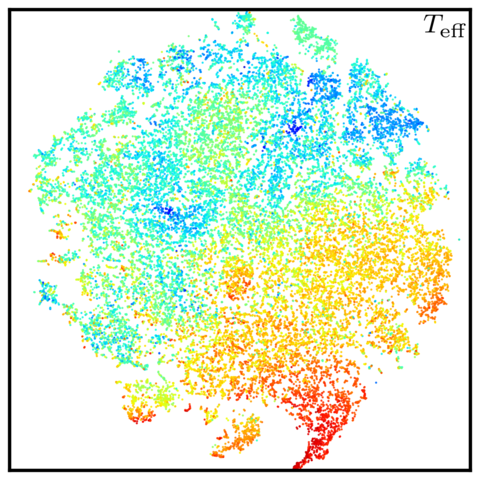}
\includegraphics[width=0.24\textwidth]{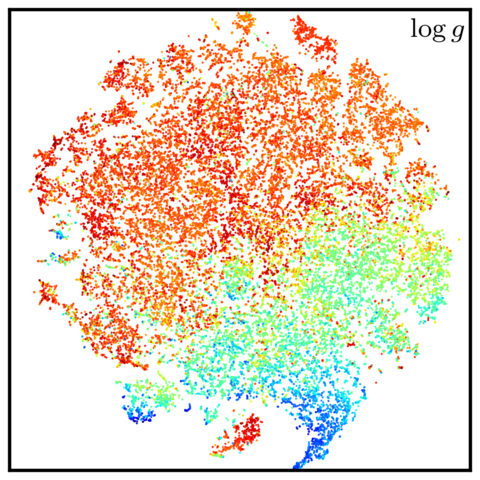}
\includegraphics[width=0.24\textwidth]{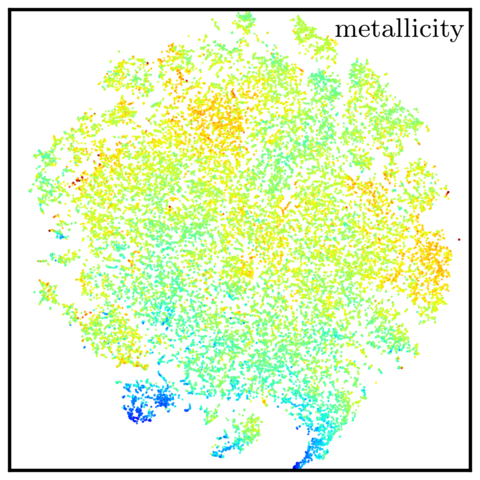}\\
\includegraphics[width=0.24\textwidth]{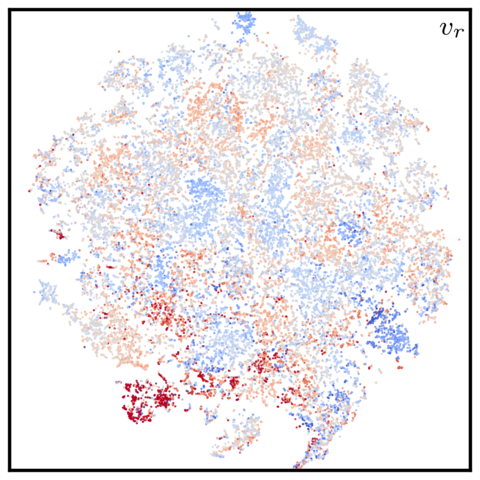}
\includegraphics[width=0.24\textwidth]{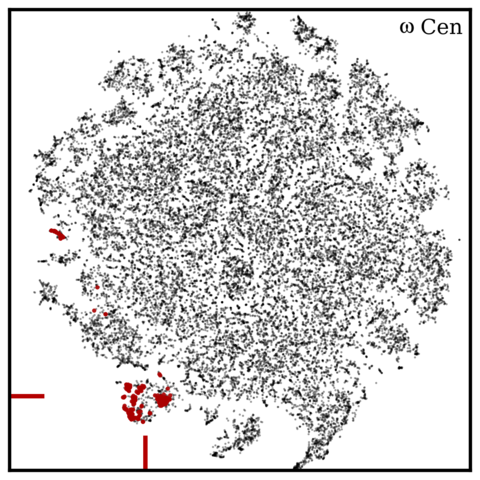}\hspace{0.5em}
\includegraphics[width=0.06\textwidth]{colorbar_abund.png}\hspace{2em}
\includegraphics[width=0.06\textwidth]{colorbar_teff.png}\hspace{2em}
\includegraphics[width=0.06\textwidth]{colorbar_logg.png}\hspace{2em}
\includegraphics[width=0.06\textwidth]{colorbar_feh.png}\hspace{2em}
\includegraphics[width=0.06\textwidth]{colorbar_rv.png}
\caption{t-SNE projection of 33,281 stars in a 30$^\circ$ radius around $\omega$~Cen. Abundances of 13 elements used to create the projection are colour-coded. $T_\mathrm{eff}$, $\log g$, metallicity and radial velocity colour-codes are also plotted. The last panel shows the stars that belong to the cluster in red and field stars in grey. 101 out of 230 $\omega$~Cen stars lie in a tight group in the bottom part of the map (marked with dashes at the edge of the plot) and additional 106 in a more sparse group next to it. }
\label{fig:proj_omegaCen}
\end{figure*}

\begin{figure*}
\begin{center}
NGC288\\
\end{center}
\includegraphics[width=0.24\textwidth]{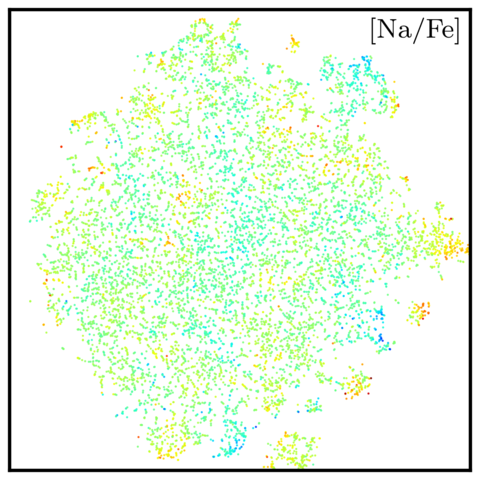}
\includegraphics[width=0.24\textwidth]{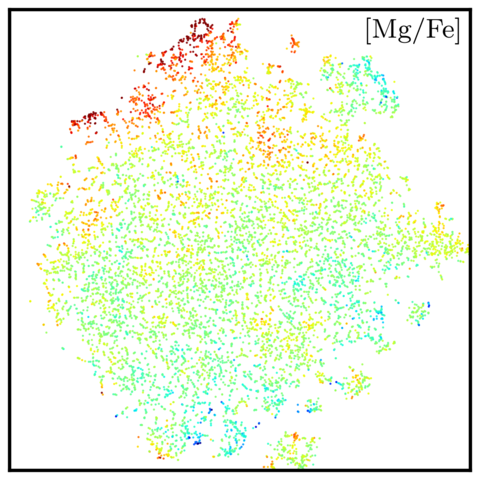}
\includegraphics[width=0.24\textwidth]{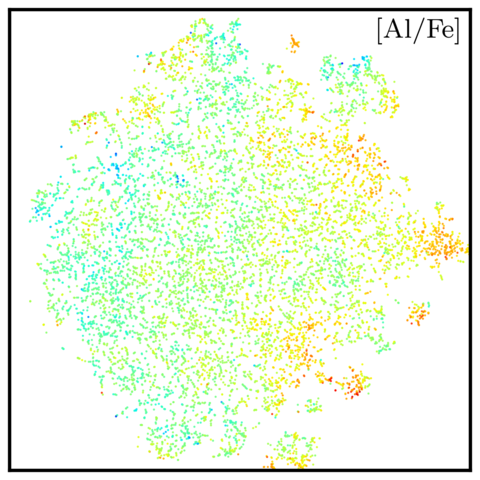}
\includegraphics[width=0.24\textwidth]{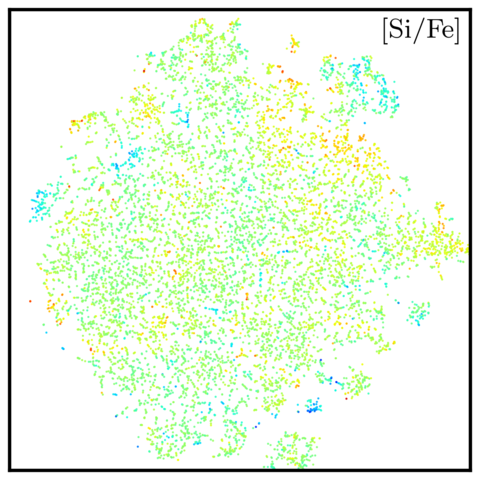}\\
\includegraphics[width=0.24\textwidth]{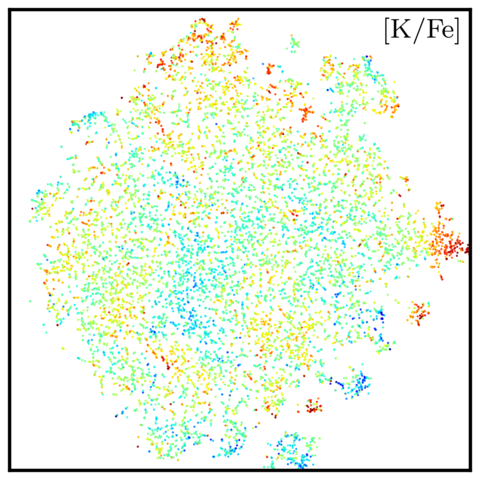}
\includegraphics[width=0.24\textwidth]{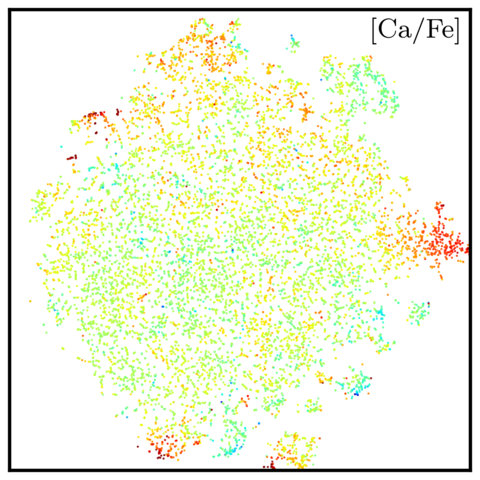}
\includegraphics[width=0.24\textwidth]{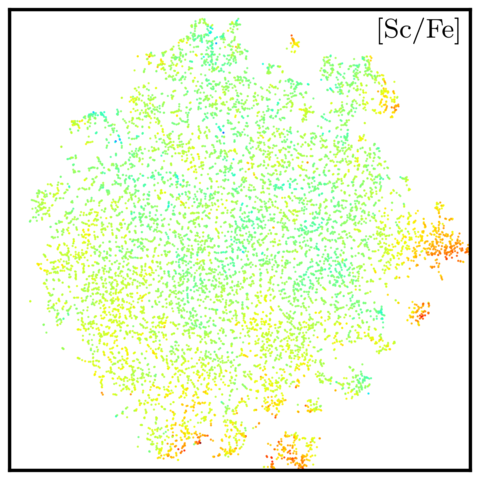}
\includegraphics[width=0.24\textwidth]{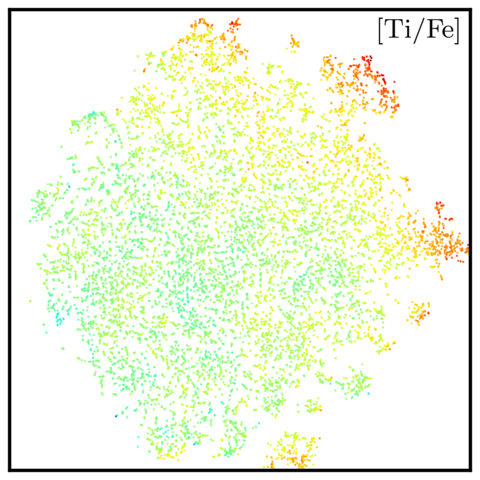}\\
\includegraphics[width=0.24\textwidth]{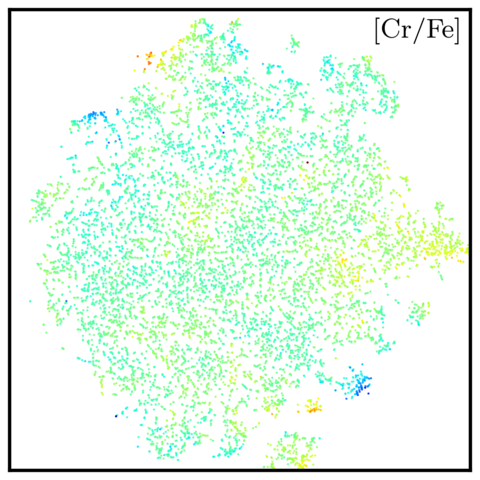}
\includegraphics[width=0.24\textwidth]{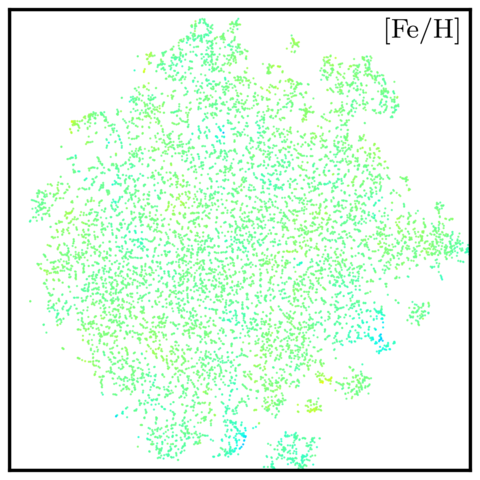}
\includegraphics[width=0.24\textwidth]{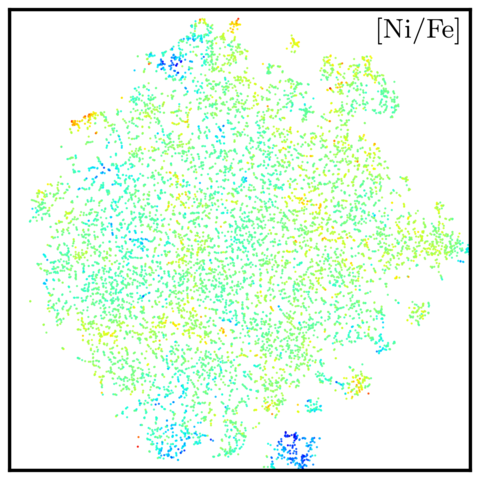}
\includegraphics[width=0.24\textwidth]{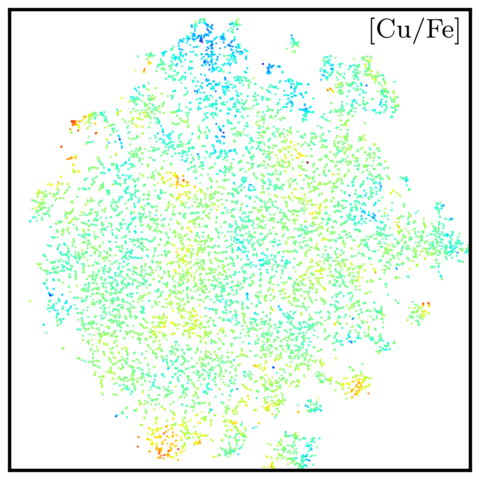}\\
\includegraphics[width=0.24\textwidth]{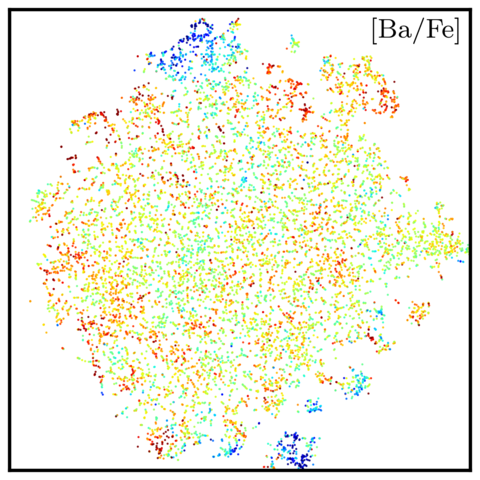}
\includegraphics[width=0.24\textwidth]{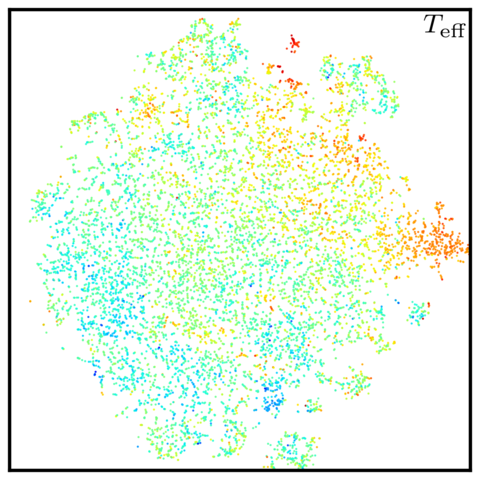}
\includegraphics[width=0.24\textwidth]{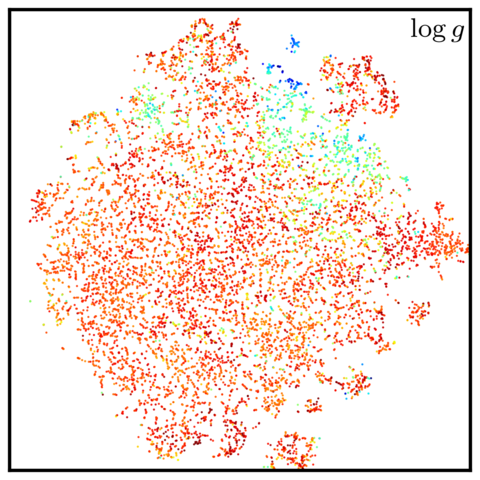}
\includegraphics[width=0.24\textwidth]{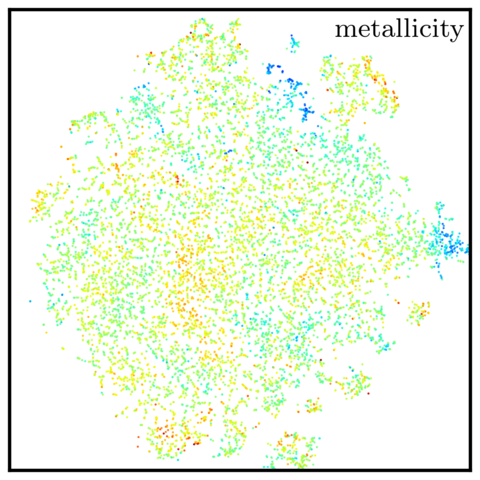}\\
\includegraphics[width=0.24\textwidth]{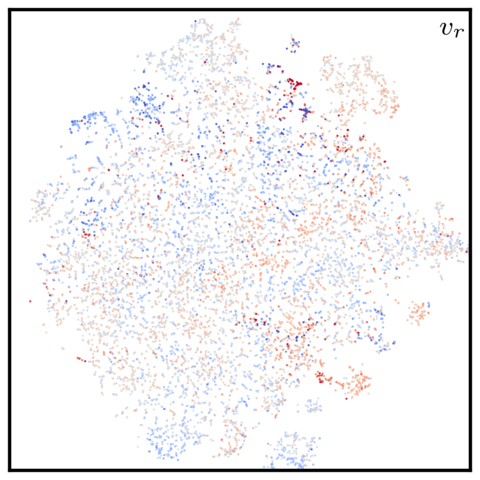}
\includegraphics[width=0.24\textwidth]{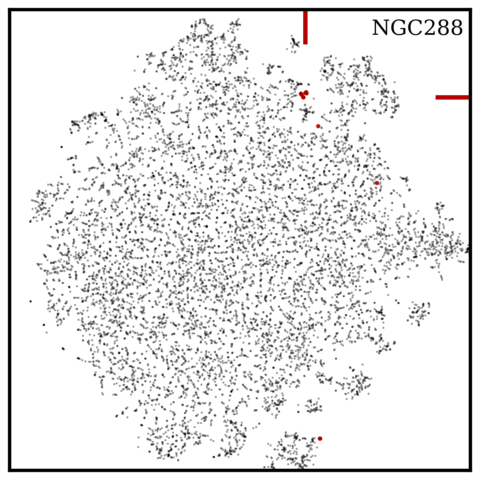}\hspace{0.5em}
\includegraphics[width=0.06\textwidth]{colorbar_abund.png}\hspace{2em}
\includegraphics[width=0.06\textwidth]{colorbar_teff.png}\hspace{2em}
\includegraphics[width=0.06\textwidth]{colorbar_logg.png}\hspace{2em}
\includegraphics[width=0.06\textwidth]{colorbar_feh.png}\hspace{2em}
\includegraphics[width=0.06\textwidth]{colorbar_rv.png}
\caption{t-SNE projection of 11,535 stars in a 45$^\circ$ radius around NGC288. Abundances of 13 elements used to create the projection are colour-coded. $T_\mathrm{eff}$, $\log g$, metallicity and radial velocity colour-codes are also plotted. The last panel shows the stars that belong to the cluster in red and field stars in grey. 10 out of 14 NGC288 stars lie in a tight group in the top part of the map (marked with dashes at the edge of the plot).}
\label{fig:proj_NGC288}
\end{figure*}

\begin{figure*}
\begin{center}
NGC362\\
\end{center}
\includegraphics[width=0.24\textwidth]{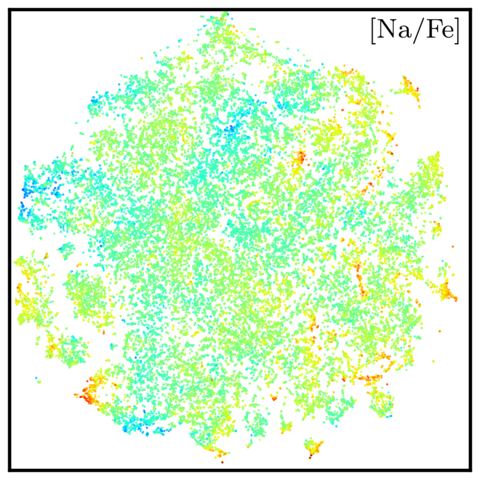}
\includegraphics[width=0.24\textwidth]{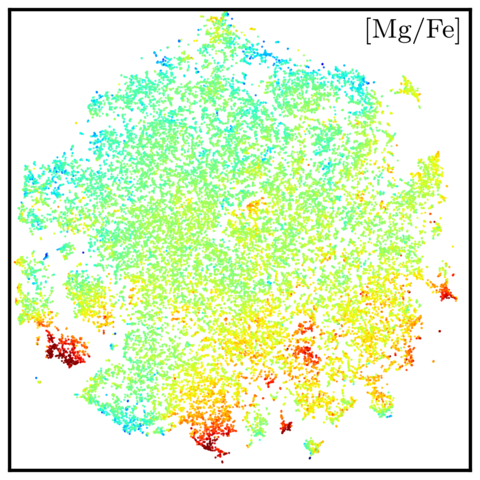}
\includegraphics[width=0.24\textwidth]{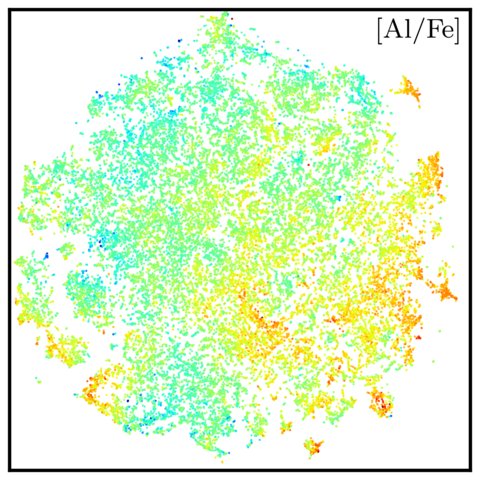}
\includegraphics[width=0.24\textwidth]{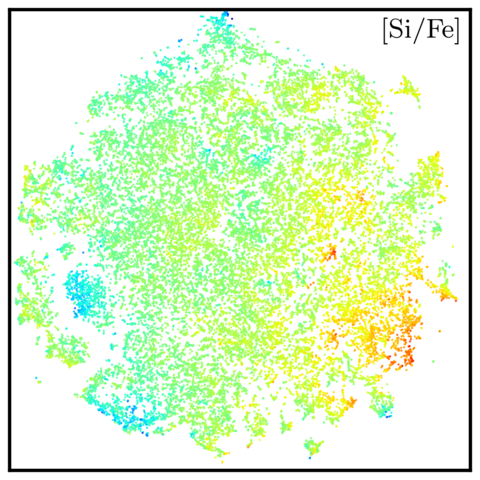}\\
\includegraphics[width=0.24\textwidth]{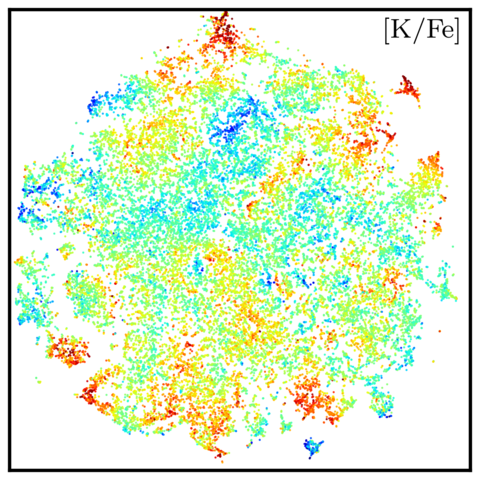}
\includegraphics[width=0.24\textwidth]{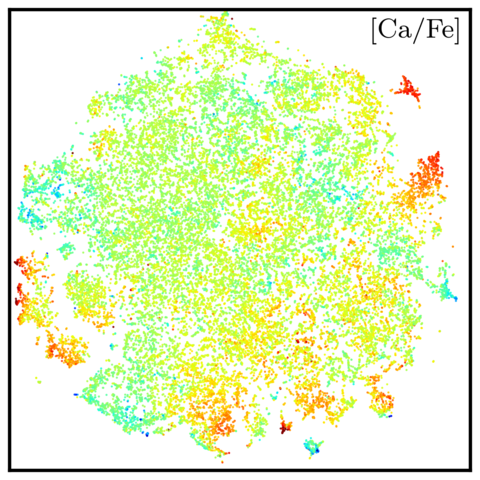}
\includegraphics[width=0.24\textwidth]{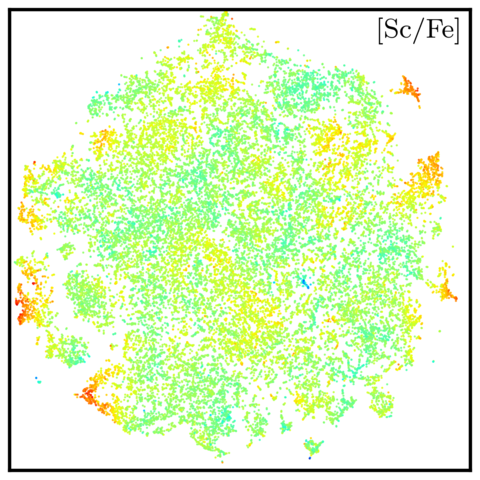}
\includegraphics[width=0.24\textwidth]{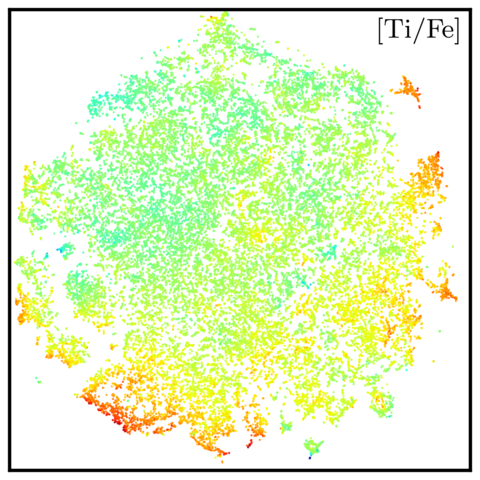}\\
\includegraphics[width=0.24\textwidth]{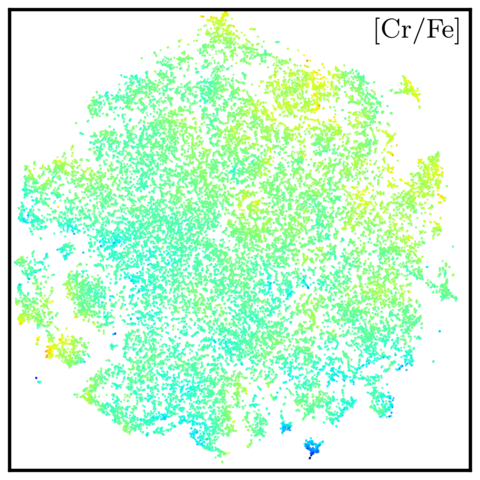}
\includegraphics[width=0.24\textwidth]{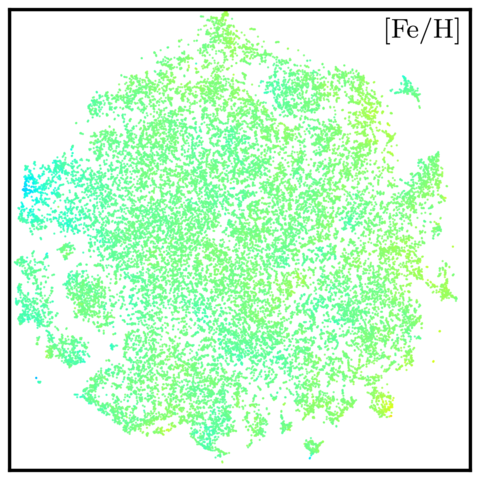}
\includegraphics[width=0.24\textwidth]{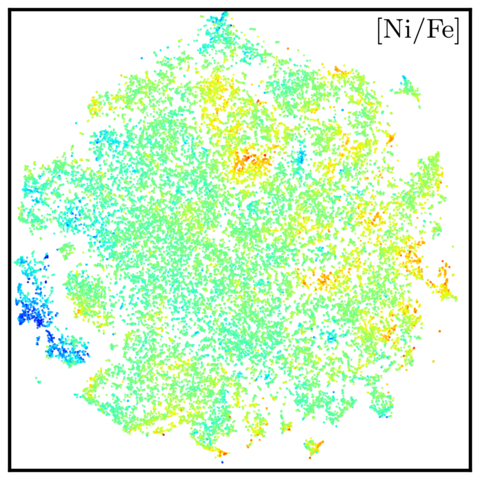}
\includegraphics[width=0.24\textwidth]{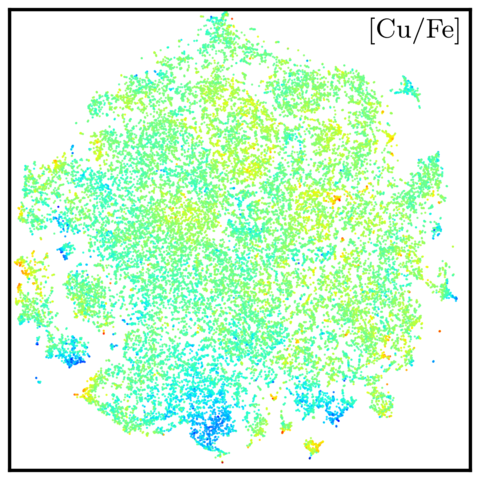}\\
\includegraphics[width=0.24\textwidth]{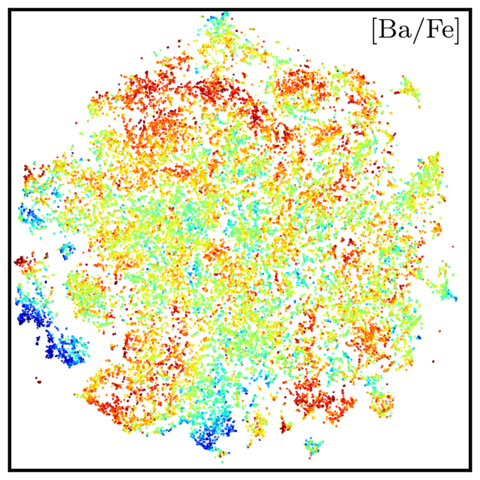}
\includegraphics[width=0.24\textwidth]{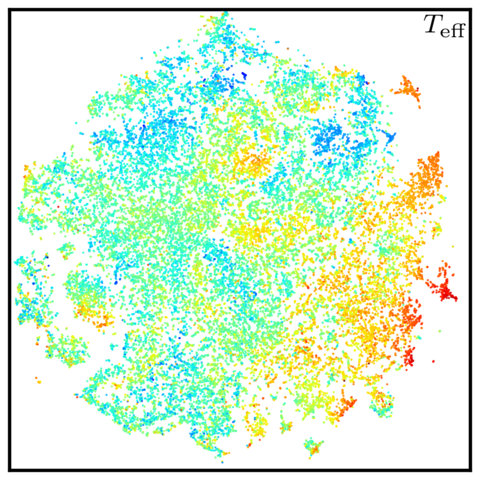}
\includegraphics[width=0.24\textwidth]{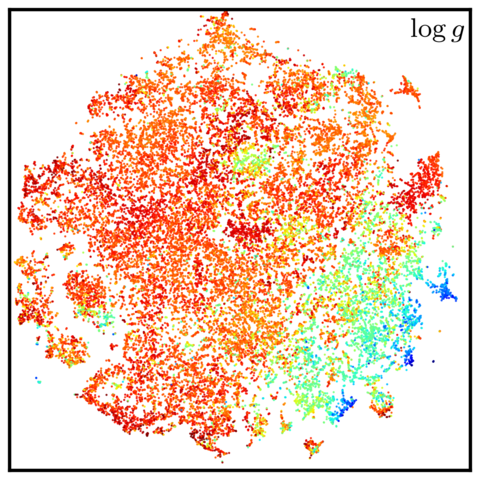}
\includegraphics[width=0.24\textwidth]{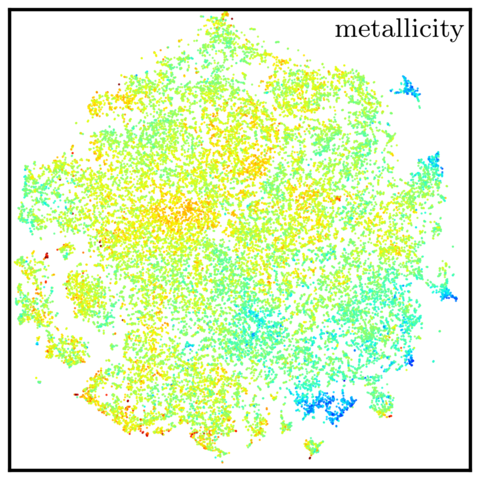}\\
\includegraphics[width=0.24\textwidth]{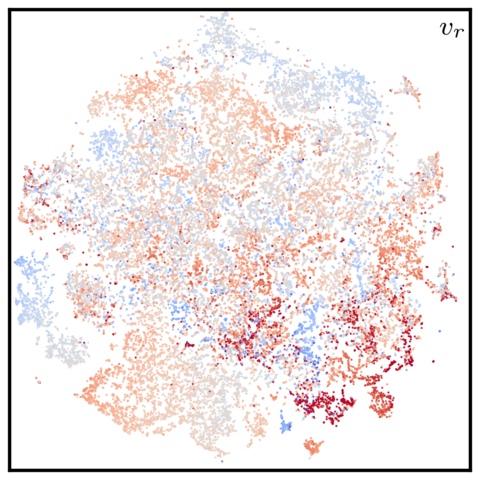}
\includegraphics[width=0.24\textwidth]{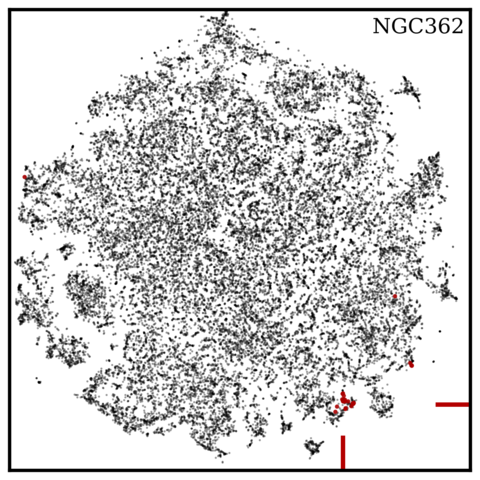}\hspace{0.5em}
\includegraphics[width=0.06\textwidth]{colorbar_abund.png}\hspace{2em}
\includegraphics[width=0.06\textwidth]{colorbar_teff.png}\hspace{2em}
\includegraphics[width=0.06\textwidth]{colorbar_logg.png}\hspace{2em}
\includegraphics[width=0.06\textwidth]{colorbar_feh.png}\hspace{2em}
\includegraphics[width=0.06\textwidth]{colorbar_rv.png}
\caption{t-SNE projection of 41,578 stars in a 35$^\circ$ radius around NGC362. Abundances of 13 elements used to create the projection are colour-coded. $T_\mathrm{eff}$, $\log g$, metallicity and radial velocity colour-codes are also plotted. The last panel shows the stars that belong to the cluster in red and field stars in grey. 23 out of 27 NGC362 stars lie in a group in the bottom part of the map (marked with dashes at the edge of the plot).}
\label{fig:proj_NGC362}
\end{figure*}

\begin{figure*}
\begin{center}
NGC1851\\
\end{center}
\includegraphics[width=0.24\textwidth]{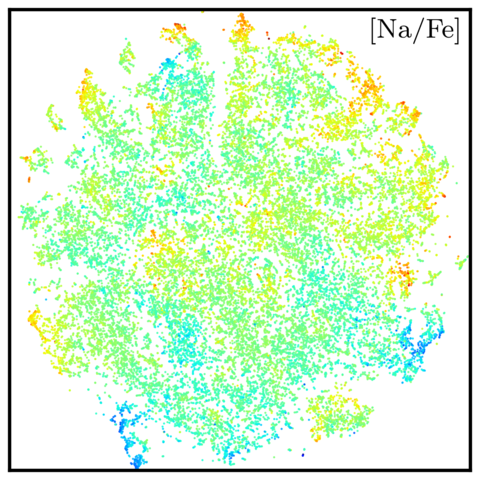}
\includegraphics[width=0.24\textwidth]{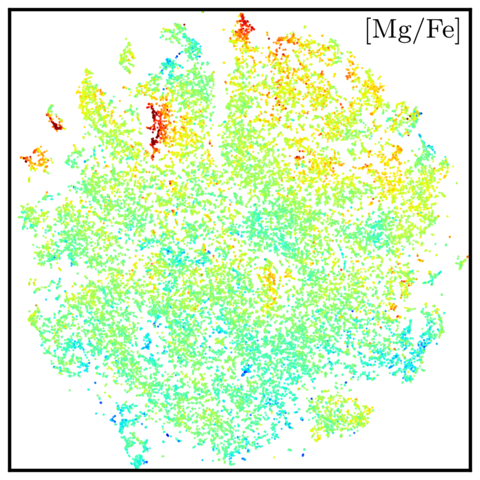}
\includegraphics[width=0.24\textwidth]{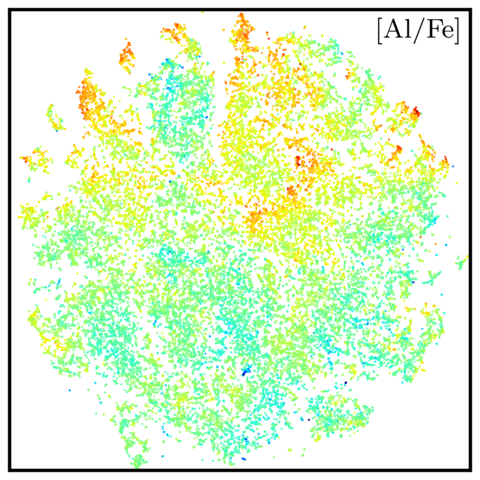}
\includegraphics[width=0.24\textwidth]{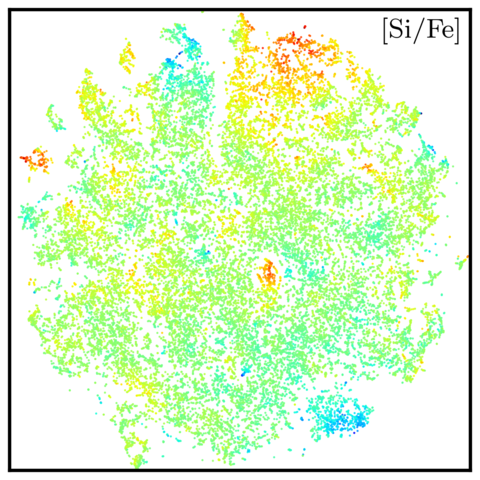}\\
\includegraphics[width=0.24\textwidth]{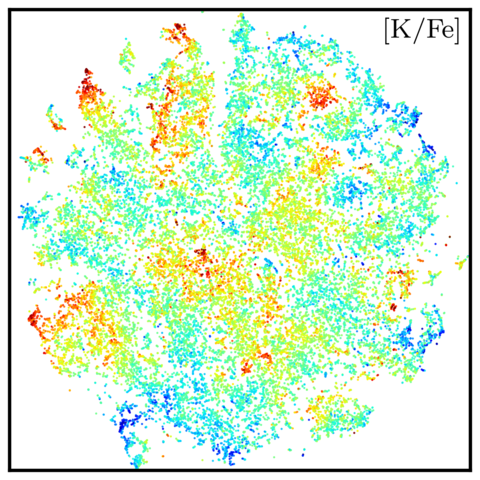}
\includegraphics[width=0.24\textwidth]{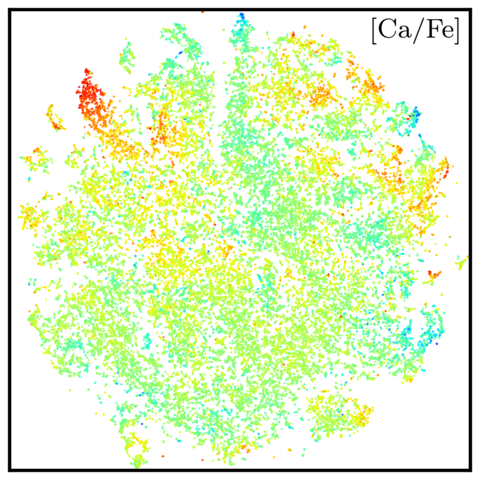}
\includegraphics[width=0.24\textwidth]{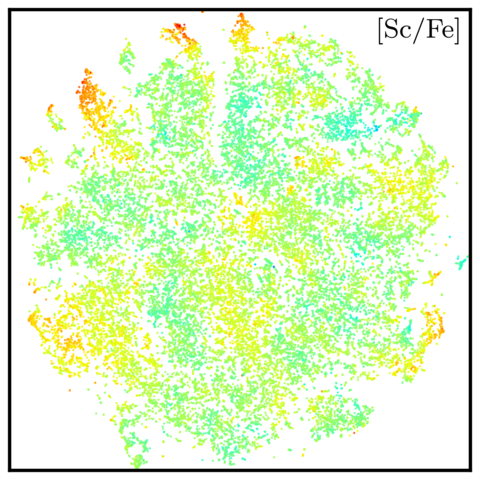}
\includegraphics[width=0.24\textwidth]{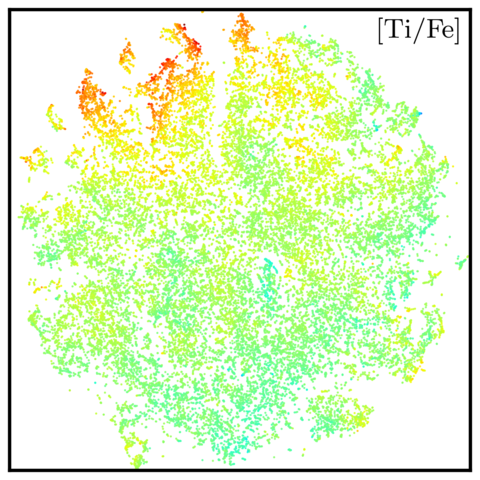}\\
\includegraphics[width=0.24\textwidth]{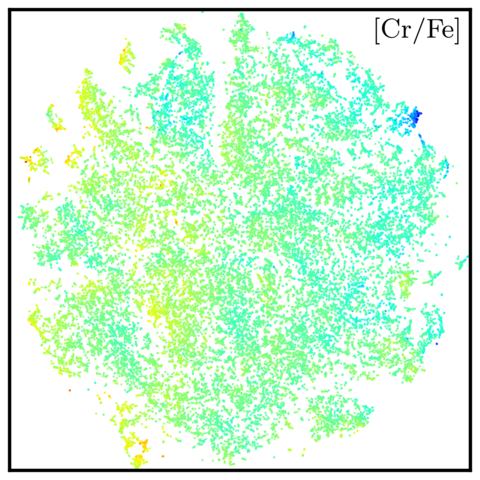}
\includegraphics[width=0.24\textwidth]{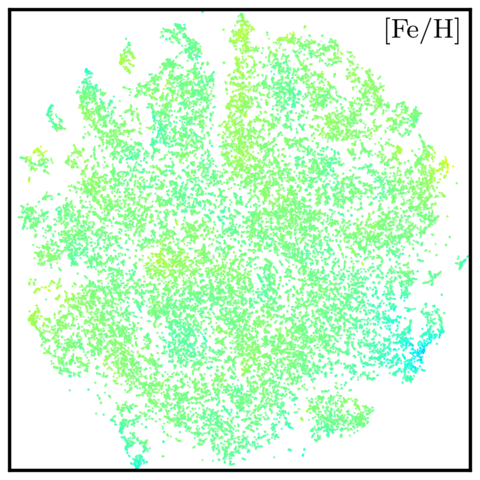}
\includegraphics[width=0.24\textwidth]{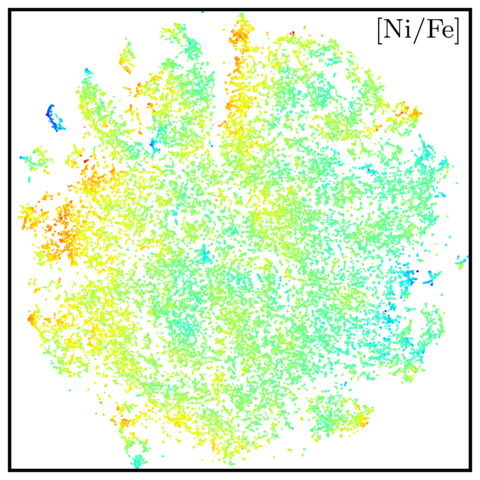}
\includegraphics[width=0.24\textwidth]{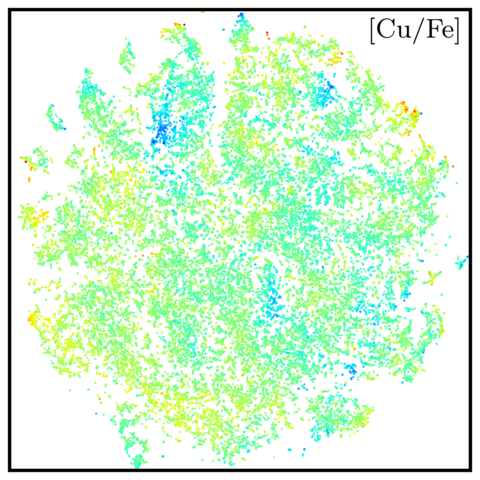}\\
\includegraphics[width=0.24\textwidth]{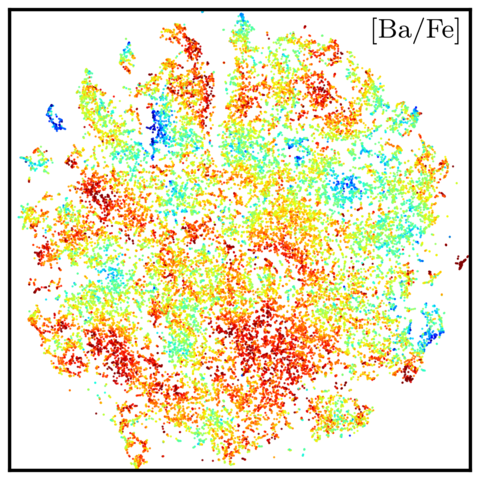}
\includegraphics[width=0.24\textwidth]{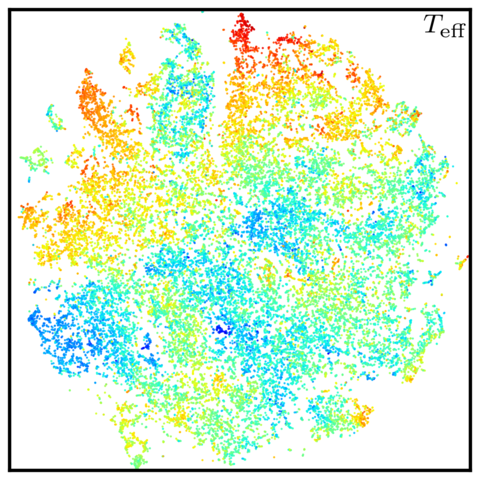}
\includegraphics[width=0.24\textwidth]{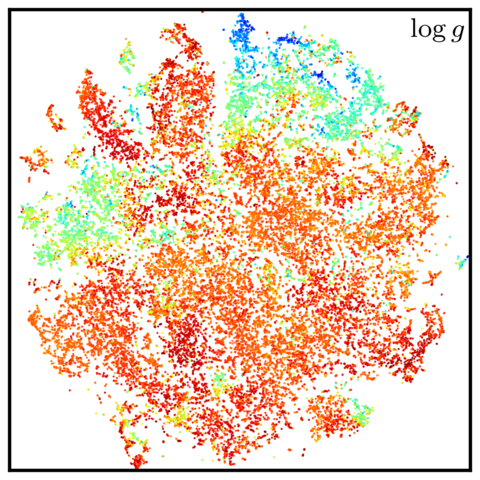}
\includegraphics[width=0.24\textwidth]{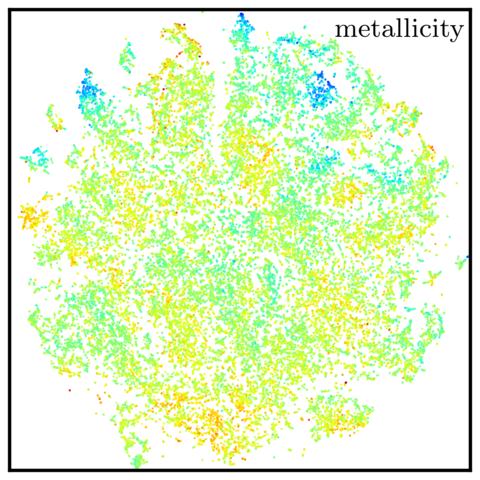}\\
\includegraphics[width=0.24\textwidth]{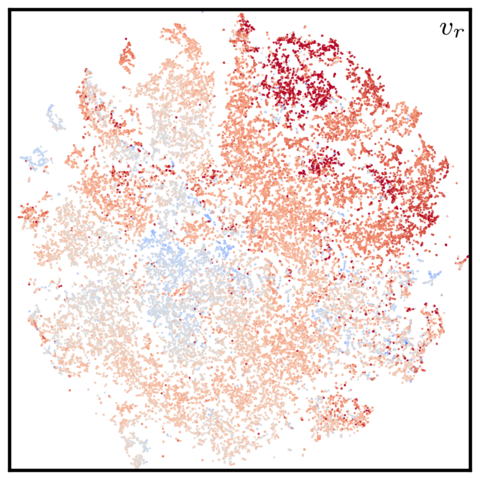}
\includegraphics[width=0.24\textwidth]{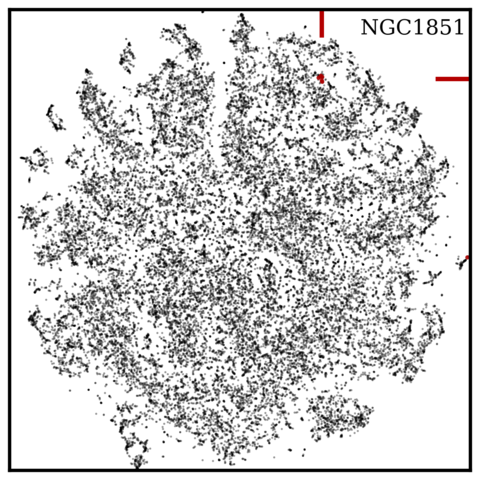}\hspace{0.5em}
\includegraphics[width=0.06\textwidth]{colorbar_abund.png}\hspace{2em}
\includegraphics[width=0.06\textwidth]{colorbar_teff.png}\hspace{2em}
\includegraphics[width=0.06\textwidth]{colorbar_logg.png}\hspace{2em}
\includegraphics[width=0.06\textwidth]{colorbar_feh.png}\hspace{2em}
\includegraphics[width=0.06\textwidth]{colorbar_rv.png}
\caption{t-SNE projection of 33,882 stars in a 35$^\circ$ radius around NGC1851. Abundances of 13 elements used to create the projection are colour-coded. $T_\mathrm{eff}$, $\log g$, metallicity and radial velocity colour-codes are also plotted. The last panel shows the stars that belong to the cluster in red and field stars in grey. 6 out of 7 NGC1851 stars lie in a tight group in the top-right part of the map (marked with dashes at the edge of the plot).}
\label{fig:proj_NGC1851}
\end{figure*}

\begin{figure*}
\begin{center}
M67\\
\end{center}
\includegraphics[width=0.24\textwidth]{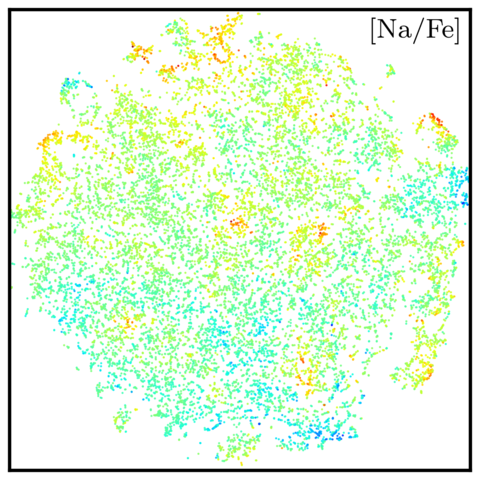}
\includegraphics[width=0.24\textwidth]{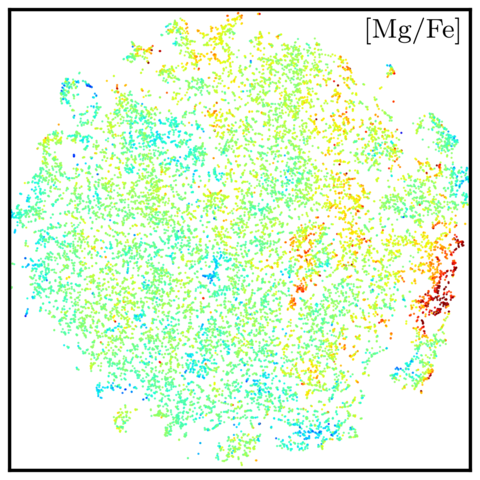}
\includegraphics[width=0.24\textwidth]{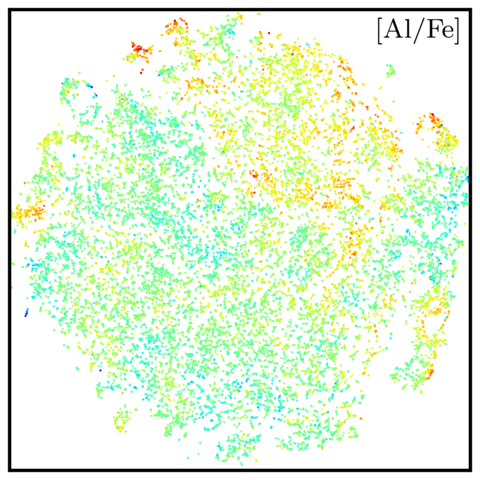}
\includegraphics[width=0.24\textwidth]{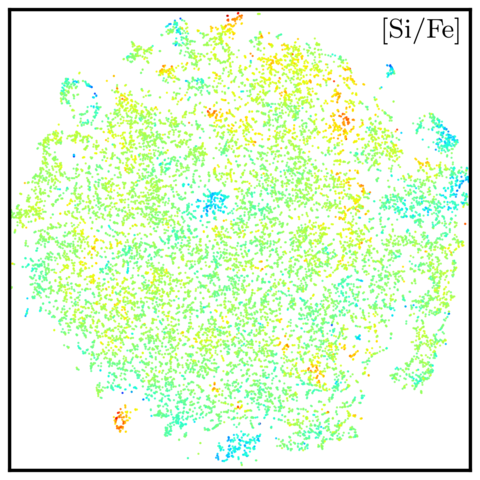}\\
\includegraphics[width=0.24\textwidth]{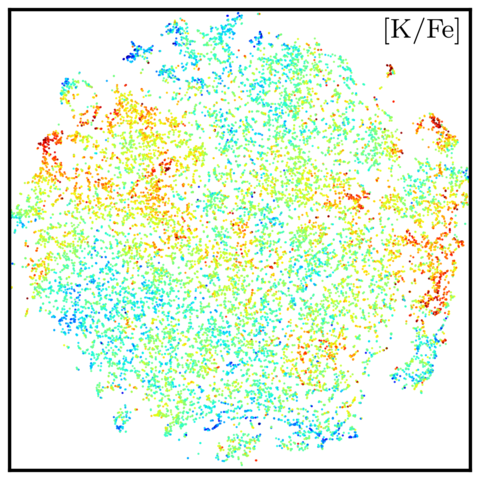}
\includegraphics[width=0.24\textwidth]{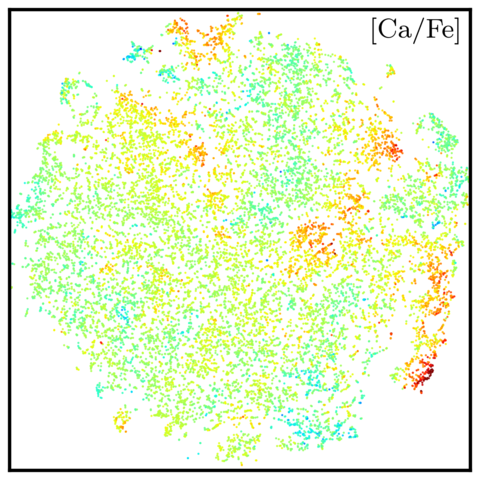}
\includegraphics[width=0.24\textwidth]{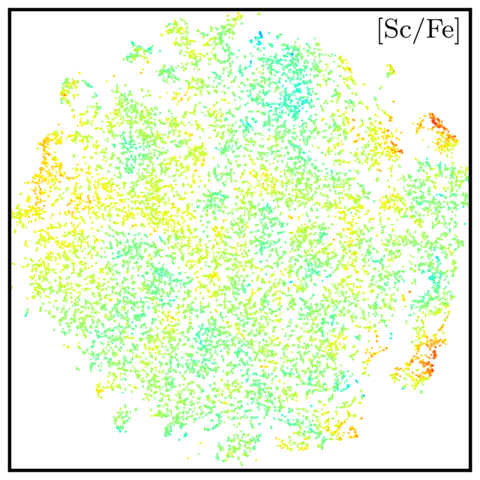}
\includegraphics[width=0.24\textwidth]{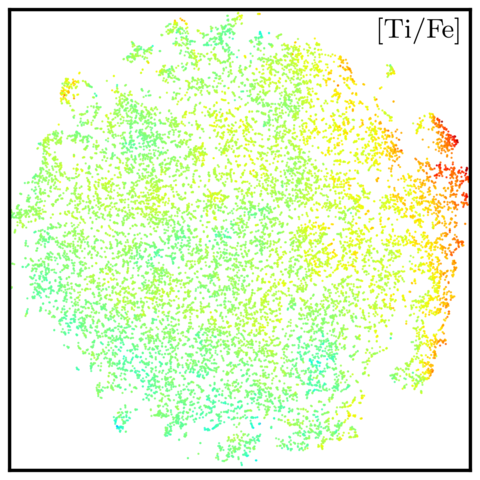}\\
\includegraphics[width=0.24\textwidth]{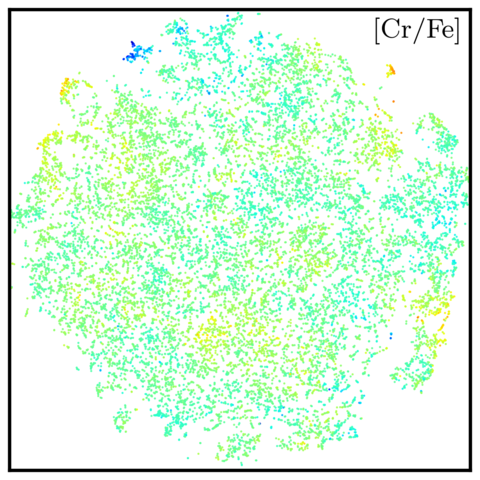}
\includegraphics[width=0.24\textwidth]{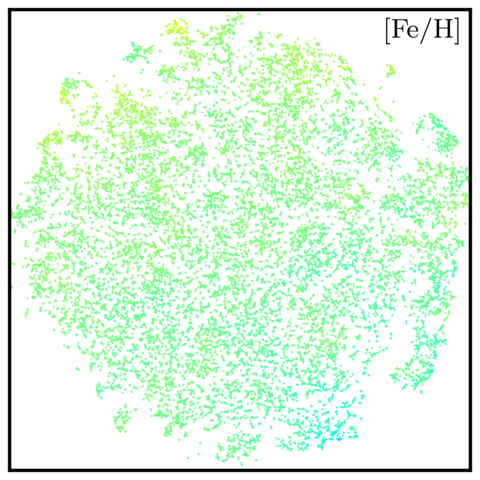}
\includegraphics[width=0.24\textwidth]{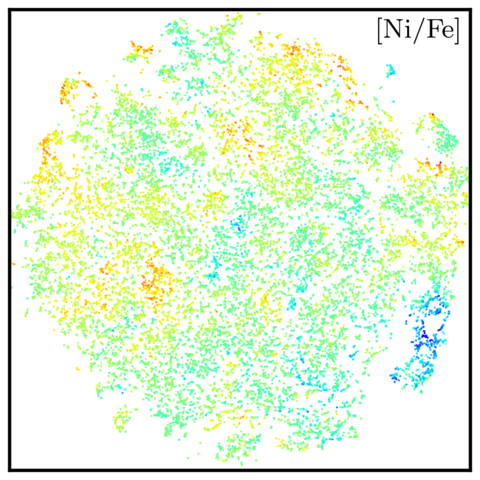}
\includegraphics[width=0.24\textwidth]{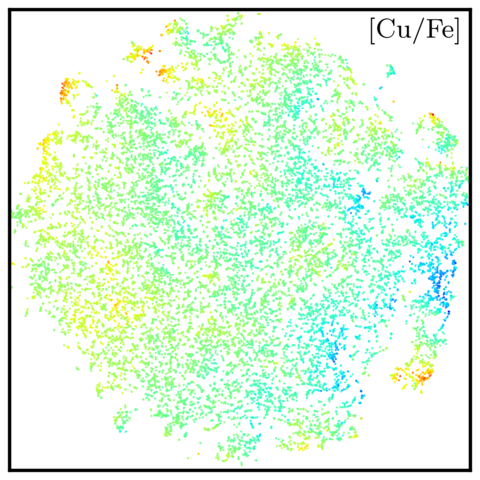}\\
\includegraphics[width=0.24\textwidth]{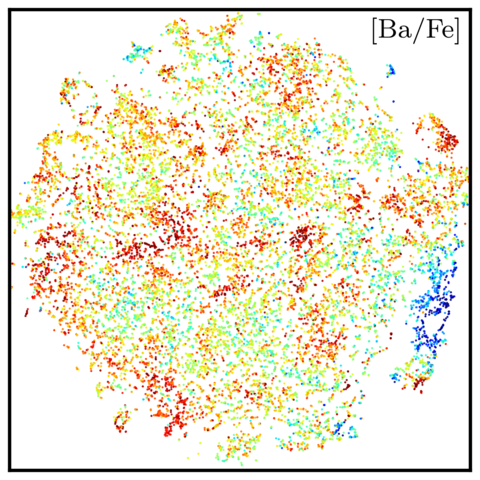}
\includegraphics[width=0.24\textwidth]{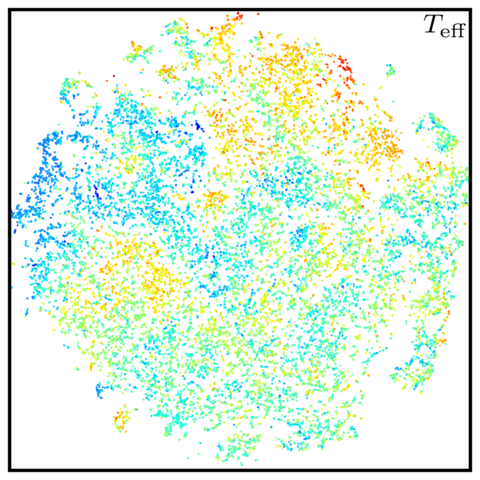}
\includegraphics[width=0.24\textwidth]{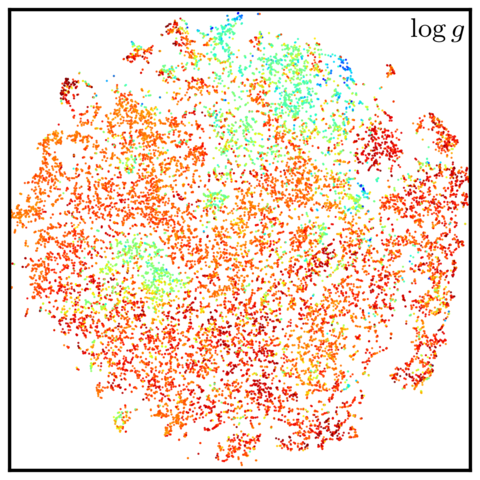}
\includegraphics[width=0.24\textwidth]{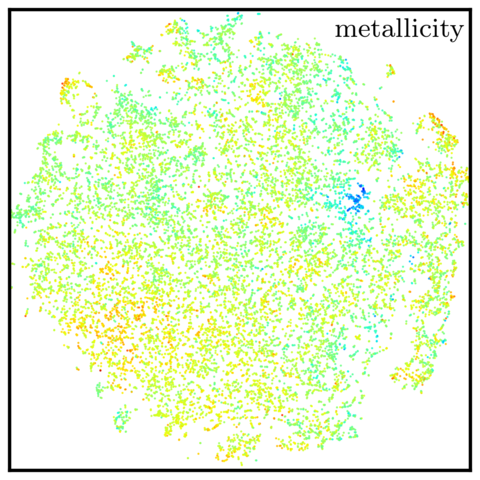}\\
\includegraphics[width=0.24\textwidth]{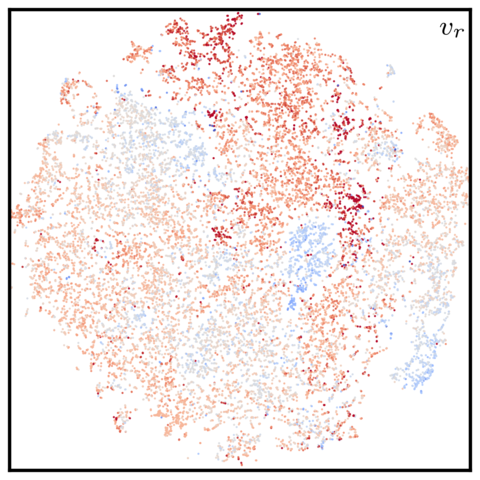}
\includegraphics[width=0.24\textwidth]{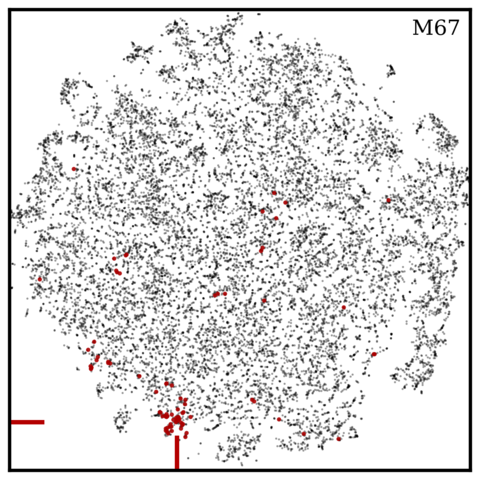}\hspace{0.5em}
\includegraphics[width=0.06\textwidth]{colorbar_abund.png}\hspace{2em}
\includegraphics[width=0.06\textwidth]{colorbar_teff.png}\hspace{2em}
\includegraphics[width=0.06\textwidth]{colorbar_logg.png}\hspace{2em}
\includegraphics[width=0.06\textwidth]{colorbar_feh.png}\hspace{2em}
\includegraphics[width=0.06\textwidth]{colorbar_rv.png}
\caption{t-SNE projection of 25,648 stars in a 45$^\circ$ radius around M67. Abundances of 13 elements used to create the projection are colour-coded. $T_\mathrm{eff}$, $\log g$, metallicity and radial velocity colour-codes are also plotted. The last panel shows the stars that belong to the cluster in red and field stars in grey. 71 of the 113 M67 stars lie in the biggest group in the bottom part of the map (marked with dashes at the edge of the plot).}
\label{fig:proj_M67}
\end{figure*}

\begin{figure*}
\begin{center}
47~Tuc\\
\end{center}
\includegraphics[width=0.24\textwidth]{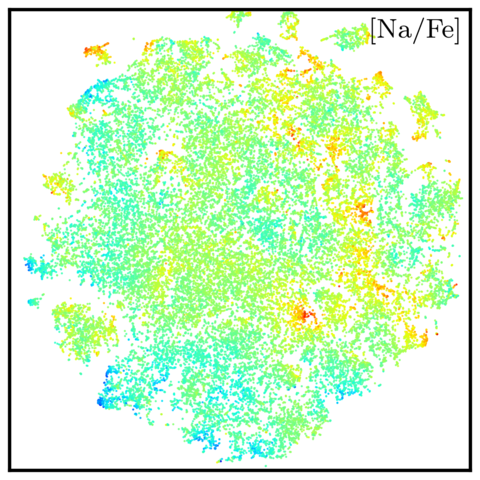}
\includegraphics[width=0.24\textwidth]{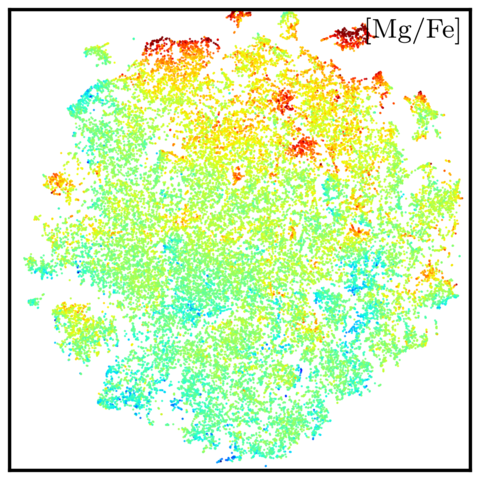}
\includegraphics[width=0.24\textwidth]{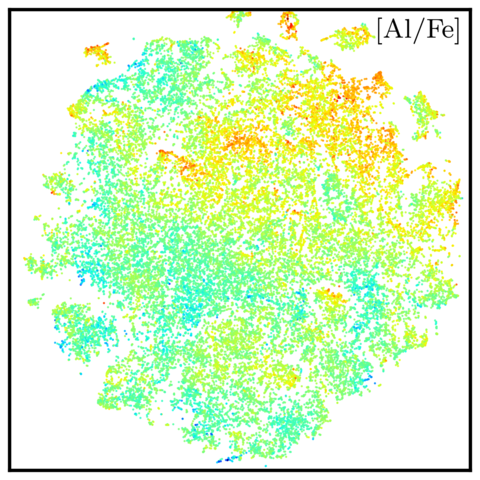}
\includegraphics[width=0.24\textwidth]{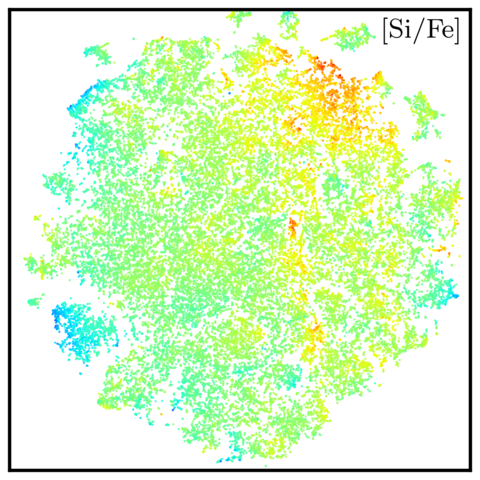}\\
\includegraphics[width=0.24\textwidth]{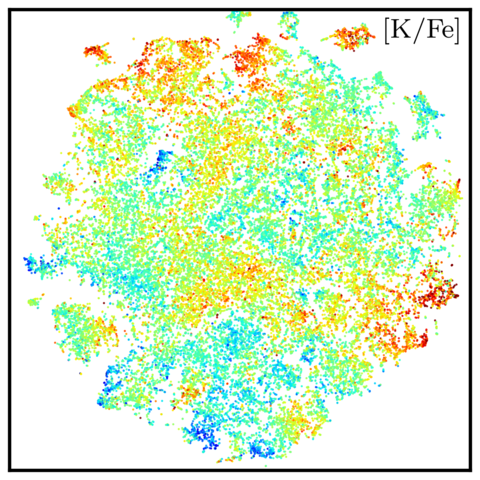}
\includegraphics[width=0.24\textwidth]{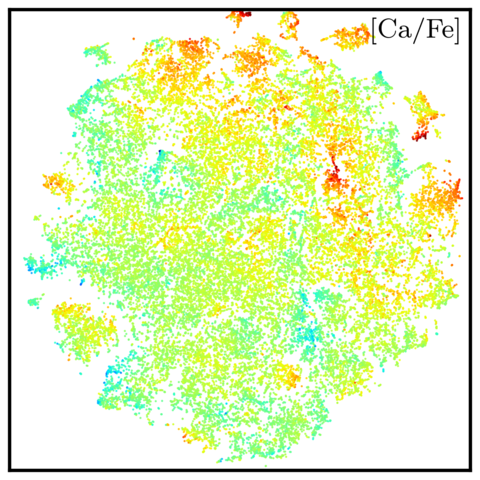}
\includegraphics[width=0.24\textwidth]{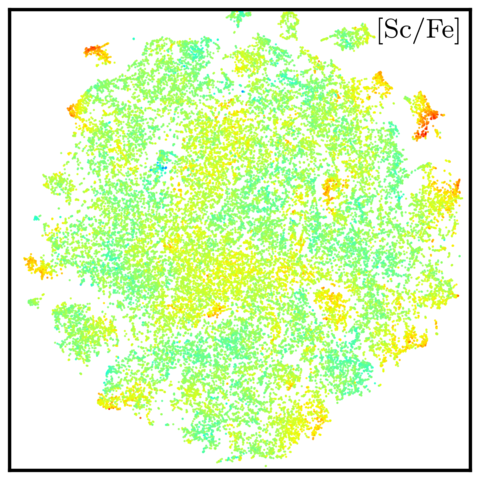}
\includegraphics[width=0.24\textwidth]{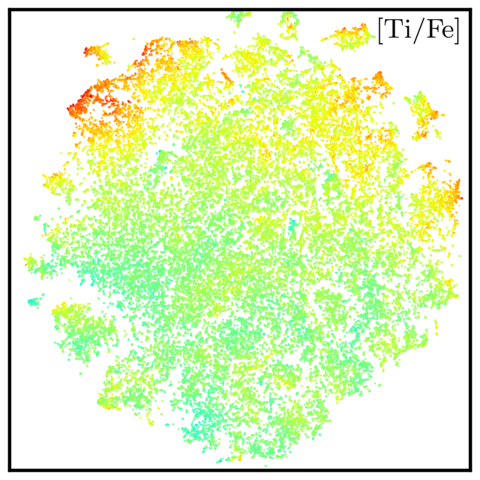}\\
\includegraphics[width=0.24\textwidth]{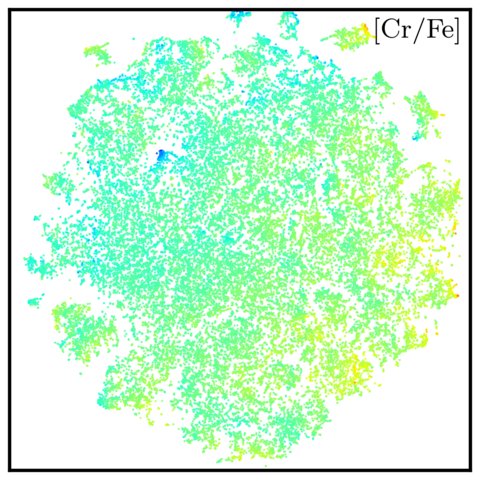}
\includegraphics[width=0.24\textwidth]{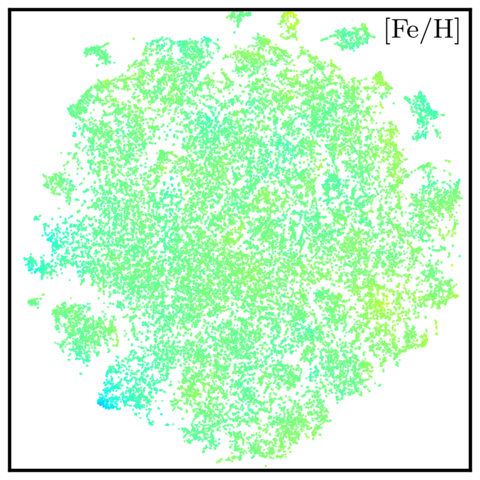}
\includegraphics[width=0.24\textwidth]{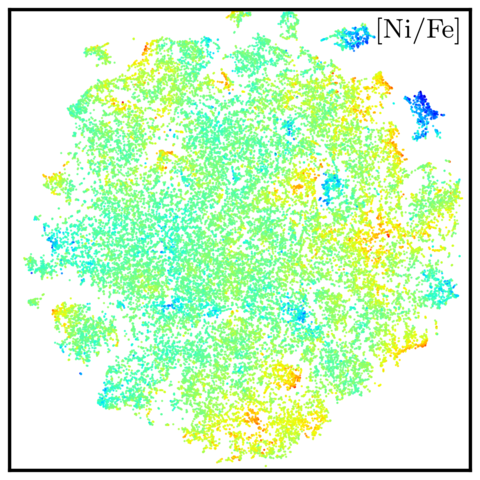}
\includegraphics[width=0.24\textwidth]{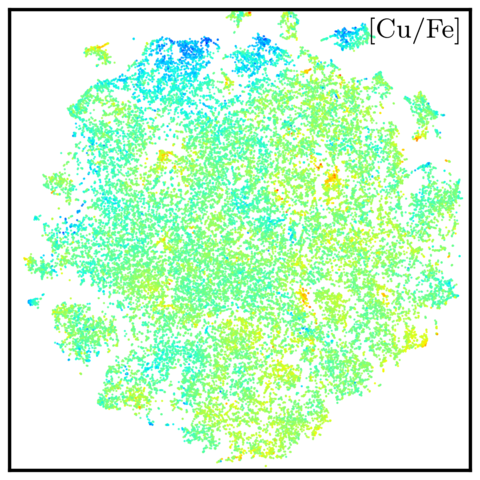}\\
\includegraphics[width=0.24\textwidth]{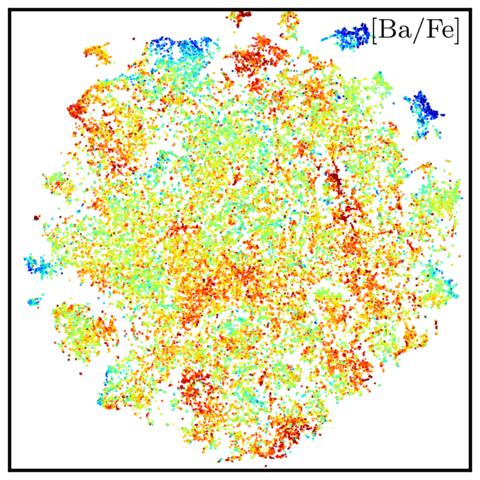}
\includegraphics[width=0.24\textwidth]{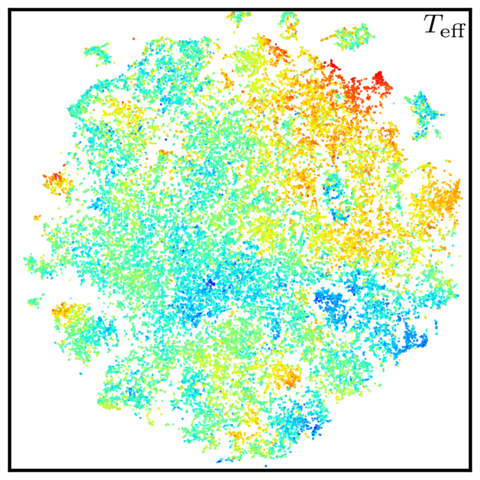}
\includegraphics[width=0.24\textwidth]{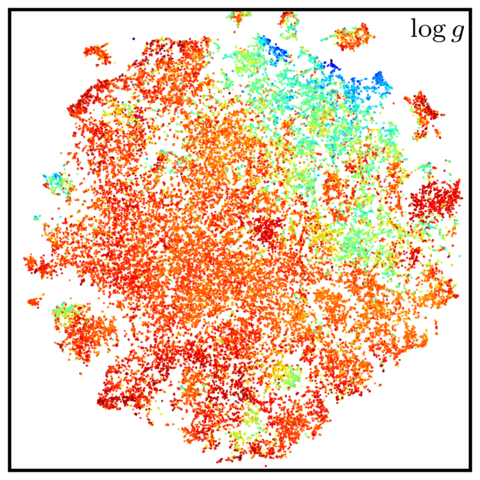}
\includegraphics[width=0.24\textwidth]{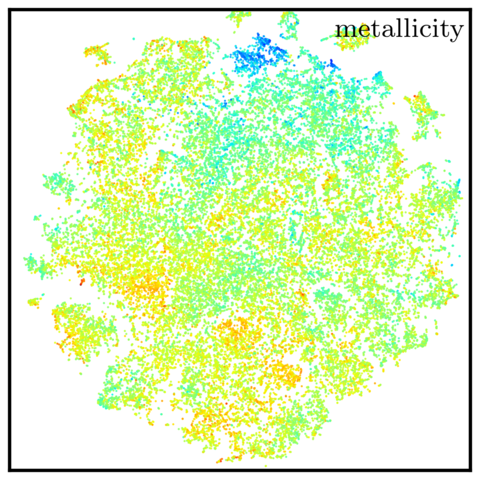}\\
\includegraphics[width=0.24\textwidth]{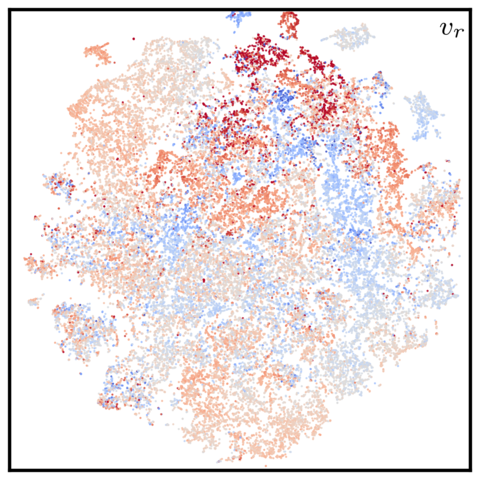}
\includegraphics[width=0.24\textwidth]{47Tuc_cluster_index_2.png}\hspace{0.5em}
\includegraphics[width=0.06\textwidth]{colorbar_abund.png}\hspace{2em}
\includegraphics[width=0.06\textwidth]{colorbar_teff.png}\hspace{2em}
\includegraphics[width=0.06\textwidth]{colorbar_logg.png}\hspace{2em}
\includegraphics[width=0.06\textwidth]{colorbar_feh.png}\hspace{2em}
\includegraphics[width=0.06\textwidth]{colorbar_rv.png}
\caption{t-SNE projection of 44,037 stars in a 35$^\circ$ radius around 47~Tuc. Abundances of 13 elements used to create the projection are colour-coded. $T_\mathrm{eff}$, $\log g$, metallicity and radial velocity colour-codes are also plotted. The last panel shows the stars that belong to the cluster in red and field stars in grey. Most 47~Tuc stars do not lie in a single group. The bigest group contains only 21 stars out of 90 47~Tuc stars.}
\label{fig:proj_47Tuc}
\end{figure*}

\begin{figure*}
\begin{center}
NGC2516\\
\end{center}
\includegraphics[width=0.24\textwidth]{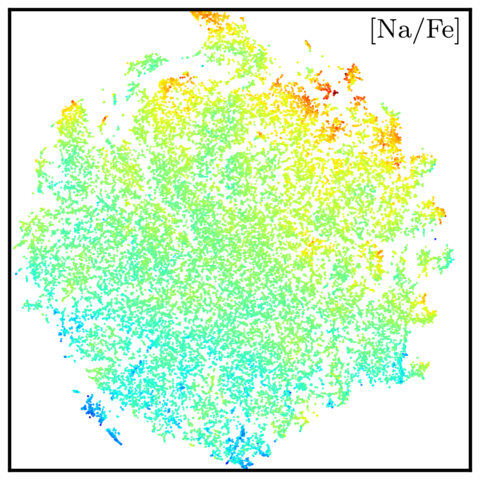}
\includegraphics[width=0.24\textwidth]{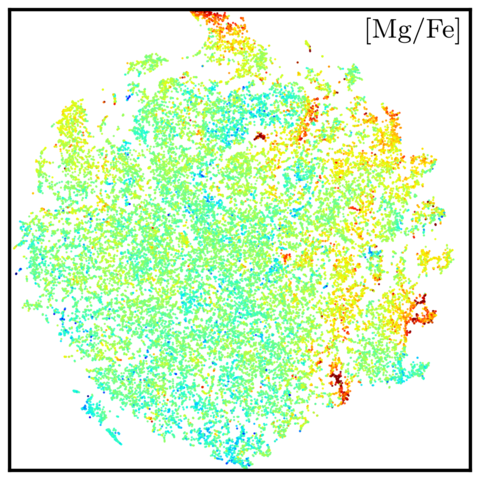}
\includegraphics[width=0.24\textwidth]{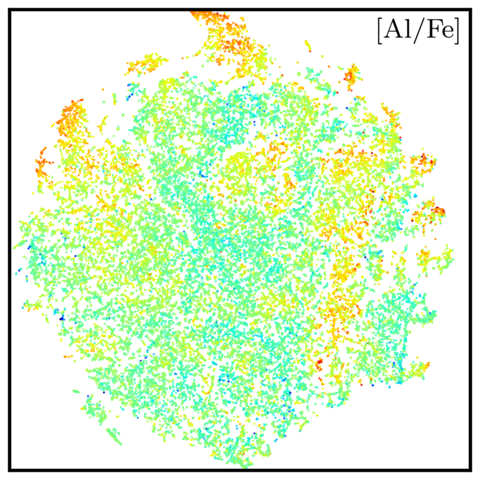}
\includegraphics[width=0.24\textwidth]{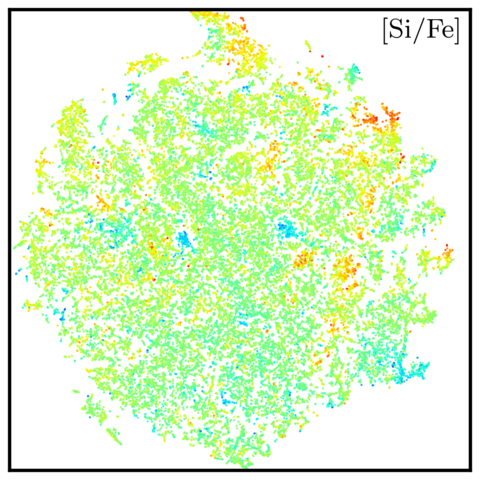}\\
\includegraphics[width=0.24\textwidth]{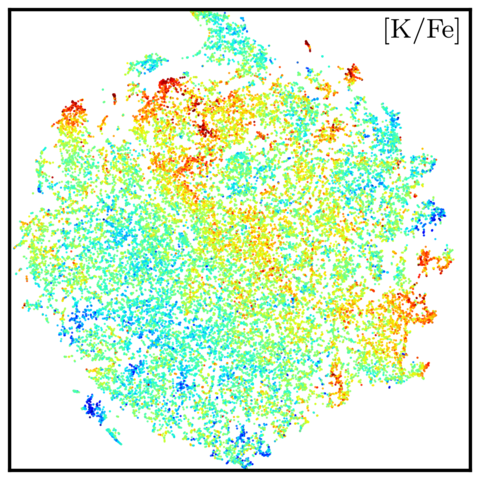}
\includegraphics[width=0.24\textwidth]{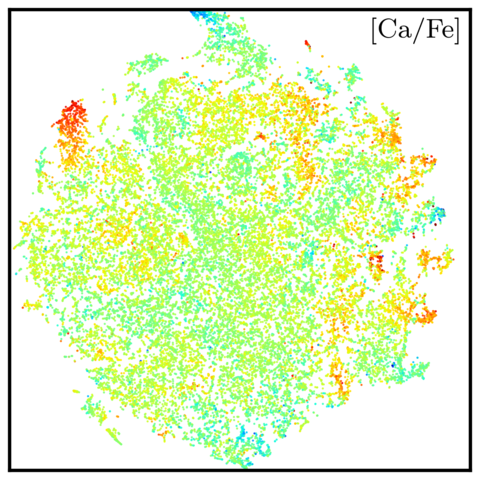}
\includegraphics[width=0.24\textwidth]{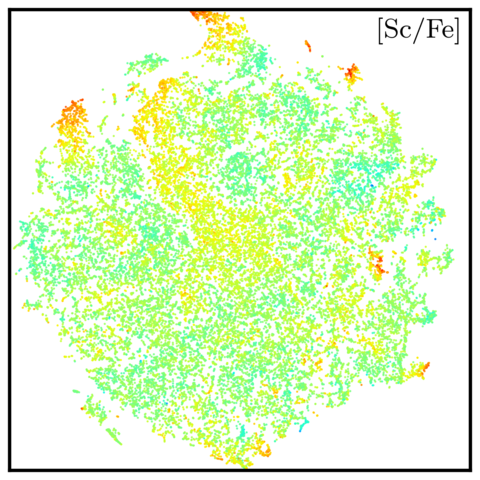}
\includegraphics[width=0.24\textwidth]{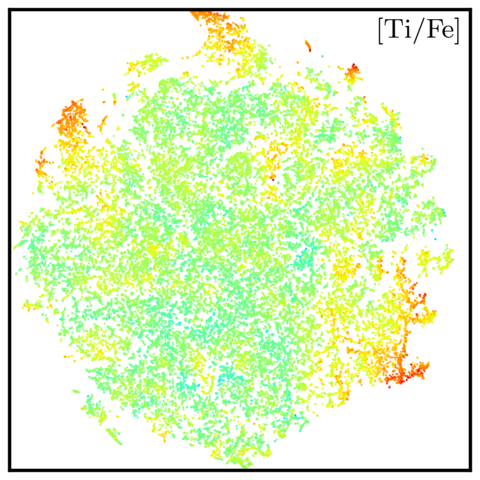}\\
\includegraphics[width=0.24\textwidth]{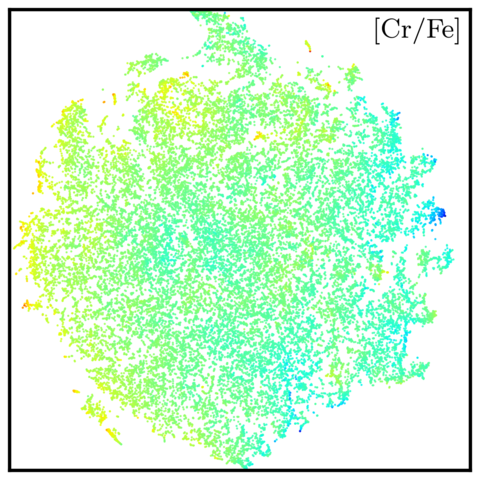}
\includegraphics[width=0.24\textwidth]{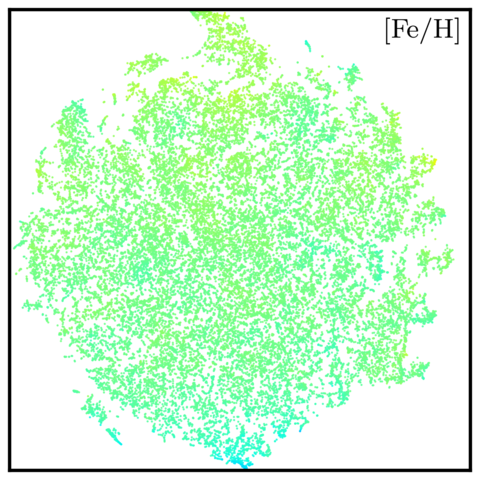}
\includegraphics[width=0.24\textwidth]{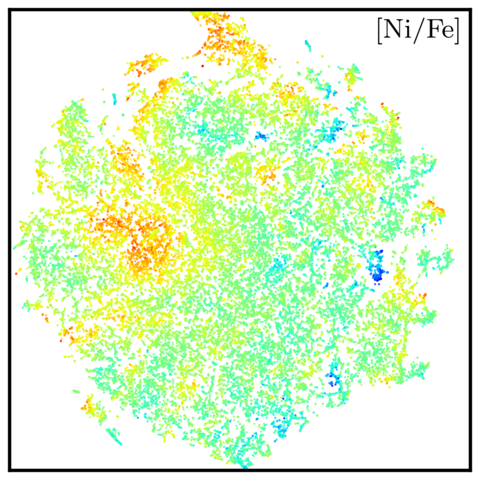}
\includegraphics[width=0.24\textwidth]{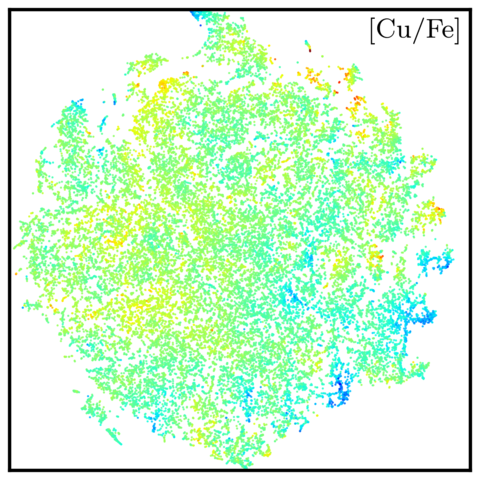}\\
\includegraphics[width=0.24\textwidth]{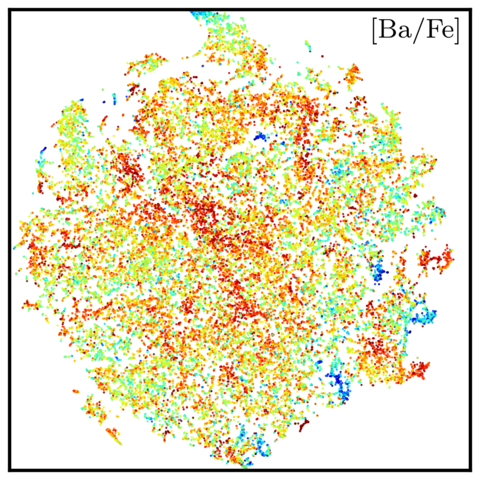}
\includegraphics[width=0.24\textwidth]{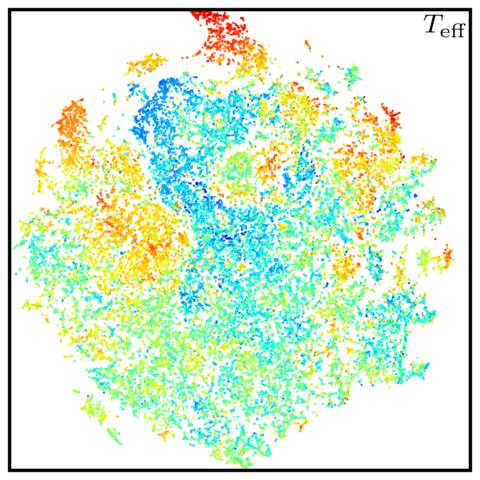}
\includegraphics[width=0.24\textwidth]{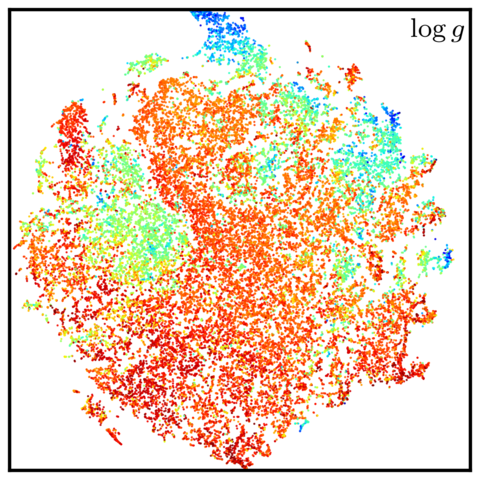}
\includegraphics[width=0.24\textwidth]{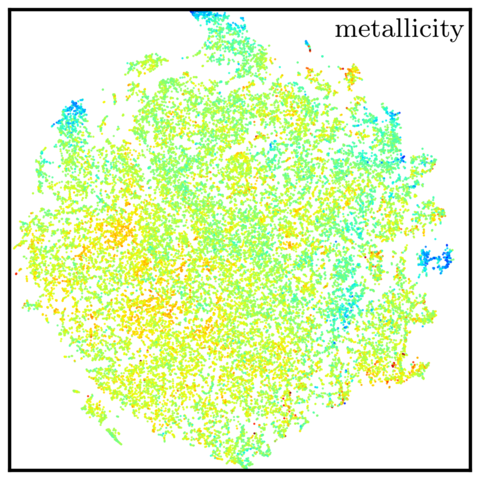}\\
\includegraphics[width=0.24\textwidth]{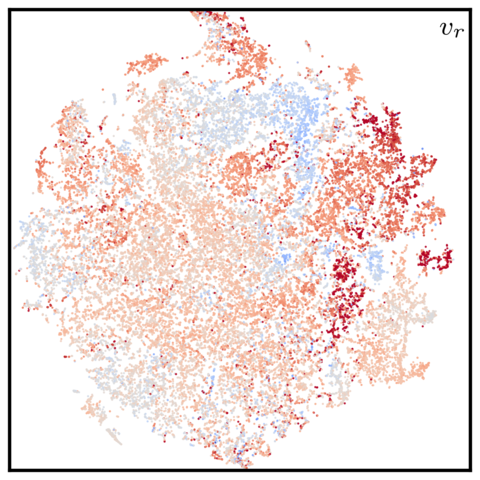}
\includegraphics[width=0.24\textwidth]{NGC2516_cluster_index_3.png}\hspace{0.5em}
\includegraphics[width=0.06\textwidth]{colorbar_abund.png}\hspace{2em}
\includegraphics[width=0.06\textwidth]{colorbar_teff.png}\hspace{2em}
\includegraphics[width=0.06\textwidth]{colorbar_logg.png}\hspace{2em}
\includegraphics[width=0.06\textwidth]{colorbar_feh.png}\hspace{2em}
\includegraphics[width=0.06\textwidth]{colorbar_rv.png}
\caption{t-SNE projection of 41,106 stars in a 30$^\circ$ radius around NGC2516. Abundances of 13 elements used to create the projection are colour-coded. $T_\mathrm{eff}$, $\log g$, metallicity and radial velocity colour-codes are also plotted. The last panel shows the stars that belong to the cluster in red and field stars in grey. We only matched 3 stars to NGC2516 of which none has a high membership probability in \citet{jeffries01}. All stars lie in a middle region of the map where stars that are hardest to classify lie (marked with dashes at the edge of the plot). }
\label{fig:proj_NGC2516}
\end{figure*}

%\end{comment}

\clearpage
\section{Cluster membership}
\label{sec:membership}
For clusters targeted in the pilot survey the observed stars were preselected based on their proper motions (47 Tuc, NGC 288, NGC 362, M 67) position on the HR diagram (47 Tuc, NGC 288, $\omega$ Cen, NGC 362, M 67) and previous spectroscopic observations (NGC 288, NGC 1851, M 30, $\omega$ Cen). See \citet{martell17} for details.

\begin{figure*}
\includegraphics[width=0.87\textwidth]{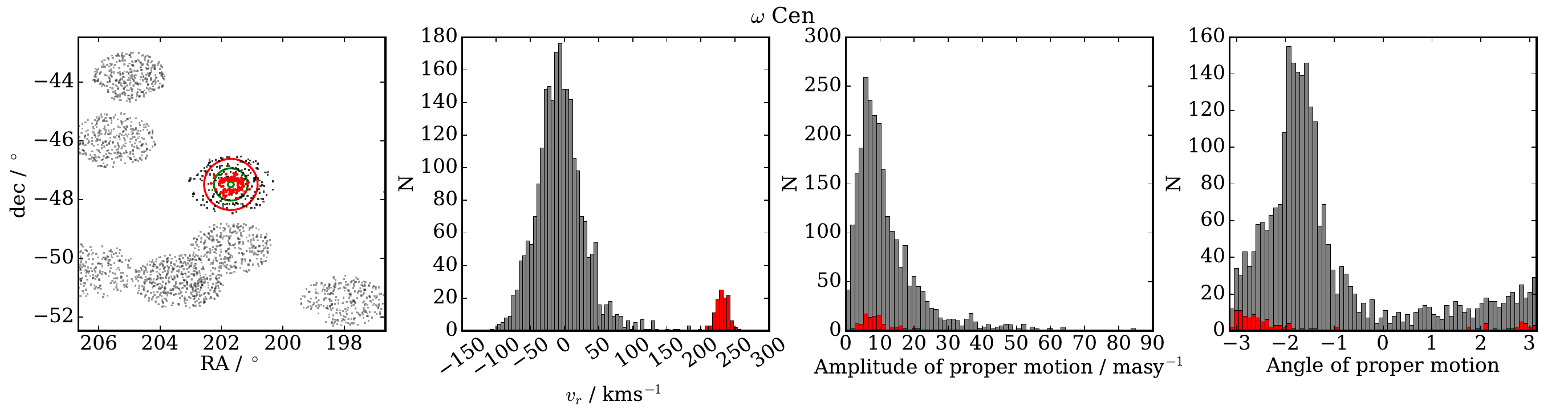}
\includegraphics[width=0.87\textwidth]{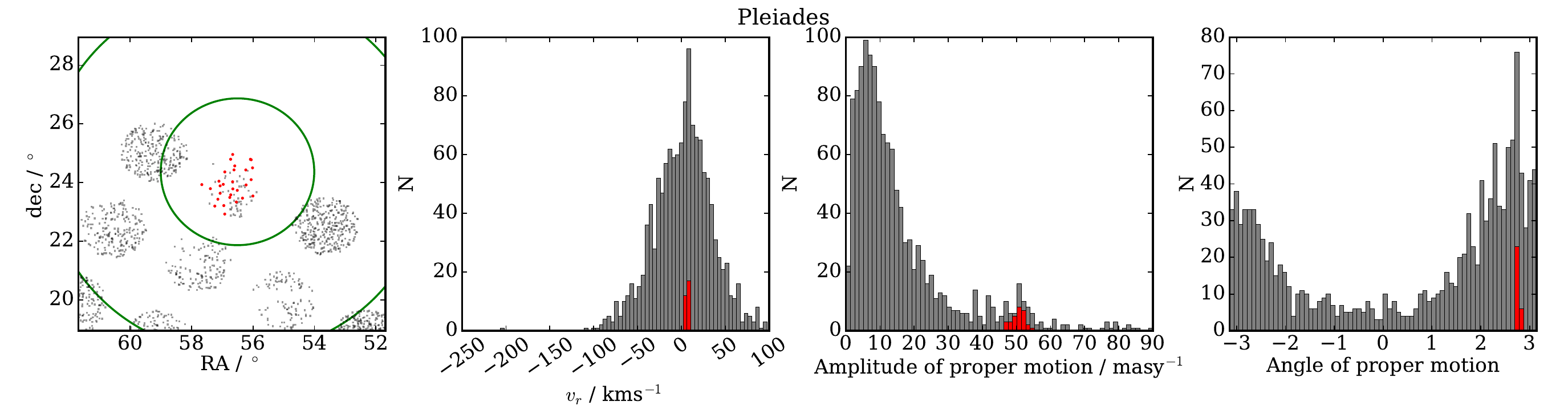}
\includegraphics[width=0.87\textwidth]{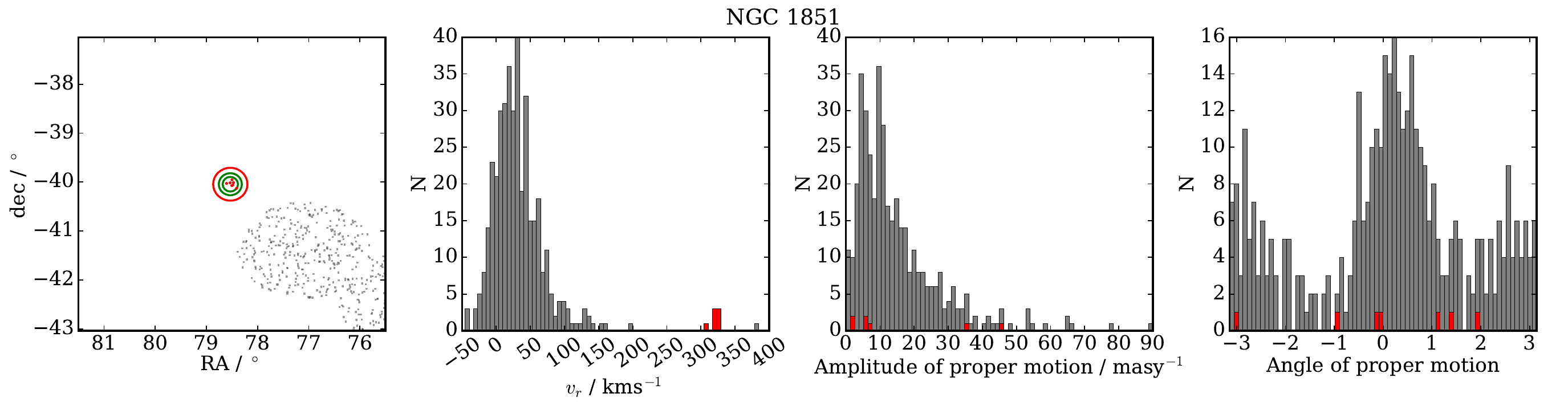}
\includegraphics[width=0.87\textwidth]{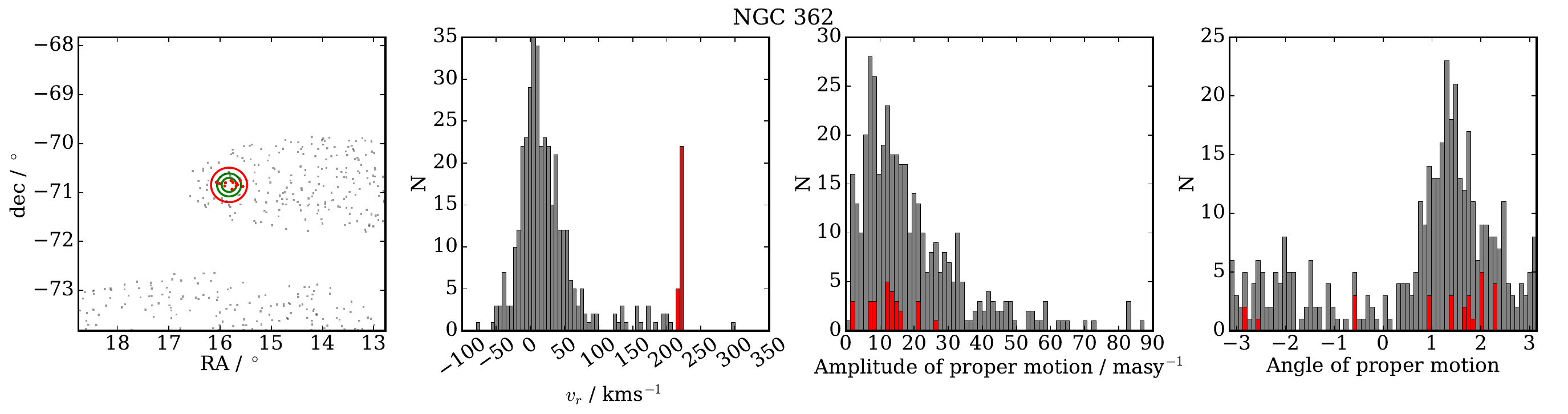}
\includegraphics[width=0.87\textwidth]{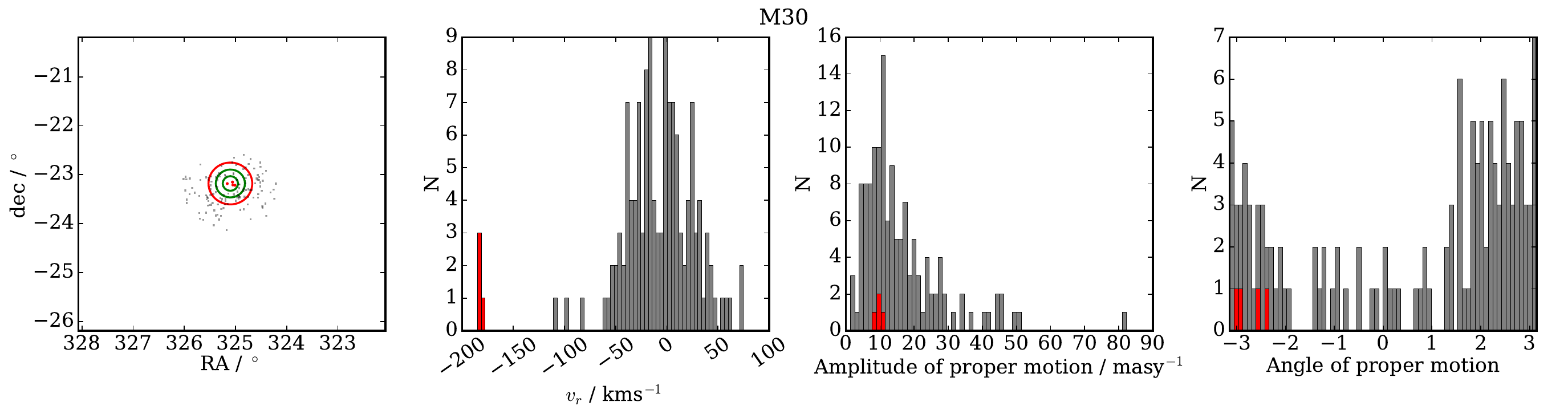}
\caption{Top to bottom: Basic information used to determine membership for 6 clusters where we did not use sources from the literature. Left to right: Position of observed stars (gray) and members (red) in a small region around each cluster centre. Red circle shows the maximum radius at which the mebers can be. Green circles show values $r_1$ (size of the cluster centre) and $r_2$ (size of the cluster) from \citet{kar13}. Second panel shows distribution of the radial velocities of all stars in the plotted field (gray) and cluster members (red). Last two panels show amplitude and angle of proper motion for all stars in the plotted field (gray) and cluster members (red). There is no panel for NGC 288, because all observed stars made the cut and there are no field stars within 5$^\circ$ of the cluster.}
\label{fig:membership}
\end{figure*}

Despite the preselection of observed stars we did a further analisys of possible members. Figure \ref{fig:membership} shows position, radial velocities and proper motions used to determine the memberships. Our conditions are simple cuts in position and radial velocity for globular clusters and additional cuts in amplitude and angle of proper motion for Pleiades. Radii $r_1$ and $r_2$ \citep{kar13} are used for the radius of the core and radius of the cluster, respectively. For globular clusters we consider all stars within $1.5r_2$, as there are members expected to be observed outside $r_2$. Any field stars are then discarded by making a cut in radial velocity. Because all our globular cluster have radial velocities that are distinct from the radial velocity of nearby field stars, we do not expect any misidentified members. This is not true for Pleiades, so we use $r_1$ as the position criterium and make additional cuts in proper motion. We might miss some members this way, but should keep the selection clear of any field stars. The criteria are conservative, as any misidentified members have more impact on the success of chemical tagging than possible missed members. The following list gives the membership criteria for each cluster:
\begin{minipage}{\textwidth}  
\begin{itemize}
\item $\omega$~Cen: All stars within $1.5r_2$ and $\left(200\ \mathrm{kms^{-1}}<v_r<260\ \mathrm{kms^{-1}}\right)$.
\item Pleiades: All stars within $r_1$ and $\left(5.0\ \mathrm{kms^{-1}}<v_r<8.0\ \mathrm{kms^{-1}}\right)$ and $\left(45\ \mathrm{masy^{-1}}<\mu<55\ \mathrm{masy^{-1}}\right)$ and $\left(2.70<\phi_\mu<2.80\right)$. $\phi_\mu$ is the angle of proper motion expressed in radians in the celestial coordinate system.
\item NGC~1851: All stars within $1.5r_2$ and $\left(300\ \mathrm{kms^{-1}}<v_r<340\ \mathrm{kms^{-1}}\right)$.
\item NGC 362: All stars within $1.5r_2$ and $\left(210\ \mathrm{kms^{-1}}<v_r<230\ \mathrm{kms^{-1}}\right)$.
\item NGC 288: All stars within $1.5r_2$ and $\left(-55\ \mathrm{kms^{-1}}<v_r<-35\ \mathrm{kms^{-1}}\right)$.
\item M 30: All stars within $1.5r_2$ and $\left(-175\ \mathrm{kms^{-1}}<v_r<-185\ \mathrm{kms^{-1}}\right)$.
\end{itemize}
\end{minipage}

\clearpage

%%%%%%%%%%%%%%%%%%%%%%%%%%%%%%%%%%%%%%%%%%%%%%%%%%

\clearpage
% Don't change these lines
\bsp% typesetting comment
\label{lastpage}
\end{document}